%% file: main.tex
\documentclass[aps,amsmath,superscriptaddress,nofootinbib,longbibliography,prx,twocolumn]{revtex4-2}
\usepackage{graphicx}
\usepackage{transparent}
\usepackage{listofitems}
\usepackage{dcolumn} 
\usepackage[mode=image|tex]{standalone}
\usepackage{relsize}
\usepackage{bbold}
\usepackage{anyfontsize}
\usepackage{xfrac}
\usepackage{bm}
\usepackage[utf8]{inputenc}
\usepackage{amsmath}
\usepackage{amsthm}
\usepackage{hyperref}
\usepackage[dvipsnames]{xcolor}
\hypersetup{
colorlinks = true,
linkcolor = [rgb]{0.12, 0.20, 0.60}, 
citecolor = [rgb]{0.13,0.55,0.13},
urlcolor = [rgb]{0.25, 0.41, 0.88}}
\usepackage{physics}
\usepackage{mymacros}
\usepackage{natbib}
\usepackage{amsfonts}
\usepackage{tikz,pgf, tikz-3dplot}
\usepackage{tabularx,stackengine}
\usepackage{soul}
\usepackage[normalem]{ulem}
\usetikzlibrary{decorations.markings,decorations.shapes,decorations.pathmorphing,svg.path, 3d}
\usetikzlibrary{chains}
\renewcommand{\selectlanguage}[1]{}

\newcommand{\tocignoredsubsection}[1]{%
  {%
    \def\addcontentsline##1##2##3{}%
    \subsection{#1}%
  }%
}

\newcommand{\Z}{\mathbb{Z}}
\renewcommand{\SS}{\mathcal{S}}
\newcommand{\hil}{\mathcal{H}}
\newcommand{\per}[1]{\pi_{(#1)}}

\definecolor{darkred}{HTML}{D40000}
\definecolor{lightblue}{HTML}{2a7fff}
\definecolor{darkblue}{HTML}{003399}

\begin{document}
\tikzset{arr/.append style={
        decoration={markings,
            mark= at position {#1} with {\arrow{>}} ,
        },
        postaction={decorate}
    }
}

\tikzset{midarr/.append style={
        decoration={markings,
            mark= at position .5 with {\arrow{>}} ,
        },
        postaction={decorate}
    }
}

\tikzset{%
  dots/.style args={#1per #2}{%
    line cap=round,
    dash pattern=on 0 off #2/#1
  }
}

\newcommand{\origami}[2]{
    \begin{scope}[shift={(0,0.866025404*#1)}]
       \foreach \x in {0,60,...,360}{
           \draw (0,0) -- (\x:#1);
           \draw (\x:#1) -- (\x+60:#1);
       }
       \draw[decorate,decoration={zigzag,segment length=3pt,amplitude=1pt},thin] (-56:#2) -- (-5:#2);
       \draw[decorate,decoration={coil,segment length=3pt,amplitude=1pt},thin] (60-5:#2) -- (5:#2);
       \draw[dotted] (65:#2) -- (115:#2);
       \draw[dashed,thin] (125:#2) -- (175:#2);
       \draw[dash dot] (182:#2) -- (238:#2);
    \end{scope}
    \begin{scope}[shift={(0,-0.866025404*#1)}]
       \foreach \x in {0,60,...,360}{
           \draw (0,0) -- (\x:#1);
           \draw (\x:#1) -- (\x+60:#1);
       }
       \draw[decorate,decoration={zigzag,segment length=3pt,amplitude=1pt},thin] (56:#2) -- (2:#2);
       \draw[decorate,decoration={coil,segment length=3pt,amplitude=1pt},thin] (-58:#2) -- (-2:#2);
       \draw[dotted] (-62:#2) -- (-118:#2);
       \draw[dashed,thin] (-122:#2) -- (-178:#2);
       \draw[dash dot] (-182:#2) -- (-238:#2);
    \end{scope}
}

\newcommand{\pizza}[4]{
\vcenter{\hbox{\begin{tikzpicture}[every node/.style={font={\scriptsize}}]
    \def\r{#1};
    \draw (0,0) circle (\r);
    \draw (0,0) -- (90:\r);
    \draw (0,0) -- (210:\r);
    \draw (0,0) -- (330:\r);
    \node at (-90:\r*.6) {$#2$};
    \node at (+30:\r*.6) {$#3$};
    \node at (180-30:\r*.6) {$#4$};
\end{tikzpicture}}}
}

\newcommand{\tpizza}[3]{
\vcenter{\hbox{\begin{tikzpicture}[every node/.style={font={\tiny}}]
    \def\r{.3};
    \draw (0,0) circle (\r);
    \draw (0,0) -- (-90:\r);
    \draw (0,0) -- (-210:\r);
    \draw (0,0) -- (-330:\r);
    \node at (90:\r/2) {$#1$};
    \node at (-180+45:\r/2) {$#2$};
    \node at (-45:\r/2) {$#3$};
\end{tikzpicture}
}}}

\title{
Probing chiral topological states with permutation defects
}
\author{Yarden Sheffer}
\email{yarden.sheffer@gmail.com}
\affiliation{Department of Condensed Matter Physics, Weizmann Institute of Science Rehovot 7610001, Israel}
\author{Ruihua Fan}
\affiliation{Department of Physics, University of California, Berkeley, CA 94720, USA}
\author{Ady Stern}
\affiliation{Department of Condensed Matter Physics, Weizmann Institute of Science Rehovot 7610001, Israel}
\author{Erez Berg}
\affiliation{Department of Condensed Matter Physics, Weizmann Institute of Science Rehovot 7610001, Israel}
\author{Shinsei Ryu}
\affiliation{Department of Physics, Princeton University, Princeton, New Jersey, 08544, USA}

\date{\today}
\begin{abstract}
     The hallmark of two-dimensional chiral topological phases is the existence of anomalous gapless modes at the spatial boundary. Yet, the manifestation of this edge anomaly within the bulk ground-state wavefunction itself remains only partially understood. In this work, we introduce a family of multipartite entanglement measures that probe chirality directly from the bulk wavefunction. Our construction involves applying different permutations between replicas of the ground state wavefunction in neighboring spatial regions, creating ``permutation defects'' at the boundaries between these regions. We provide general arguments for the robustness of these measures and develop a field-theoretical framework to compute them systematically. While the standard topological field theory prescription misses the chiral contribution, our method correctly identifies it as the chiral conformal field theory partition function on high-genus Riemann surfaces. This feature is a consequence of the bulk-edge correspondence, which dictates that any regularization of the theory at the permutation defects must introduce gapless boundary modes. We numerically verify our results with both free-fermion and strongly-interacting chiral topological states and find excellent agreement. Our results enable the extraction of the chiral central charge and the Hall conductance using a finite number of wavefunction replicas, making these quantities accessible to Monte-Carlo numerical techniques and noisy intermediate-scale quantum devices. 
\end{abstract}
\maketitle

{
\hypersetup{linkcolor=black}
\def\l@f@section{%
  \addpenalty{\@secpenalty}%
  \addvspace{0.2em}
}
\tableofcontents
}

\section{Introduction}

Two-dimensional gapped quantum many-body systems can host topologically ordered phases with robust edge modes and anyonic bulk excitations~\cite{xiaogangreview}. Such phases evade description by local order parameters. Instead, their defining signatures are encoded in global properties of the ground-state wavefunction, especially its entanglement structure~\cite{Zeng:2015pxf}. 

Entanglement measures for topologically ordered states are thus of interest for both fundamental and practical reasons. Fundamentally, such measures reveal essential topological properties of these phases, including bulk-boundary correspondence~\cite{Kitaev_Preskill_2006,li2008entanglement, qi:2012prl, siva_universal_2022}, the algebraic structure of anyons~\cite{Kitaev_Preskill_2006,Levin_Wen_2006,wen2016topological, wen2016edge, Shi_Kato_Kim_2020, Liu_2024, sheffer2025extracting}, and quantized response functions~\cite{kim_chiral_2022, Fan_2023_extracting}. Practically, the entanglement of the ground state has become a valuable tool in both numerical simulations~\cite{hastings2010measuring} and experiments on quantum devices~\cite{satzinger2021realizing}. Furthermore, since the low-energy physics of these phases is described by topological quantum field theories (TQFTs), they offer a simple playground for studying entanglement in general quantum field theories, a subject of broad interest across condensed matter and high-energy physics~\cite{nishioka2018entanglement, Witten_2018, casini_lectures_2023}.  

Among topologically ordered states, a class of particular importance is that of chiral states, which carry protected gapless edge modes characterized by a nonzero chiral central charge $c_-$. These states appear in most examples of fractional quantum Hall states~\cite{prange1987quantum,Nayak_2008}, as well as in many spin liquid states~\cite{kitaev2006anyons, broholm2020quantum, savary2016quantum}. Despite their ubiquity, such states are often challenging to study from the entanglement perspective. The presence of $c_-\neq0$ induces multiple complications, which manifest both in the bulk and boundary descriptions. In the bulk, a non-zero $c_-$ precludes the existence of a commuting projector parent Hamiltonian~\cite{kitaev2006anyons, kapustin2020thermal}. Practically, this means that calculations on such exactly-solvable models, which proved useful in the context of studying entanglement measures of topological order~\cite{Levin_Wen_2006,flammia_topological_2009,sheffer2025extracting}, cannot capture the full entanglement properties of chiral phases. In the boundary, the presence of a nonzero $c_-$ means that the boundary low-energy theory suffers from a gravitational anomaly: it cannot arise from any purely one-dimensional system~\cite{alvarez1984gravitational,hellerman2021quantum, Fan_2022}.

The challenge of identifying the topological data from entanglement measures has been addressed for both chiral and non-chiral states. The prime example of such a measure is the topological entanglement entropy~\cite{Kitaev_Preskill_2006, Levin_Wen_2006}. This measure extracts the ``total quantum dimension" of the phase, but does not carry direct information about the chirality. In the specific context of chiral states, the work of~\cite{kim_chiral_2022} presented the ``modular commutator," which extracts $c_-$ directly from the ground-state bulk entanglement. This measure (defined below) relies on the modular Hamiltonian $K=-\log\rho$, where $\rho$ is the reduced density matrix in some region. For both practical and theoretical applications, however, it is desirable to have a ``R\'enyi-like" quantity, which can be calculated using a finite number of replicas of the wavefunction. For example, this would enable the extraction of the chiral central charge in Monte-Carlo simulations and experiments on quantum devices, similar to the topological entanglement entropy~\cite{hastings2010measuring, satzinger2021realizing}. More recently, in Ref.~\cite{sheffer2025extracting}, some of us proposed a set of multipartite entanglement measures that extract information about topological spins. This measure, however, was calculated only for non-chiral states.

In this work, we construct a series of multipartite entanglement measures, including a R\'enyi generalization of the modular commutator, its ``charged" counterpart that extracts the quantized Hall conductivity~\cite{Fan_2023_extracting}, and the lens-space measure proposed in Ref.~\cite{sheffer2025extracting}. These quantities are defined via the expectation values of permutation operators acting on replicas of the wavefunction: take multiple copies of the wavefunction, partition a disk subregion into three pieces (see Fig.~\ref{fig: pizza}), and create permutation defects by assigning a different replica permutation to each subregion. We call them ``topological multi-entropy measures". To evaluate these measures, we develop a unified field-theoretical framework that integrates the contributions of bulk anyons and edge gapless modes. This framework offers a general methodology for combining bulk and edge effects, whose applicability extends beyond the examples analyzed here. 

Specifically, our framework maps each entanglement measure into a partition function on a three-manifold, where the permutation defects give rise to a high-genus boundary surface when the theory is properly regularized.
We demonstrate that, at long distances, this partition function factorizes into two components. 
The first is a TQFT partition function on a closed three-manifold, obtained by shrinking the boundary, which captures anyon data through the bulk \emph{topology}~\cite{dong_topological_2008, wen2016topological}.
This component recovers previous results for non-chiral phases, e.g., the topological entanglement entropy~\cite{Kitaev_Preskill_2006, Levin_Wen_2006}.
The second component is the partition function of a conformal field theory (CFT) residing on the boundary. This term is related to the chiral central charge through the \emph{geometry} of the high-genus surface. Consequently, the entanglement measures we discuss serve, in essence, as probes of this emergent surface geometry.

\section{Overview and summary of results}
\label{sec: overview}

In this section, we provide an overview of the main results and outline the structure of the rest of the paper. We begin by defining ``topological multi-entropy measures" and outlining their general properties. We specifically discuss results on three instances: the R\'enyi modular commutator, the lens-space multi-entropy, and the charged R\'enyi modular commutator. Subsequently, we summarize the field-theoretic framework developed to compute these quantities.

\subsection{Topological multi-entropy measures}
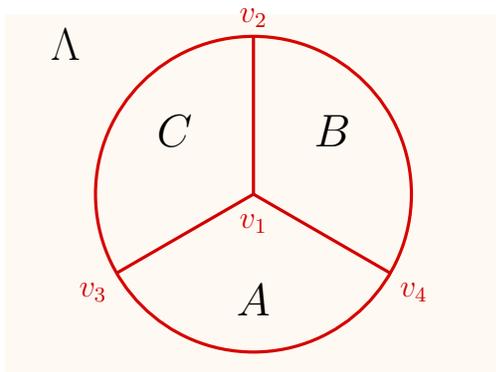
\begin{figure}
  \begin{center}
  \begin{tikzpicture}
    \def\r{2.1};
      \draw[draw=darkred,very thick] (0,0) circle (\r);
      \fill[fill=orange,fill opacity=.05] (-3.3,-2.4) rectangle (3.3,2.4);
      \begin{scope}[every node/.append style={font={\large}}]
      \draw[very thick,darkred] (0,0) node[below, yshift=-4pt] {$v_1$} -- (-30:\r) node[below right] {$v_4$};
      \draw[very thick,darkred] (0,0) -- (-150:\r) node[below left] {$v_3$};
      \draw[very thick,darkred] (0,0) -- (90:\r) node[above] {$v_2$};
      \end{scope}
      \node at (-.0,-1.4) {\LARGE \textbf{$A$}};
      \node at (-2.5,2) {\LARGE \textbf{$\Lambda$}};
      \node at (+\r/2,+\r*.4) {\LARGE \textbf{$B$}};
      \node at (-\r/2,+\r*.4) {\LARGE \textbf{$C$}};
  \end{tikzpicture}
  \end{center}
  \caption{The geometry considered in this work. Three permutation operators are applied on three regions $A,B,C$ of the plane.}
  \label{fig: pizza}
\end{figure}

In this work, we characterize multipartite entanglement in quantum states using ``multi-entropy" measures. 
The construction takes multiple replicas of a state, assigns replica permutations to different spatial regions, and defines the multi-entropy as the expectation value of the resulting product of permutation operators.
Such measures were first introduced in the context of holography \cite{gadde_new_2022, penington_fun_2023} and were later used to study 2+1D topologically ordered states \cite{liu_multi_2024, sheffer2025extracting}. Note that these quantities are generally not entanglement monotones~\cite{plenio2005introduction}.

Specifically, consider a spatial partitioning of the system into three adjacent regions $A$, $B$, and $C$, surrounded by an external region $\Lambda$, as illustrated in Fig.~\ref{fig: pizza}.
Taking $R$ replicas and three permutation operators $\pi_A$, $\pi_B$, and $\pi_C$ acting on the respective regions, we define: 
\begin{equation}
    \mathcal{M}(\psi) = \langle\psi^{\otimes R}| \pi_A\pi_B\pi_C |\psi^{\otimes R}\rangle \, . 
    \label{eq: multi-ent}
\end{equation}
We will make a slight abuse of notation throughout and employ $\pi_I$ as both the abstract permutation and its realization as an operator acting on the degrees of freedom in the region $I$. We always set the boundary lengths of the subregions to be much larger than the correlation length of the state. Note that, while no operator is applied in the region $\Lambda$, it is necessary that $A,B,C$ do not cover the entire plane. Otherwise, the entanglement measure captures only tripartite entanglement, rather than entanglement between four regions.

The assignment of distinct permutations to adjacent spatial regions introduces permutation defects in replica space. Given a generic gapped state (not a product state), we expect the magnitude of these measures to decay exponentially with the total length $L$ of these defects, following the scaling $|\mathcal{M}|\sim \alpha e^{-\beta L}$. Here, $\beta > 0$ is a non-universal factor determined by microscopic details. 

While the prefactor $\alpha$ might capture topological information (for example, the total quantum dimension), the phase $\arg(\mathcal{M})$ is the more natural probe for chirality. 
This is evident from its transformation under time-reversal $\mathcal{T}$: since permutation operators are time-reversal invariant, we have $\mathcal{M}(\mathcal{T}\psi) = \overline{\mathcal{M}(\psi)}$.
Namely, a nonzero phase $\arg(\mathcal{M})$ signals the absence of time-reversal symmetry.
To ensure $\arg(\mathcal{M})$ is a genuine topological quantity, App.~\ref{app: topological-invariance} provides conditions on the permutation operators that lead to two more necessary properties: (1) \textit{Robustness:} $\arg(\mathcal{M})$ is invariant under smooth deformations of the region boundaries and local unitaries acting on the state. (2) \textit{$\mathcal{RT}$ invariance:} $\arg(\mathcal{M})$ is invariant under the combination of time-reversal and mirror reflection with respect to an arbitrary plane perpendicular to the system. 
The second condition is expected for topological phases described by TQFTs, such as Chern-Simons theory~\cite{sheffer2025extracting}. Both properties are expected to hold under generic conditions, with corrections that are exponentially small in $L$. We will also comment on the ``spurious" contributions at the end of the manuscript.
We designate measures satisfying these criteria as ``topological multi-entropy" (TME) measures and restrict our analysis to this class.

Even under these conditions, the residual freedom of choosing permutations allows us to access a wide range of properties. Here we focus on two choices that are related to the chiral central charge and the topological spins. These two examples will be the main focus of our work, but the method we present applies more generally. As an example, we extend our discussion to cases with global U(1) symmetries, where the symmetry-enriched TME yields quantum Hall conductance.

\subsubsection{Rényi modular commutator}
Ref.~\cite{kim_chiral_2022} introduced a new entanglement measure, the ``modular commutator", conjectured to extract the chiral central charge $c_-$ from a single ground-state wavefunction. Defining the modular Hamiltonian in region $I$ as $K_I=-\log\rho_I$, the modular commutator reads~\cite{kim_chiral_2022}
\begin{equation}
    J=i\Tr \rho_{ABC}[K_{AC},K_{AB}]=\frac{\pi}{3}c_-.
    \label{eq: modular-commutator}
\end{equation}
Refs.~\cite{Zou_Shi_Sorce_Lim_Kim_2022,Fan_2022} support this conjecture by assuming that the reduced density matrices are fully described by 
a CFT. Ref.~\cite{Fan_Zhang_Gu_2023} provided a rigorous proof in free fermion systems by generalizing the real-space spectral Chern number formula introduced in Appendix C of \cite{kitaev2006anyons}.

Here we construct a ``R\'enyi-$n$" version of the modular commutator $J$ that recovers $J$ in the replica limit
\begin{align}
    \label{eq: j-lim}
    J&=\lim_{n\to0}\frac{i}{n^2}\qty(J_n-\overline{J_n}),\\
    \label{eq: j_n-def}
    J_n&=\mel{\psi}{\rho^n_{AC}\rho^n_{AB}}{\psi},
\end{align}
where $J_n$ is a TME on $2n+1$ replicas 
\begin{equation}
    J_n:\;\begin{cases}
    \pi_A=(1,\ldots,2n+1) & \\
    \pi_B=(n+1,\ldots,2n+1) & \\
    \pi_C=(1,\ldots,n+1) & 
    \end{cases},
\end{equation}
and $\overline{J_n}$ is its complex conjugate.
That is, $\pi_A$ cyclically permutes all $2n+1$ replicas, $\pi_B$, and $\pi_C$ permute the last and first $n+1$, respectively. 
One of our main goals in this work is to show
\begin{equation}
    J_n\propto \exp(-\frac{2\pi i c_-}{24}\frac{2n^2}{(2n+1)(n+1)})
    \label{eq: j_n phases}
\end{equation}
Throughout this paper, ``$\propto$" means ``equal up to a nonuniversal, real, and positive proportionality factor". 
We expect this non-universal term to decrease exponentially in the region size $L$ and the R\'enyi index $n$.
The original conjecture \eqref{eq: modular-commutator} follows immediately from \eqref{eq: j_n phases} by taking the replica limit, thus providing an independent proof.
Moreover, the phase of $J_n$ yields $c_-$ directly, resulting in a method to extract $c_-$ numerically by evaluating an operator on the ground-state wavefunction. For example, we can measure $c_-$ (mod 72) from $J_1$ with only 3 replicas of the bulk wavefunction~\cite{hastings2010measuring}. We comment that the measures $J_n$ are sensitive to edge physics, and not only to bulk anyon statistics, and can therefore access $c_-$ itself, and not only to $c_-$ mod an integer. As such, they are capable of detecting invertible states with $c_-\neq 0$, such as the $p+ip$ superconductor.

\subsubsection{Lens-space multi-entropy}
In Ref.~\cite{sheffer2025extracting}, some of us defined the lens-space multi-entropy (LeSME) measure using permutations on $R=2r$ replicas, for $r\ge2$. Indexing the replicas with a tuple $(s,t)$, with $s=1,2,\,t=1,\ldots,r$, we define
\begin{equation}
    \Phi_r:\;\begin{cases}
    \pi_A(1,t) = (1,t-1); & \pi_A(2,t) = (2,t+1), \\
    \pi_B(s,t) = (s+1,t), \\ 
    \pi_C(1,t) = (2,t+1); & \pi_C(2,t) = (1,t-1),    
    \end{cases}
    \label{eq: phi_r-def}
\end{equation}
where addition is defined mod 2 for $s$ and mod $r$ for $t$. For \textit{non-chiral} topological order, Ref.~\cite{sheffer2025extracting} shows
\begin{equation}
    \Phi_r\propto \sum_ad_a^2\theta_a^r,
\end{equation}
where $d_a,\theta_a$ are the anyon quantum dimension and topological spin, respectively, and the sum is carried over all anyons in the theory. The quantity on the RHS is proportional to the ``higher central charge" \cite{kaidi_higher_2022}. 

In this work, we extend the result to chiral topological orders. Using a combination of analytical arguments and numerical evidence, we demonstrate that
\begin{equation}
    \Phi_r\propto e^{\frac{2\pi i c_-}{24}\qty(-r-\frac{2}{r})}\sum_a d_a^2\theta_a^r
    \label{eq: phi_r}
\end{equation}
A nontrivial check is provided by the case $r=2$. Since $\pi_A,\pi_B,\pi_C$ are Hermitian in this case, $\Phi_2$ must be real.
Although not immediately apparent from Eq.~\eqref{eq: phi_r}, we prove the reality of this expression at $r=2$ in App.~\ref{app: real-phi_r}. Special care is needed when the sum in \eqref{eq: phi_r} is zero (for example, in the bosonic $\nu=1/2$ Laughlin state for $r$ even), see Sec.\ \ref{sec: bulk-edge}. Interestingly, the phase factor is the same as that obtained in \cite{kobayashi_extracting_2024}, which presents another entanglement measure related to the lens space but relies on translation invariance. Since the phase is universal, \eqref{eq: phi_r} together with \eqref{eq: j_n phases} enable the extraction of the higher central charge. Similarly, the procedure outlined in \cite{sheffer2025extracting} can be used to cancel the non-universal contributions and extract the magnitude $\abs{\sum_a d_a^2\theta_a^r}$.

\subsubsection{Charged R\'enyi modular commutator}

The general principles naturally extend to symmetry-protected \cite{chen2013symmetry} and symmetry-enriched \cite{barkeshli_symmetry_2019} topological orders. In addition to permutation operators, we can act on the wavefunction with partial symmetry transformations. Conceptually, both operations create defects: the former introduces geometric defects in the replica space, while the latter inserts symmetry defects.

An important and illustrative example is the case of phases with a global $U(1)$ symmetry.
Ref.~\cite{Fan_2023_extracting} defined a ``charged modular commutator" to extract the quantum Hall conductance. In the same geometry of Fig.\ \ref{fig: pizza}, it was argued that
\begin{equation}
    \SS=\Tr\rho_{ABC}[Q_{AC}^2,K_{AB}]=-2i\sigma_{xy},
    \label{eq: charged-modular-commutator}
\end{equation}
where $Q_{AC}$ is the total charge operator in the region $AC$,  and $\sigma_{xy}$ is the Hall conductance. We use natural units, where $\sigma_{xy}$ of a single Chern band is $1/2\pi$. 

Similar to the R\'enyi modular commutator \eqref{eq: j_n-def}, we can ``R\'enyi-ize" Eq.~\eqref{eq: charged-modular-commutator} to obtain
\begin{align}
    \SS&=\lim_{n\to0,\mu\to 0} \frac{-2}{\mu^2 n} (\SS_{\mu,n}-\overline{\SS_{\mu,n}}), \\
    \SS_{\mu,n}&=\mel{\psi}{e^{\mu Q_{AC}}\rho^n_{AB}}{\psi}.
\end{align}
We call $\SS_{\mu,n}$ the charged R\'enyi modular commutator.
We can write $\SS_{\mu,n}$ using permutation operators on $n+1$ replicas as
\begin{equation}
    \SS_{\mu,n}=\mel{\psi^{\otimes n+1}}{e^{\mu Q_{AC}}\pi_{AB}}{\psi^{\otimes n+1}},
    \label{eq: s-mu-n-def}
\end{equation}
where $\pi_{AB}$ is a cyclic permutation, and $e^{\mu Q_{AC}}$ acts only on replica $1$. Below, we will show that
\begin{equation}
    \SS_{\mu,n}\propto \exp(i\sigma_{xy}\frac{n\mu^2}{2(n+1)}),
    \label{eq: s-mu-n-phases}
\end{equation}
for $\mu\ll L/\xi$. In App.~\ref{app: topological-invariance} we argue that the resulting phase is universal.

\subsection{Method of calculation}
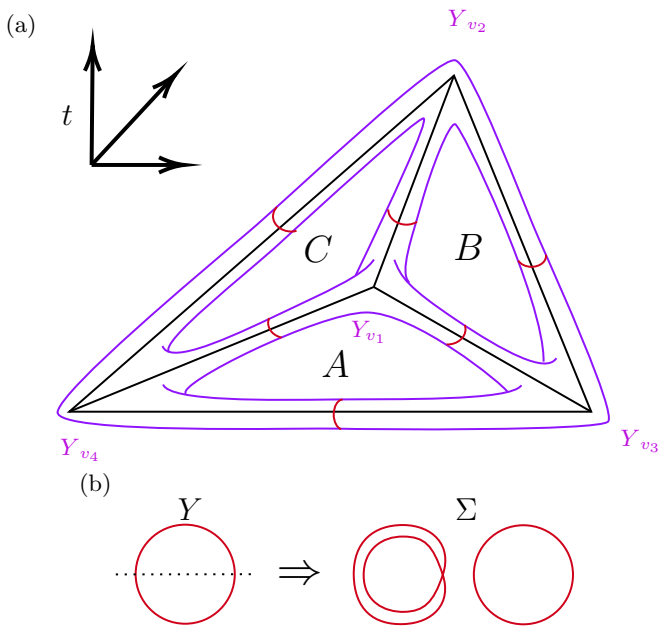
\begin{figure}
    \centering
    
\tikzset{every picture/.style={line width=0.75pt}} 

\begin{tikzpicture}[x=0.75pt,y=0.75pt,yscale=-.9,xscale=.9]

\node at (-5,20) {(a)};

\draw   (237.61,47.75) -- (314.5,236.25) -- (22.01,236.25) -- cycle ;
\draw    (192.64,166.31) -- (314.5,236.25) ;
\draw    (237.62,47.75) -- (192.64,166.31) ;
\draw    (22,236.25) -- (192.64,166.31) ;
\draw  [color={rgb, 255:red, 144; green, 19; blue, 254 }  ,draw opacity=1 ] (324,241.75) .. controls (319.5,247.25) and (191.5,246.25) .. (172.5,245.75) .. controls (153.5,245.25) and (13,249.25) .. (15.5,236.25) .. controls (18,223.25) and (93.75,159.75) .. (121.5,136.25) .. controls (149.25,112.75) and (229,35.75) .. (238.5,39.25) .. controls (248,42.75) and (276,118.75) .. (286,141.25) .. controls (296,163.75) and (328.5,236.25) .. (324,241.75) -- cycle ;
\draw [color={rgb, 255:red, 144; green, 19; blue, 254 }  ,draw opacity=1 ]   (78,202.75) .. controls (72,197.25) and (121,159.25) .. (140,141.75) .. controls (159,124.25) and (217,71.25) .. (220.5,71.75) .. controls (224,72.25) and (201.5,120.25) .. (182.5,159.75) ;
\draw [color={rgb, 255:red, 144; green, 19; blue, 254 }  ,draw opacity=1 ]   (74.25,199.38) .. controls (77.75,209.13) and (103.5,198.75) .. (135,183.75) .. controls (166.5,168.75) and (187,160) .. (192.75,150.88) ;
\draw [color={rgb, 255:red, 144; green, 19; blue, 254 }  ,draw opacity=1 ]   (74.6,221) .. controls (78.1,230.75) and (135.8,229.8) .. (170.2,229.4) .. controls (204.6,229) and (269.65,230.53) .. (275.4,221.4) ;
\draw [color={rgb, 255:red, 144; green, 19; blue, 254 }  ,draw opacity=1 ]   (87.4,226.2) .. controls (81.4,220.7) and (161.4,181) .. (188.2,180.6) .. controls (215,180.2) and (265.4,221) .. (267.8,225.4) ;
\draw [color={rgb, 255:red, 144; green, 19; blue, 254 }  ,draw opacity=1 ]   (212.6,163.8) .. controls (206.6,158.3) and (214.2,138.2) .. (219.4,119.4) .. controls (224.6,100.6) and (234.7,74.5) .. (238.2,75) .. controls (241.7,75.5) and (288.6,174.2) .. (287.8,208.2) ;
\draw [color={rgb, 255:red, 144; green, 19; blue, 254 }  ,draw opacity=1 ]   (204.2,151) .. controls (207.7,160.75) and (211.8,166.6) .. (238.6,184.6) .. controls (265.4,202.6) and (289.25,215.33) .. (295,206.2) ;
\draw  [draw opacity=0][line width=0.75]  (149.36,134.84) .. controls (148.42,135.07) and (147.43,135.2) .. (146.4,135.2) .. controls (140.77,135.2) and (136.2,131.39) .. (136.2,126.7) .. controls (136.2,124.7) and (137.03,122.86) .. (138.42,121.4) -- (146.4,126.7) -- cycle ; \draw  [color={rgb, 255:red, 208; green, 2; blue, 27 }  ,draw opacity=1 ][line width=0.75]  (149.36,134.84) .. controls (148.42,135.07) and (147.43,135.2) .. (146.4,135.2) .. controls (140.77,135.2) and (136.2,131.39) .. (136.2,126.7) .. controls (136.2,124.7) and (137.03,122.86) .. (138.42,121.4) ;  
\draw  [draw opacity=0][line width=0.75]  (216.91,129.74) .. controls (215.24,130.75) and (213.17,131.35) .. (210.95,131.35) .. controls (205.31,131.35) and (200.75,127.54) .. (200.75,122.85) .. controls (200.75,122.82) and (200.75,122.79) .. (200.75,122.76) -- (210.95,122.85) -- cycle ; \draw  [color={rgb, 255:red, 208; green, 2; blue, 27 }  ,draw opacity=1 ][line width=0.75]  (216.91,129.74) .. controls (215.24,130.75) and (213.17,131.35) .. (210.95,131.35) .. controls (205.31,131.35) and (200.75,127.54) .. (200.75,122.85) .. controls (200.75,122.82) and (200.75,122.79) .. (200.75,122.76) ;  
\draw  [draw opacity=0][line width=0.75]  (141.07,194.7) .. controls (136.7,193.75) and (133.47,190.43) .. (133.47,186.48) .. controls (133.47,185.51) and (133.67,184.58) .. (134.03,183.71) -- (143.67,186.48) -- cycle ; \draw  [color={rgb, 255:red, 208; green, 2; blue, 27 }  ,draw opacity=1 ][line width=0.75]  (141.07,194.7) .. controls (136.7,193.75) and (133.47,190.43) .. (133.47,186.48) .. controls (133.47,185.51) and (133.67,184.58) .. (134.03,183.71) ;  
\draw  [draw opacity=0][line width=0.75]  (243.63,187.62) .. controls (243.91,188.39) and (244.05,189.2) .. (244.05,190.05) .. controls (244.05,194.74) and (239.49,198.55) .. (233.85,198.55) .. controls (233.71,198.55) and (233.57,198.54) .. (233.42,198.54) -- (233.85,190.05) -- cycle ; \draw  [color={rgb, 255:red, 208; green, 2; blue, 27 }  ,draw opacity=1 ][line width=0.75]  (243.63,187.62) .. controls (243.91,188.39) and (244.05,189.2) .. (244.05,190.05) .. controls (244.05,194.74) and (239.49,198.55) .. (233.85,198.55) .. controls (233.71,198.55) and (233.57,198.54) .. (233.42,198.54) ;  
\draw  [draw opacity=0][line width=0.75]  (174.35,246.04) .. controls (171.98,245.25) and (170.2,241.86) .. (170.2,237.81) .. controls (170.2,233.3) and (172.39,229.62) .. (175.15,229.41) -- (175.4,237.81) -- cycle ; \draw  [color={rgb, 255:red, 208; green, 2; blue, 27 }  ,draw opacity=1 ][line width=0.75]  (174.35,246.04) .. controls (171.98,245.25) and (170.2,241.86) .. (170.2,237.81) .. controls (170.2,233.3) and (172.39,229.62) .. (175.15,229.41) ;  
\draw  [draw opacity=0][line width=0.75]  (289.1,148.19) .. controls (289.15,148.44) and (289.17,148.69) .. (289.17,148.95) .. controls (289.17,152.29) and (285.52,155) .. (281.02,155) .. controls (277.74,155) and (274.91,153.56) .. (273.62,151.49) -- (281.02,148.95) -- cycle ; \draw  [color={rgb, 255:red, 208; green, 2; blue, 27 }  ,draw opacity=1 ][line width=0.75]  (289.1,148.19) .. controls (289.15,148.44) and (289.17,148.69) .. (289.17,148.95) .. controls (289.17,152.29) and (285.52,155) .. (281.02,155) .. controls (277.74,155) and (274.91,153.56) .. (273.62,151.49) ;  
\draw [line width=1.5]    (34.53,97.87) -- (35.11,33.87) ;
\draw [shift={(35.13,30.87)}, rotate = 90.51] [color={rgb, 255:red, 0; green, 0; blue, 0 }  ][line width=1.5]    (14.21,-4.28) .. controls (9.04,-1.82) and (4.3,-0.39) .. (0,0) .. controls (4.3,0.39) and (9.04,1.82) .. (14.21,4.28)   ;
\draw [line width=1.5]    (34.53,97.87) -- (79.63,49.2) ;
\draw [shift={(81.67,47)}, rotate = 132.82] [color={rgb, 255:red, 0; green, 0; blue, 0 }  ][line width=1.5]    (14.21,-4.28) .. controls (9.04,-1.82) and (4.3,-0.39) .. (0,0) .. controls (4.3,0.39) and (9.04,1.82) .. (14.21,4.28)   ;
\draw [line width=1.5]    (34.53,97.87) -- (83.53,97.87) ;
\draw [shift={(86.53,97.87)}, rotate = 180] [color={rgb, 255:red, 0; green, 0; blue, 0 }  ][line width=1.5]    (14.21,-4.28) .. controls (9.04,-1.82) and (4.3,-0.39) .. (0,0) .. controls (4.3,0.39) and (9.04,1.82) .. (14.21,4.28)   ;

\draw (178.4,183.8) node [anchor=north west][inner sep=0.75pt]    {$\textcolor[rgb]{0.74,0.06,0.88}{Y_{v_{1}}}$};
\draw (232.4,9.4) node [anchor=north west][inner sep=0.75pt]    {$\textcolor[rgb]{0.74,0.06,0.88}{Y}\textcolor[rgb]{0.74,0.06,0.88}{_{v_{2}}}$};
\draw (329.2,244.8) node [anchor=north west][inner sep=0.75pt]    {$\textcolor[rgb]{0.74,0.06,0.88}{Y}\textcolor[rgb]{0.74,0.06,0.88}{_{v_{3}}}$};
\draw (15.2,249.8) node [anchor=north west][inner sep=0.75pt]    {$\textcolor[rgb]{0.74,0.06,0.88}{Y}\textcolor[rgb]{0.74,0.06,0.88}{_{v_{4}}}$};
\draw (16.13,62.93) node [anchor=north west][inner sep=0.75pt]    {\large $t$};
\draw (161.6,200.8) node [anchor=north west][inner sep=0.75pt]  [font=\Large]  {$A$};
\draw (235.6,134.8) node [anchor=north west][inner sep=0.75pt]  [font=\Large]  {$B$};
\draw (151.6,136) node [anchor=north west][inner sep=0.75pt]  [font=\Large]  {$C$};

\end{tikzpicture}

\begin{tikzpicture}[x=0.75pt,y=0.75pt,yscale=-1,xscale=1]

\draw  [color={rgb, 255:red, 208; green, 2; blue, 27 }  ,draw opacity=1 ][line width=0.75]  (180.86,55.1) .. controls (180.86,41.29) and (192.05,30.1) .. (205.86,30.1) .. controls (219.66,30.1) and (230.86,41.29) .. (230.86,55.1) .. controls (230.86,68.9) and (219.66,80.1) .. (205.86,80.1) .. controls (192.05,80.1) and (180.86,68.9) .. (180.86,55.1) -- cycle ;
\draw  [color={rgb, 255:red, 208; green, 2; blue, 27 }  ,draw opacity=1 ][line width=0.75]  (125.29,55.5) .. controls (125.2,42.23) and (134.54,35.68) .. (145.29,35.81) .. controls (156.04,35.93) and (160.14,40) .. (165.09,54.81) .. controls (170.04,69.62) and (160.16,79.5) .. (145.29,80) .. controls (130.41,80.5) and (120.69,72.7) .. (120.29,55.5) .. controls (119.89,38.3) and (130.54,29.87) .. (145.29,30) .. controls (160.04,30.12) and (170.04,40.25) .. (165.29,54.81) .. controls (160.54,69.37) and (155.54,73.56) .. (145.7,73.81) .. controls (135.87,74.06) and (125.37,68.77) .. (125.29,55.5) -- cycle ;
\draw  [color={rgb, 255:red, 208; green, 2; blue, 27 }  ,draw opacity=1 ][line width=0.75]  (10.29,54.81) .. controls (10.29,41) and (21.48,29.81) .. (35.29,29.81) .. controls (49.09,29.81) and (60.29,41) .. (60.29,54.81) .. controls (60.29,68.62) and (49.09,79.81) .. (35.29,79.81) .. controls (21.48,79.81) and (10.29,68.62) .. (10.29,54.81) -- cycle ;
\draw [line width=0.75]  [dash pattern={on 0.84pt off 2.51pt}]  (0.29,54.92) -- (70.29,54.81) ;

\node at (-10,10) {(b)};
\draw (80,50) node [anchor=north west][inner sep=0.75pt]   [align=left] {\LARGE $\displaystyle \Rightarrow $};
\draw (30,15) node [anchor=north west][inner sep=0.75pt]    {\large $Y$};
\draw (170,15) node [anchor=north west][inner sep=0.75pt]    {\large $\Sigma $};
\end{tikzpicture}
    \caption{ (a) Permutation defect (black) and the enclosing regularization surface $Y$ (purple). Cutting along the red lines yields four three-punctured spheres. In this figure these are the POPs around each vertex, depicted as $Y_{v_i}$. The handles of $Y$ have circumferences $\epsilon \ll L$. 
    (b) Cross section of the tube on the boundary of region $C$, in the example of $J_1$. Unfolding the permutations around this tube creates two disjoin tubes, one corresponding to replicas 1,2 and one corresponding to replica 3.
    }
    \label{fig: regulating-surface}
\end{figure}

In this work, we focus on ``generic" states, which are homogeneous and can be captured by a Chern-Simons field theory at long distances.  Under this assumption, we can calculate the TME within the field-theory framework. It is then given by the 2+1D Euclidean partition function on $R$ replicas with permutation and symmetry defects inserted at the imaginary time $t=0$. The standard TQFT approach represents density matrices as path integrals on three-manifolds, glues them according to the permutation operators, and maps each measure to a partition function on a smooth three-manifold $M$~\cite{dong_topological_2008}. However, this approach is insufficient for our purpose. For example, it reduces the R\'enyi modular commutator $J_n$ to a partition function on $S^3$, a positive number, thereby yielding a null result! 
One crucial missing element is that the field-theoretic calculation requires regularization, which must introduce gapless edge modes for a chiral theory ~\cite{Ohmori:2014eia,wong_note_2018, Jafferis:2019wkd, Belin:2019mlt}. Below, we outline the procedure for incorporating these modes to correctly evaluate the measures of the form \eqref{eq: multi-ent} and \eqref{eq: s-mu-n-def}.  

The first step is geometrical regularization. 
The permutation defects on the region boundaries introduce conical singularities to the path integral, which we regulate by excising a tubular neighborhood $W$ surrounding the defects.  This yields a manifold with boundary $M\backslash W$, where $M$ is the closed three-manifold obtained via the standard gluing procedure. Consequently, the measure is given by the partition function
\begin{equation}
    \mathcal{M} \propto Z(M\backslash W)\,.
\end{equation}
In the case of TMEs, the boundary $\Sigma = \partial(M\backslash W) = \partial W$ is a high-genus surface. It carries gapless modes described by the edge CFT. As explained in Fig.\ \ref{fig: regulating-surface}, we construct $\Sigma$ in two steps: first, we introduce the regularization surface $Y$ that encloses the one-dimensional region boundaries in each replica. The lengths of the handles of $Y$ scale as $L$, and their circumferences scale as the cutoff $\epsilon \ll L$. Then, we construct the surface $\Sigma$ by taking $R$ copies of $Y$ (one for each replica), cutting and gluing them according to the permutations $\pi_{A,B,C}$. In short, the surface $Y$ encodes information about the region boundaries, while $\Sigma$ encodes additional information about the applied permutations.

Next, we evaluate the partition function $Z(M\backslash W)$. In the limit $\epsilon/L \rightarrow 0$, Sec.~\ref{sec: bulk-edge} shows that this partition function can be separated into bulk and edge contributions:
\begin{equation}
    \mathcal{M}\propto Z_{\rm topo}(M) Z_{\rm CFT}(\Sigma) \,.
    \label{eq: separated-pf}
\end{equation}
Specifically, $Z_{\rm topo}(M)$ is the bulk TQFT partition function and $Z_{\rm CFT}(\Sigma)$ is a chiral CFT partition function evaluated at the surface $\Sigma$. In the following, we omit the subscripts ``topo" and ``CFT", and simply write $Z(M)$ and $Z(\Sigma)$. The standard TQFT approach only covers $Z(M)$. We will focus on $Z(\Sigma)$ from now on.

Although the magnitude of $Z(\Sigma)$ can be of independent interest~\cite{Liu_Kusuki_Kudler-Flam_Sohal_Ryu_2024, liu_multi_2024}, it is the phase of $Z(\Sigma)$ that possibly encodes the necessary information to diagnose chiral topological phases. 
Besides the general symmetry argument presented earlier, the insensitivity of the magnitude to chirality can also be understood from the regularization perspective. Consider the ``doubled" theory formed by taking the tensor product of the original theory with its time-reversal conjugate. The magnitude $\abs{\mathcal{M}}$ corresponds to the square root of the measure evaluated in this doubled theory. Crucially, the doubled theory is non-chiral and always admits a gapped boundary~\cite{kitaev2012models}. Consequently, the permutation defects can be regulated without introducing a gapless edge (for example, by defining the path integral in discrete spacetime). This implies that the magnitude is largely blind to the chiral central charge, leaving the phase as the primary probe~\footnote{A subtle point is that, while not strictly universal, the corner contributions discussed in \cite{siva_universal_2022} are argued to satisfy a universal lower bound. The lower bound associated with the original state and the square root of the lower bound for the doubled state could be different.}.

Indeed, we find that the phase of $Z(\Sigma)$ is proportional to $c_-$ in the limit $\epsilon/L \rightarrow 0$. To demonstrate this, it is convenient to consider a pair-of-pants (POP) decomposition of $\Sigma$. This procedure decomposes a closed surface into a union of multiple three-punctured spheres (see Fig.\ \ref{fig: regulating-surface}). As detailed in Sec.\ \ref{sec: fn_coordinates}, if the surface $\Sigma$ has a genus $g$, we can characterize it, more precisely its conformal structure, by $3g-3$ length parameters $l_i$, which describe the perimeter of the holes on each pair of pants, and $3g-3$ twist parameters $\tau_i$, which describe the angle by which two POPs are twisted when glued to each other. These are the Fenchel-Nielsen coordinates~\cite{farb2011primer,imayoshi2012introduction, hubbard2016teichmuller}. The partition function is evaluated by summing over the CFT states at the holes between each POP. Given a pair of glued holes with the length parameter $l_i$ and twist parameter $\tau_i$, the contribution from a CFT state propagating between the two POPs takes the form $e^{-E_s/l_i+iP_s\tau_i}$, where $E_s,P_s$ are the energy and momentum of the state. In the current problem, all length parameters $l_i$ approach zero as $\epsilon/L \rightarrow 0$. As a result, only the CFT vacuum state $|0\rangle$ propagates between any two POPs. The vacuum states of two adjacent POPs are glued with a twist $\tau_i$, giving a contribution to the partition function of the form \cite{di_francesco_conformal_1997, tu2013momentum}
\begin{equation}
    \mel{0}{e^{i \tau_i P}}{0}=e^{-\frac{i\tau_i c_-}{24}},
    \label{eq: single-twist-contribution}
\end{equation}
where $P$ is the momentum operator. The total phase of the partition function is then given by
\begin{equation}
    Z(\Sigma)\propto \exp(-\frac{ic_-}{24}\sum_i \tau_i).
    \label{eq: sum-of-twists}
\end{equation}
The above equation is the main tool for our calculations. It relates the phase $\arg\mathcal{M}$ to the geometry of the surface $\Sigma$ and the chiral central charge. In Sec. \ref{sec: fn_coordinates} we show how the twist angles can be computed given the permutations $\pi_{A,B,C}$.

An important subtlety of the discussion above is that both the bulk and edge partition functions in Eq.\ \eqref{eq: separated-pf} suffer from the ``framing anomaly" \cite{witten1989quantum, Witten_1990, gromov_framing_2015}. The framing anomaly is present since, for chiral theories, the partition function $Z(M)$ does not depend exclusively on the topology of $M$, but also on a choice of framing of $M$. A framing of a manifold $M$ is a choice of basis for the tangent bundle of $M$ at each point. Any two framing choices are related by a relative winding number. The framing anomaly means that, when passing between two choices of framing, the partition function transforms as
\begin{equation}
    Z(M)\mapsto e^{\frac{2\pi i k c_-}{24}} Z(M)
    \label{eq: framing-change}
\end{equation}
where $k\in\mathbb{Z}$ counts the winding number of the new framing with respect to the old one. A similar anomaly exists in the partition function $Z(\Sigma)$, as we will discuss in Sec.~\ref{sec: fn_coordinates}. We will not deal with the framing anomaly directly here but will rather evaluate the quantities of interest up to a phase factor of the form $\eqref{eq: framing-change}$ and rely on physical arguments and numerics to fix the ambiguity.

\subsection{Outline of the paper}
The remainder of the manuscript is organized as follows. 
In Sec.\ \ref{sec: bulk-edge}, we provide the general prescription for calculating R\'enyi entanglement measures in Chern-Simons theories and show that the result separates into bulk and edge contributions. 
In Sec.\ \ref{sec: fn_coordinates}, we provide the essential technical tools for the calculation of the twist angles $\tau_i$ and, as a result, the edge contribution to the partition function. 
In Sec.\ \ref{sec: renyi-mc} and \ref{sec: LSME}, we apply these tools to calculate the R\'enyi modular commutator and the Lens-space multi-entropy. 
In Sec.\ \ref{sec: quantum-hall}, we extend our framework to systems with $U(1)$ symmetry and obtain the charged R\'enyi modular commutator for quantum Hall conductance. 
We also numerically calculate these TME measures in three lattice models and collect the results in Sec.\ \ref{sec: numerics}. The first two are the Kitaev honeycomb model \cite{kitaev2006anyons} and a Chern insulator, for which the calculation can be done in polynomial time, resulting in very accurate estimates of the entanglement measures we considered above. 
The third model is a trial wavefunction describing the $\nu=1/2$ Laughlin state \cite{nielsen2012laughlin}, for which we evaluate $J_1$ and $\SS_{1,\mu}$ via a Monte Carlo method. 
The numerical results are in agreement with our analytical predictions in all these cases.
In Sec.\ \ref{sec: conclusions}, we conclude and discuss open questions.

\section{Bulk-edge separation of entanglement measures}
\label{sec: bulk-edge}

In this section, we provide the general field-theoretical framework for calculating entanglement measures that explicitly account for edge state contributions. We begin with the simplest case, bipartite entanglement, to illustrate the main ingredients. While this case has been analyzed by Ref.~\cite{Fliss:2017wop,wong_note_2018}, we introduce a modified prescription that generalizes naturally to the multipartite setting. The main goal is to derive Eq.\ \eqref{eq: separated-pf}. In essence, we show that TME measures factorize into distinct components: a topological contribution from the bulk TQFT and a geometric contribution governed by the edge CFT.

\subsection{Bipartite entanglement}

To properly define entanglement measures in a field-theoretical framework, we must first resolve a conceptual obstacle: the Hilbert space of a quantum field theory does not naturally factorize across spatial boundaries~\cite{nishioka2018entanglement, Witten_2018, casini_lectures_2023}. This is related to the fact that, if the theory is regularized on the lattice, the entanglement entropy is UV divergent. To remedy this issue, we embed the physical Hilbert space into a tensor-product space using the so-called ``cutting map"~\cite{Ohmori:2014eia}. As we demonstrate below, this procedure inherently requires a short-distance regularization. We will later show that the universal information we extract from TME is independent of the microscopic details of this regularization.

Consider the bipartite entanglement between a subregion $A$ and its complement $A^c$, both having the topology of a disk. The cutting map is an embedding of the form
\begin{equation}
    i_\epsilon:\hil\to\hil_A\otimes\hil_{A^c}.
\end{equation}
where $\hil$ is the Hilbert space of the entire system, and $\hil_{A}$/$\hil_{A^c}$ is the Hilbert space in the region $A$/$A^c$. For Chern-Simons theory, $\hil_A,\hil_{A^c}$ are the Hilbert spaces of the associated edge CFT on the boundaries of the disks~\cite{elitzur1989remarks}. The map $i_\epsilon$ depends on a small regularization parameter, which is the cutoff scale of the theory. The most natural way to define the cutting map is by using a Euclidean path integral, starting from the state in $\hil$ and adding a boundary condition on a tubular neighborhood of $\partial A$ at the final time slice to obtain a state in $\mathcal{H}_A\otimes \mathcal{H}_{A^c}$. Schematically, we have
\begin{equation}
    \begin{tikzpicture}[scale=1.,baseline = {(current bounding box.center)}]
\def \x {0.6};
\filldraw[gray!15] (0,0) -- (4,0) -- (4,\x) -- (0,\x) -- cycle;
\filldraw[white] (1,\x+0.01) arc (180:360:0.25 and 0.25);
\filldraw[white] (2.5,\x+0.01) arc (180:360:0.25 and 0.25);
\draw[thick] (0,0) -- (4,0) node[right]{$\ket{\psi}\in\mathcal{H}$};
\draw[thick] (0,\x) -- (1, \x);
\draw[thick] (1.5, \x) -- (2.5, \x);
\draw[thick] (3, \x) -- (4, \x) node[right]{$\ket{i_\epsilon \psi} \in \mathcal{H}_A \otimes \mathcal{H}_{A^c}$};
\draw[thick,lightblue] (1,\x) arc (180:360:0.25 and 0.25);
\draw[thick,lightblue] (2.5,\x) arc (180:360:0.25 and 0.25);
\node[above] at (0.6,\x) {$A^c$};
\node[above] at (2,\x) {$A$};
\node[above] at (3.5,\x) {$A^c$};
\draw[<->,>=stealth] (1,\x+0.05) --node[above]{\scriptsize$\epsilon$} (1.5,\x+0.05);
\draw[<->,>=stealth] (2.5,\x+0.05) --node[above]{\scriptsize$\epsilon$} (3,\x+0.05);
\end{tikzpicture}  
\label{eq: i_xi-map}
\end{equation}
The shaded area between $\psi$ and $i_\epsilon\psi$ represents the Euclidean path integral, and the blue arcs represent the edges where we assign boundary conditions for the field.\footnote{In general, the cutting map should satisfy a ``shrinkable" condition: any two states $\psi,\psi'$ under the embedding obey
$ \langle i_\epsilon \psi | i_\epsilon \psi' \rangle = e^{-S_{\text{cl}}} \langle \psi | \psi' \rangle \,$ when $\epsilon \ll 1$, where $S_{\text{cl}}$ is localized on the regularized entangling surface $\partial A \times S^1$, a thin torus in the case of bipartite entanglement~\cite{Jafferis:2019wkd}. Importantly, $S_{\text{cl}}$ is independent of $\psi,\psi'$ and is a local counter term. Therefore, the boundary condition on $\partial A \times S^1$ ``shrinks" to a trivial codimension-2 operator in the limit $\epsilon \rightarrow 0$. The Dirichlet boundary conditions for the Chern-Simons theory obey this condition.} To visualize the above picture in 3D, one should imagine the region $A$ inside the $A^c$ plane, and a ``half-torus" excised around the boundary $\partial A$. The resulting state will depend on the regularization parameter $\epsilon$.
To give a physical picture of the construction of $\ket{i_\epsilon\psi}$, we can imagine cutting the system along the boundary of $A$, creating two disks with counter-propagating edge modes. We then mend the cut by coupling the two edge modes, generating an energy gap of order $\mathcal{O}(1/\epsilon)$.

\begin{figure}
    \centering
    \begin{tikzpicture}
        \node at (0,0) {
        \begin{tikzpicture}
        \foreach \yy in {0,1.5,2.7}{
        \node at (0,\yy) {
        \begin{tikzpicture}
        \def\hh{.3};
        \foreach \sss in {-1,1}{
        \begin{scope}[yscale=\sss]
            \draw[semithick] (-.6,\hh) --++ (1.2,0);
            \draw[semithick] (1.,\hh) --++ (.8,0);
            \draw[semithick] (-1.,\hh) --++ (-.8,0);
            \draw[semithick,lightblue] plot [smooth,tension=2] coordinates {(-.6,\hh) (-.8,\hh-.12) (-1,\hh)};
            \draw[semithick,lightblue] plot [smooth,tension=2] coordinates {(.6,\hh) (.8,\hh-.12) (1,\hh)};
        \end{scope}
        }
        \foreach \sss in {-1,1}{
        \begin{scope}[xscale=\sss]
        \draw[thick,orange,->] plot [smooth,tension=.8] coordinates {(1.6,\hh) (1.8,\hh+.1) (2.0,0) (1.8,-\hh-.1) (1.6,-\hh)};
        \draw[thick,orange,->] plot [smooth,tension=.9] coordinates {(1.3,\hh+.02) (1.8,\hh+.2) (2.2,0) (1.8,-\hh-.2) (1.3,-\hh-.02)};
        \end{scope}
        }
        \end{tikzpicture}
        };
        }
        \draw[thick,dotted] (-1.1,.75) node[left=.2] {\small $n$ times} -- (1.2,.75);
        \foreach \xx in {-.3,.3}{
        \begin{scope}[every path/.style={thick,orange,->}]
            \draw (\xx,-.65) --++ (0,.3);
            \draw (\xx,.35) --++ (0,.3);
            \draw (\xx,.8) --++ (0,.3);
            \draw (\xx,1.85) -- (\xx,2.35);
            \draw (\xx,3.05) --++ (0,.3);
        \end{scope}}
        \end{tikzpicture}
        };
        \node at (2.9,0) {\huge $\Rightarrow$};
        \node at (4.7,0) {\includegraphics[width=.35\linewidth]{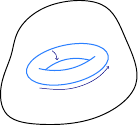}};
        \node at (5,-.7) {\small $L$};
        \node at (4.6,.25) {\tiny $n\epsilon$};
        \node at (5,1.7) {\Large $S^3$};
    \end{tikzpicture}
    \caption{Calculation of the $n$th R\'enyi entropy using the extended Hilbert space approach. Gluing $n$ copies of $\rho$ results in the partition function on a single solid torus, or, equivalently, a single copy of $S^3$ with a solid torus excised.}
    \label{fig: torus-edge}
\end{figure}
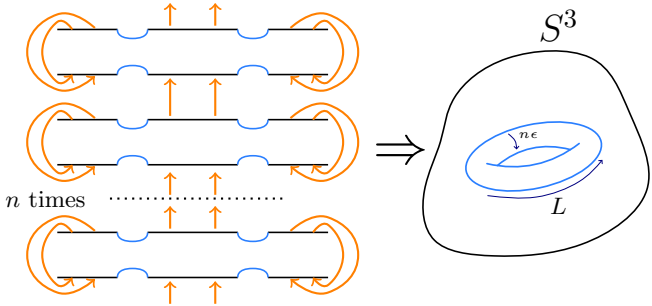

We can use this approach to compute the $n$-th R\'enyi  of the region $A$. The $n$-th moment $\Tr\rho^n_A$ is calculated by taking $n$ copies of $\ket{i_\epsilon\psi}$, gluing the region $A$ of replica $i$ to $i+1$ (mod $n$), and the region $A^c$ to itself. The resulting path integral should be carried out on $S^3$ with a torus-shaped boundary, whose handles have circumferences $L$ and $n\epsilon$  (see Fig.\ \ref{fig: torus-edge}). This partition function can be calculated directly by employing modular transformations. We present a different method which will be useful in the general case. 
The partition function is 
\begin{equation}
    \Tr \rho^n_A = \frac{1}{N^n}Z\qty(\vcenter{\hbox{\includegraphics[width=.3\linewidth]{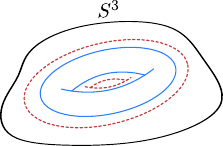}}})
    \label{eq: renyi-part-fun}
\end{equation}
where the solid blue torus is the boundary and $N = \langle i_\epsilon \psi | i_\epsilon \psi \rangle$ is the wavefunction normalization factor. We ``cut" the manifold along the dotted red torus by inserting a resolution of identity of the TQFT, where each state $\ket{\psi_a}$ is represented by a solid torus with a Wilson loop inserted
\begin{equation}
    \ket{\psi_a}=\vcenter{\hbox{\includegraphics[width=.25\linewidth]{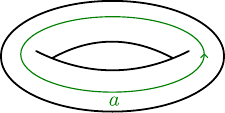}}}
\end{equation}
We thus have 
\begin{equation}
    \Tr\rho_A^n=\frac{1}{N^n}\sum_a C_a Z\qty(\vcenter{\hbox{\includegraphics[width=.28\linewidth]{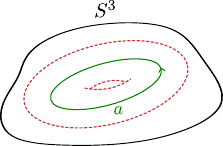}}})Z\qty(\vcenter{\hbox{\includegraphics[width=.18\linewidth]{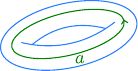}}})
    \label{eq: renyi-entropy-sum}
\end{equation}
The terms $C_a$ are normalization constants accounting for the path integral representation of inserted states $|\psi_a\rangle$. The last term is the partition function of the solid torus with open boundary conditions and a Wilson line of charge $a$ inserted. It is given by the character $\chi_a(iL/n\epsilon)$ \cite{elitzur1989remarks}. In the limit $L/\epsilon \gg 1$, we have $\chi_a(iL/n\epsilon)\approx e^{2\pi\frac{L}{n\epsilon} (\frac{c_+}{24}-h_a)}$ with $c_+ = c + \bar{c}$ being the total central charge and $h_a \geq 0$ the conformal dimension. We will take $c_+ = c_-$ in the case of a purely chiral CFT. The vacuum sector (with $a=0$) dominates, and we have
\begin{equation}
    \Tr \rho_A^n \approx \frac{1}{N^n}C_0Z\qty(S^3)Z\qty(\vcenter{\hbox{\includegraphics[width=.2\linewidth]{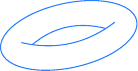}}})
\end{equation}
which can be compared with \eqref{eq: separated-pf} in the case of a genus $g=1$. Here $C_0=Z(S^2\times S^1)=1$ and $Z(S^3)=D^{-1}$, we therefore get
\begin{equation}
    \Tr \rho_A^n=\frac{1}{N^n}D^{-1}e^{\frac{2\pi c_+}{24}\frac{L}{n\epsilon}}.
\end{equation}
The computation of the wavefunction normalization factor $N$ is identical to the above procedure except that the boundary torus has circumferences $L$ and $2\pi \epsilon$. We obtain the R\'enyi entropies as
\begin{align}
    S_n(A)=-\frac{1}{n-1}\log\Tr\rho_A^n \approx \frac{2\pi c_+}{24\epsilon} \frac{n+1}{n} L-\log D
    \label{eq: Sn}
\end{align}
up to corrections that are exponentially small in $L/\epsilon$. The second term gives the topological entanglement entropy 
\footnote{Note that the discussion above shows that the entanglement-cut picture is \textit{equivalent} to the Li-Haldane conjecture \cite{li2008entanglement}. Indeed, \eqref{eq: renyi-part-fun} can alternatively be written as $\Tr\rho_A^n=\frac{1}{N^n}\chi_0(n\epsilon/L)$, which is, by definition, the sum over $\sum_i e^{-n\epsilon E_i/L}$, where $E_i$ is the spectrum of the CFT in the vacuum sector. Since the traces match for any $n$, we conclude that the spectrum of $K=-\log\rho$ is $\epsilon E_i/L$, up to a constant shift. This provides an alternative argument to \cite{qi:2012prl}, under the assumption of a Chern-Simons description.}.

\subsection{Multipartite entanglement}
We now extend our analysis to multipartite entanglement measures, specifically the TME~\eqref{eq: multi-ent}. The generalization of the cutting map is straightforward: we define the embedding by regularizing the neighborhoods along all region boundaries shown in Fig.~\ref{fig: pizza}. This maps the entanglement measure to a path integral on a three-manifold $M\backslash W$, where $W$ is an excised tubular neighborhood around the permutation defect. Compared to the bipartite case, two complications arise. (1) The closed bulk $M$ can be more complicated than $S^3$, such as the lens space $L(r,1)$ in the case of $\Phi_r$. (2) the boundary surface $\Sigma = \partial W$ is not a torus, but rather a higher-genus surface. For example, even calculating the wavefunction normalization $\langle i_\epsilon \psi | i_\epsilon \psi \rangle$  involves a genus-3 surface. 

Nevertheless, the general principles still hold. 
To compute the TME $\mathcal{M}$, we now perform the cut along a surrounding genus-$g$ surface (drawn below as a genus-3 surface for the sake of clarity) by inserting the complete set of states of the form
\begin{equation}
        \ket{\psi_\qty{a_i}}=\vcenter{\hbox{\includegraphics[width=.5\linewidth]{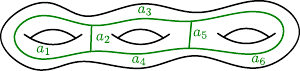}}}\,,
\end{equation}
where $a_i$ label the CFT states propagating along the handles.
Inserting the resolution of identity, we obtain
\begin{equation}
\begin{aligned}
    \mathcal{M}&=\frac{1}{N^R}Z\qty(\vcenter{\hbox{\def\svgwidth{.4\linewidth}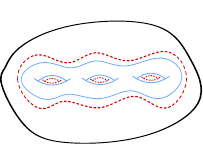}}) \\
    &=\frac{1}{N^R}\sum_\qty{a_i}C_\qty{a_i}Z\qty(\vcenter{\hbox{\relscale{.6}\def\svgwidth{.35\linewidth}\input{g_3_boundary_with_charge.pdf_tex}}}) \\
    &\qquad\qquad\qquad\times Z\qty(\vcenter{\hbox{\def\svgwidth{.3\linewidth}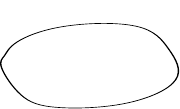}})
\end{aligned}
    \label{eq: high-g-bulk-edge-sep}
\end{equation}
The propagation of a state $a_i$ along each handle of $\Sigma$ yields a contribution of the form $e^{2\pi\frac{L}{\epsilon}(\frac{c_+}{24}-h_{a_i})}$, where $L$ and $\epsilon$ denote the length and circumference of the handle, respectively. As $\epsilon/L\to0$, the term with all $a_i=0$ dominates the sum. This leads to the factorized form in Eq.~\eqref{eq: separated-pf}
\begin{equation}
    \mathcal{M}=\frac{1}{N^R}C_{\qty{0}}Z(M)Z(\Sigma)\,.
    \label{eq: M-separates-bulk-edge}
\end{equation}
The normalization coefficient $C_\qty{0}$ is given as the partition function of the manifold $H_g$ obtained by gluing two genus-$g$ handlebodies (the surface of genus $g$ with the inside filled) with the identity map on their boundaries. It can alternatively be written as
\begin{equation}
    H_g=\underset{g\,\text{times}}{\underbrace{(S^{2}\times S^{1})\#(S^{2}\times S^{1})\#\,\dots\,\#(S^{2}\times S^{1})}},
\end{equation}
where $\#$ denotes a connected sum of manifolds, defined by deleting a ball around a point in each manifold and gluing together the resulting boundary spheres (for example, $S^3 \# S^3 = S^3$). Using the TQFT formula \cite{witten1989quantum}
\begin{equation}
    Z(M_1\#M_2)=Z(M_1)Z(M_2)/Z(S^3)
\end{equation}
we get
\begin{equation}
    C_\qty{0}=Z(H_g)^{-1}=(Z(S^3))^{g-1}=D^{1-g}.
\end{equation}
If $Z(M)$ vanishes exactly (for example, Lens-space multi-entropy \eqref{eq: phi_r} for the $\nu=1/2$ bosonic Laughlin state), we instead get the contribution with the lowest scaling dimensions $h_{a_i}$ other than the vacuum. We focus here on the generic non-vanishing case.

The calculation above relies on a subtle assumption: all handles of $\Sigma$ are degenerate. That is, the only handles are the thin, long tubes around the defects, and $\Sigma$ does not have additional ``small" handles around the vertices. Otherwise, the approximation of taking only the vacuum sector from Eq.~\eqref{eq: high-g-bulk-edge-sep} to \eqref{eq: M-separates-bulk-edge} breaks down. In App.\ \ref{app: m-manifold-topology} we show that the calculation is valid provided that the space $M$ obtained by gluing is a manifold (rather than a CW complex in the general case). Specifically, we require
\begin{equation}
    \abs{\pi_{IJ}}+\abs{\pi_{JK}}+\abs{\pi_{IK}}=2+R
    \label{eq: m-is-manifold-req}
\end{equation}
for any three distinct regions $I,J,K\in\qty{A,B,C,\Lambda}$, where $\pi_{IJ}=\pi_I\pi_J^{-1}$ and $\abs{\pi}$ is the number of independent cycles in the permutation $\pi$. \footnote{We also assume that the group generated by $\pi_{IJ},\pi_{JK}$ acts transitively on the replicas for any distinct $I,J,K$.} This condition holds for all entanglement measures considered in this work. An example where the condition does not hold is the higher multi-entropies considered by \cite{gadde_new_2022, harper_multi-entropy_2024} in the context of holography. These were defined as follows: for $q^2$ replicas, with $q\ge2$, label the replicas with indices $(i,j)\in\Z_q^2$ and set $\pi_A(i,j)=(i+1,j),\pi_B=(i,j+1)$. These measures violate the condition above for any $q\ge3$. A related issue arises when calculating R\'enyi entanglement measures in 1+1d CFTs \cite{calabrese2012entanglement}.

We remark that our results are fully consistent with the previous results obtained within the TQFT picture~\cite{dong_topological_2008, wen2016topological}. The consistency rests on two observations. First, prior works focused on time-reversal invariant entanglement measures, for which the edge contribution $Z(\Sigma)$ derived in our framework reduces to a non-universal area-law term, leaving the bulk piece $Z(M)$ as the sole universal contribution. Second, one might question whether the factor $C_\qty{0} = D^{1-g}$ introduces a discrepancy. However, we show in App.~\ref{app: m-manifold-topology} that
\begin{equation}
    \frac{\log C_\qty{0}}{\log D}=1-g=\sum_{v_i}o_{IJK}-\frac{1}{2}\qty(\abs{\pi_{IJ}}+\abs{\pi_{JK}}+\abs{\pi_{IK}}),
    \label{eq: sigma-genus}
\end{equation}
where $I,J,K$ label the three regions that border $v_i$, and $o_{IJK}$ is the number of independent orbits of the group generated by $\pi_{IJ},\pi_{JK}$ on the $R$ replicas. Namely,  $C_\qty{0}$ effectively behaves as a product of corner contributions. Therefore, in any subtraction scheme designed to isolate topological invariants (such as the Kitaev-Preskill and Levin-Wen prescriptions~\cite{Kitaev_Preskill_2006, Levin_Wen_2006}), these corner terms cancel out, preserving the agreement with standard TQFT results.\footnote{From a field-theoretical perspective, $\mathcal{M}$ includes a multiplicative local geometric counterterm $e^{-S_{\text{cl}}}$ that can absorb $C_\qty{0} = D^{1-g}$. Specifically, the integral of the scalar curvature $S_{\text{cl}} \ni \int_\Sigma R_\epsilon \sqrt{g_\epsilon} d^2 x$ yields a contribution proportional to $1-g$ via the Gauss-Bonnet theorem.}

Finally, we address the sensitivity of our results to the specific regularization scheme. As an example, in physical applications, the cutoff scale $\epsilon\ll L$ might be position dependent. Given the topological nature of the TME argued in Sec.~\ref{sec: overview}, the universal results must be robust against such microscopic details. 
In App.~\ref{app: sigma-complex-structure}, we provide a more specific discussion of this question within the field-theoretic framework. By mapping the choice of regularization to a choice of complex structure on $\Sigma$, we show explicitly that the details of the regularization procedure do not affect the universal results considered in this paper.

\section{Calculation of the edge partition function}
\label{sec: fn_coordinates}

In this section, we detail the calculation of the edge partition function $Z(\Sigma)$, focusing on extracting the phase determined by the permutations $\pi_{A,B,C}$. 
The calculation relies on decomposing the high-genus surface $\Sigma$ into pair-of-pants (POPs).
We then utilize Fenchel-Nielsen coordinates to characterize the gluing geometry, specifically the twist angles.
As established previously, these twist parameters are the essential quantities that dictate the phase of $Z(\Sigma)$.

\subsection{General strategy}
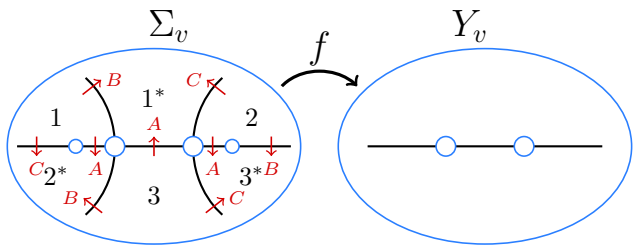
\begin{figure}
    \centering
    \begin{tikzpicture}
        \node at (-2.2,0) {
        \begin{tikzpicture}[scale=1.3]
            \draw[semithick, lightblue] (0,0) ellipse (1.5 cm and 1cm);
            \draw[thick] (-1.4,0) -- (1.4,0);
            \foreach \fl in {-1,1}
            {
                \draw[thick] plot [smooth,tension=1.2] coordinates {(-.7*\fl,.7) (-.4*\fl,0) (-.7*\fl,-.7)};
                \draw[fill=white, draw=lightblue,semithick] (-.4*\fl,0) circle (.1);
                \draw[fill=white, draw=lightblue,semithick] (-.8*\fl,0) circle (.07);
                }
            \begin{scope}[every node/.style={font={\normalsize}}]
                \node at (0,-.5) {$3$};
                \node at (0,.5) {$1^*$};
                \node at (1.,-.3) {$3^*$};
                \node at (+1.,.3) {$2$};
                \node at (-1.,-.3) {$2^*$};
                \node at (-1.,.3) {$1$};
            \end{scope}
            \begin{scope}[every path/.style={semithick,darkred,->},every node/.style={darkred,font={\scriptsize}}]
                \draw (0,-.1) -- (0,.1) node[above=-.03] {$A$};
                \draw (.6,.1) -- ++(0,-.2) node[below=-.03] {$A$};
                \draw (-.6,.1) -- ++(0,-.2) node[below=-.03] {$A$};
                \draw (1.2,.1) -- ++(0,-.2) node[below=-.03] {$B$};
                \draw (-1.2,.1) -- ++(0,-.2) node[below=-.03] {$C$};
                \draw (.53,-.68) --++ (.19,.15) node[right=-.06] {$C$};
                \draw (-.53,-.68) --++ (-.19,.15) node[left=-.06] {$B$};
                \draw (.73,.53) --++ (-.19,.15) node[left=-.06] {$C$};
                \draw (-.73,.53) --++ (.19,.15) node[right=-.06] {$B$};
            \end{scope}
        \end{tikzpicture}
        };
        \node at (2.2,0) {
        \begin{tikzpicture}[scale=1.3]
            \draw[semithick, lightblue] (0,0) ellipse (1.5 cm and 1cm);
            \draw[thick] (-1.2,0) -- (1.2,0);
            \foreach \fl in {-1,1}
                \draw[fill=white, draw=lightblue,semithick] (-.4*\fl,0) circle (.1);
        \end{tikzpicture}
        };
        \draw[very thick,->] plot [smooth,tension=1.2] coordinates {(-.5,.8) (0,1.0) (.5,.8)};
        \node at (0,1.3) {\Large $f$};
        \node at (-2,1.6) {\Large $\Sigma_v$};
        \node at (2,1.6) {\Large $Y_v$};
    \end{tikzpicture}
    \caption{The structure of $\Sigma_v$ and its mapping to $Y_v$. Here we used the example of $v_1$ and the permutations giving $J_1$, with 3 replicas. Each of the surfaces $Y_v$ is a three-punctured sphere, drawn here as the complex plane with two holes (ramification points) at $z=0$,$1$, and a hole at infinity (blue curves). There are also two additional holes in $\Sigma$ with no ramification (i.e., $f'(z)\neq 0$), which will be ignored later in the paper. The horizontal line is the $t=0$ line, which divides the $t<0$ side of the surface, labeled with the replica index $i$, and the $t>0$ side, labeled with $i^*$. The covering surface $\Sigma_v$ covers a generic point of $Y_v$ $R$ times, and is obtained from the permutation operators between the replicas. The black lines in $\Sigma_v$ map to the $t=0$ line in $Y_v$. We indicated with red labels which replica permutation is acted by when the $t=0$ lines are crossed. 
    }
    \label{fig: covering-map}
\end{figure}

For a R\'enyi-$R$ TME measure, the surface $\Sigma$ is identified as a $R$-sheet covering surface of the regularization surface $Y$. 
Related to our discussion in the previous section, $Y$ is associated with the norm squared of the wavefunction $|i_\epsilon \psi \rangle$ after the embedding. 
To construct $\Sigma$, we take $R$ copies of $Y$, cut, and glue them according to the permutations $\pi_{A,B,C}$. The general strategy for analyzing the geometry of $\Sigma$ is to break it up into sub-surfaces $\Sigma_{v_i}$ around each vertex, analyze each of these sub-surfaces separately, then obtain the twist angles $\tau_i$ that describe how they are glued together. This is achieved as follows:
\begin{enumerate}
    \item Cut the surface $Y$ along the cross sections of the thin long tubes around the boundary of the regions (the red lines in Fig.\ \ref{fig: regulating-surface}). 
    This decomposes $Y$ into a set of vertex surfaces $Y_v$, each of which is a three-punctured sphere.
    As drawn on the right side of Fig.~\ref{fig: covering-map}, we can conformally map $Y_v$ into a Riemann surface $\bar{\mathbb{C}}$ with three punctures located at the reference points $z = 0,1,\infty$. The radii of these punctures on the Riemann surface are exponentially small in $L/\epsilon$\footnote{The map that takes the region of the puncture to a cylinder of circumference $2\pi \epsilon$ and length $L$ will be of the form $w=\epsilon\ln z$. See e.g. \cite{polchinski2005string}}.
    \item Find the covering vertex surface $\Sigma_v$, i.e., the part of $\Sigma$ that covers $Y_v$. Provided that the condition \eqref{eq: m-is-manifold-req} holds, each surface $\Sigma_v$ is topologically a sphere with multiple (at least 3) punctures (see Fig.~\ref{fig: covering-map}). Consequently, characterizing $\Sigma_v$ amounts to constructing an analytic `uniformization" map $f:\bar{\mathbb{C}}\to\bar{\mathbb{C}}$, which maps the $R$-sheet cover onto the base sphere, and maps punctures on $\Sigma_v$ to those on $Y_v$ at $z = 0,1,\infty$. The structure of $f$ near the punctures is constrained by the permutation. Specifically, near a pre-image $z_0$ of one of the three reference points, we have $f(z) \approx f(z_0) + (z - z_0)^n$, the ramification index $n$ is the length of the permutation cycle. The problem of finding such covering maps was studied in the context of correlation functions of twist defects in CFT, for example in \cite{lunin_correlation_2001}, and we use their results in multiple occasions. 
    \item Glue the vertex covering surfaces. Using the explicit description of the surfaces $\Sigma_v$, we can calculate the twists $\tau_i$ between two glued surfaces, and obtain the partition function using \eqref{eq: sum-of-twists}. The calculation detail is described below.
\end{enumerate}

\subsection{Fenchel-Nielsen Coordinates}
\begin{figure}
    \centering
    \begin{tikzpicture}
    \node at (0,0) {
    \def\svgscale{1.3}
    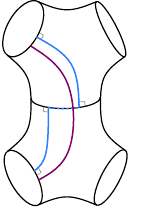};
    \node at (-2,2) {(a)};
    \node at (2,2) {(b)};
    \node at (3.5,0){
    \def\svgscale{1.3}
    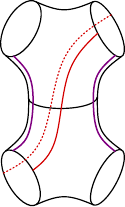
    };
    \end{tikzpicture}
    \caption{(a) Using seams to measure the twist angles $\tau_i$. The seam (purple line) is first homotoped such that it agrees with the geodesics (blue lines) inside each of the POPs. The angle $\tau_i$ is then the angle traveled by the seam between the two geodesics. (b) In our cases of interest, we might have four different ``seams" between the two geodesics. Among them, two will measure the same value of $\tau_i$ (the purple lines), and two (the red lines, the solid one of which is on the front and the dashed one on the back) will measure the values $\tau_i\pm\pi$. We take the value as measured by the two seams that agree.
    }
    \label{fig: twist-parameters}
\end{figure}

To quantify the angles between the glued surfaces $\Sigma_v$, we use the Fenchel-Nielsen coordinates, a canonical coordinate system for the Teichm\"uller space of hyperbolic surfaces.
In this section, we review the definition of these coordinates, with a focus on the twist parameters relevant to our partition function calculation.
Our discussion will mostly follow Ref.~\cite{farb2011primer}. The reader can refer to standard textbooks on Teichm\"uller theory for more details~\cite{farb2011primer, imayoshi2012introduction, hubbard2016teichmuller}.

The space of conformal structures of all closed oriented surfaces of genus $g>1$ (the Teichm\"uller space) has a real dimension $6g-6$. 
The Fenchel-Nielsen (FN) coordinates provide an explicit parametrization of this space based on a pair-of-pants (POP) decomposition.
The main idea relies on the uniformization theorem: for any surface $\Sigma$ with negative Euler characteristic equipped with a conformal structure, there is a unique hyperbolic metric of constant curvature that is compatible with this structure. 
In this metric, the conformal structure of each individual POP is uniquely specified by the geodesic length $l_i$ of its three boundaries. They provide the $3g-3$ length parameters of the FN coordinates.
Our problem is defined in the ``thin-tube" limit, where every puncture is pinched $l_i \rightarrow 0$. Therefore, we focus exclusively on the remaining degrees of freedom.

The remaining $3g-3$ coordinates are the twist parameters $\tau_i$, which characterize the angle at which two adjacent POPs are glued. To define them, we note that inside each POP, any two boundary punctures are connected by a unique geodesic of minimal length, the orthogeodesic~\cite{farb2011primer}. 
A naive definition of the twist parameters $\tau_i$ is the angular offset between the orthogeodesics of adjacent POPs at the gluing interface. However, this procedure only defines $\tau_i$ mod $2\pi$, whereas the phase of the partition function \eqref{eq: sum-of-twists} depends on $\tau_i$ as real parameters. 
To resolve this ambiguity, we fix a global frame of reference by introducing a set of \textit{seams} on the surface. These are nonintersecting closed curves $\qty{\b_i}$ on $\Sigma$ such that exactly three seams traverse each POP, one connecting each pair of the three punctures. 
We use these seams to define the twists as follows: consider a seam passing between the two POPs, homotope it to align with the orthogeodesics inside the POPs; the twist $\tau_i$ is the angle the seam traverses along the boundary of the two POPs (see Fig.~\ref{fig: twist-parameters}a).
Our sign convention will be that $\tau_i$ is positive when the seam turns right as it touches the boundary between the POPs.

The fact that an additional choice of seams is needed to fix the angles $\tau_i$ can be seen as the manifestation of the framing anomaly in the partition function $Z(\Sigma)$. Namely, we see that without fixing the seams, the result is only defined up to an integer power of $\exp(2\pi i c_-/24)$. We do not have a unique choice for fixing this phase factor.\footnote{Mathematically, this stems from the fact that $Z(\Sigma)$ depends on $\Sigma$ as a point in the Teichm\"uller space of genus $g$ surfaces, rather than as a point in the moduli space \cite{Witten_1990}.} We will now suggest a prescription that fixes the ambiguity in the examples we consider, and show below that it matches the numerical calculations. Note that it is sufficient to resolve the ambiguity for a single model, as this integer phase factor is model-independent.

In our case, a natural choice of seams will be the preimages of the $t=0$ lines (the black lines in Fig.\ \ref{fig: covering-map}). This is motivated by the fact that, when no permutations are applied, they are flat and therefore represent no twist. The issue with this choice is that, generally, more than three $t=0$ lines traverse each POP. We propose to remedy this as follows: since the lines are always non-intersecting, between any adjacent POPs there are at most four such homotopically distinct lines. In the cases of interest here, two will measure the same value of $\tau_i$ (the purple lines in Fig.\ \ref{fig: twist-parameters}b), and the other two will measure $\tau_i\pm \pi$ (the red lines). We will take $\tau_i$ as measured by the two lines whose values agree. 

\section{R\'enyi modular commutator}
\label{sec: renyi-mc}
\begin{figure}
    \centering
    \includegraphics[scale=.9]{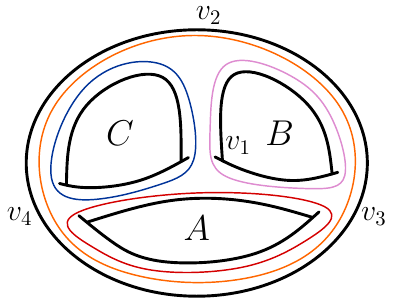}
    \caption{Illustration of the seams used to calculate $J_n$, as drawn in Fig.\ \ref{fig: j_n_pops}. Handles corresponding to punctures with ramification 1 are not included in the figure. 
    }
    \label{fig: j_n_seams}
\end{figure}
In the TQFT picture, each density matrix in \eqref{eq: j_n-def} is represented as a ball~\cite{dong_topological_2008, sheffer2025extracting}. The manifold $M$ obtained by gluing the the density matrices and taking the trace is topologically $S^3$. Choosing the framing on $M$ such that $Z(M)=Z(S^3)=D^{-1}$, the only contribution to the phase of $J_n$ is $Z(\Sigma)$. In this section, we calculate the phase contribution of $J_n$, deriving the formula \eqref{eq: j_n phases}. 

\begin{figure}
    \centering
    \newcommand{\jtwo}[3]{
    \begin{tikzpicture}[scale=.75]
    \setsepchar{,}
    \readlist\reg#1
    \readlist\ver#2
    \readlist\cols#3
        \begin{scope}[every path/.style={thick}]
        \draw (-2.4,0) -- (2.4,0);
        \foreach \rrr in {1,-1}{
        \foreach \xxx in {-1,1}{
        \begin{scope}[yscale=\rrr,xscale=\xxx]
        \draw plot [smooth,tension=1.1] coordinates {(1,0) (1.4,.7) (1.6,1.5)};
        \draw plot [smooth,tension=.8] coordinates {(1,0) (.6,-.7) (.4,-1.8)};
        \end{scope}
        }}
        \end{scope}
        region labels
        \begin{scope}[every node/.style={font=\normalsize}]
            \node at (0,-1) {$\reg[1]$};
            \node at (1,-1.3) {$\reg[2]$};
            \node at (1.9,-.5) {$\reg[3]$};
            \node at (1.9,.5) {$\reg[4]$};
            \node at (1,1.3) {$\reg[5]$};
            \node at (0,1) {$\reg[6]$};
            \node at (-1,1.3) {$\reg[7]$};
            \node at (-1.9,.5) {$\reg[8]$};
            \node at (-1.9,-.5) {$\reg[9]$};
            \node at (-1,-1.3) {$\reg[10]$};
        \end{scope}
        \begin{scope}[every path/.append style={dotted,line width=2.5}]
        \draw[{\cols[1]}] (-1,0) -- (1,0);
        \draw[{\cols[2]}] plot [smooth,tension=.8] coordinates {(-1,0) (-.6,-.7) (-.4,-1.8)};
        \draw[{\cols[3]}] plot [smooth,tension=.8] coordinates {(1,0) (.6,.7) (.4,1.8)};
        \end{scope}
        
        \foreach \xx in {1,-1}
            \draw[fill=white, draw=lightblue,thick] (\xx,0) circle (.15);

        \begin{scope}[every node/.style={lightblue}]
            \node at (-.55,.2) {$v_\ver[1]$};
            \node at (.55,.2) {$v_\ver[2]$};
            \node at (-2.8,1) {$v_\ver[3]$};
        \end{scope}
        \draw[thick,lightblue] plot [smooth cycle,tension=.8] coordinates {(-2.7,0) (-1.5,1.6) (1.5,1.6) (2.7,0) (1.5,-1.6) (-1.5,-1.6)};

    \end{tikzpicture}
    }
    \begin{tikzpicture}
    \definecolor{lavender}{HTML}{de8cd0ff}
    \node at (0,0) {\Large $\Sigma_{v_1}$:};
    \node at (3,0) {\jtwo{{5,5^*,4,4^*,3,1^*,1,2^*,2,3^*}}{{3,4,2}}{{darkred,lavender,darkblue}}};
    \node at (0,-3.0) {\Large $\Sigma_{v_2}$:};
    \node at (3,-3.0) {\jtwo{{3,4^*,4,5^*,5,3^*,2,2^*,1,1^*}}{{4,3,1}}{{orange,darkblue,lavender}} };
    \node at (0,-3.0*2) {\Large $\Sigma_{v_3}$:};
    \node at (3,-3.0*2) {\jtwo{{5,5^*,4,4^*,3,3^*,2,2^*,1,1^*}}{{1,2,4}}{{lavender,darkred,orange}}};
    \node at (0,-3.0*3) {\Large $\Sigma_{v_4}$:};
    \node at (3,-3.0*3) {\jtwo{{3,4^*,4,5^*,5,1^*,1,2^*,2,3^*}}{{2,1,3}}{{darkblue,orange,darkred}}};
    \end{tikzpicture}

    \caption{
    The covering surfaces $\Sigma_v$ used in the calculation of $J_n$, presented here for $n=2$. The orthogeodesics of the POPs are the real (horizontal) lines, and the gluing angles are measured with respect to that line. The angles are measured using the colored seams in the picture, which are chosen according to the prescription described in Sec.\ \ref{sec: fn_coordinates}. Not drawn in this figure are additional holes corresponding to the pre-images of $w=0,1$ with ramification order $1$. 
    }
    \label{fig: j_n_pops}
\end{figure}
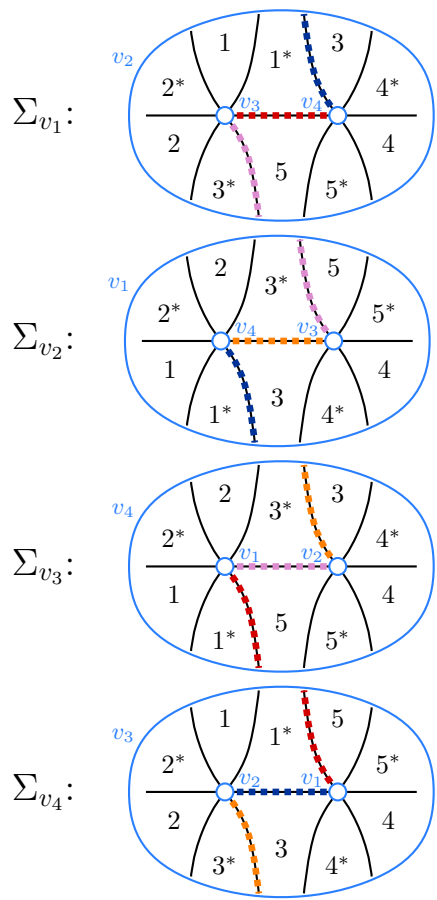

We first find the covering map $\Sigma_v\to Y_v$ in the vertices. This map does not depend directly on the permutations $\pi_1,\pi_2,\pi_3$ around the vertex, but rather on the form of the defect lines around the vertex, which are given by $\pi_1\pi_2^{-1},\pi_2\pi_3^{-1},\pi_3\pi_1^{-1}$. As a result, the same covering map can be used for all $\Sigma_v$. In general, we need a map that has a ramification point of order $n+1$ around $0,1$, and a ramification point of order $2n+1$ around infinity. That is, we want
\begin{equation}
\begin{aligned}
    f(w)\sim \begin{cases}
        w^{n+1} & w\to 0 \\
        (w-1)^{n+1}+1 & w\to 1 \\
        w^{2n+1} & w\to \infty \\
    \end{cases}
\end{aligned}
\label{eq: j_n-uniformization}
\end{equation}
See Ref.~\cite{lunin_correlation_2001} for the general result and this simpler version \eqref{eq: j_n-uniformization} is sufficient to draw the regions $\Sigma_v$ and understand how they are glued together. We draw the covering surfaces around each vertex in Fig.\ \ref{fig: j_n_pops}. Note that there are additional holes, corresponding to $n$ pre-images of $w=0,1$ with ramification order 1. These holes do not contribute any twists and are ignored in the calculation. 

The surfaces $\Sigma_v$ in these examples are POPs, and their orthogeodesics are placed on the real line (the horizontal line in Fig.\ \ref{fig: j_n_pops}). Each replica index continues between any two POPs at the edges where they are glued. We then choose the seams according to the prescription in Sec.~\ref{sec: fn_coordinates} and measure the twist angles. The topology of the seams is illustrated in Fig. \ref{fig: j_n_seams}. We find
\begin{equation}
    \begin{aligned}
    &
        \tau_{v_1v_2}=\tau_{v_3v_4}=-\frac{2\pi n}{2n+1}, \\
    &    \tau_{v_1v_3}=\tau_{v_1v_4}=\tau_{v_2v_3}=\tau_{v_2v_4}=\frac{\pi n}{n+1}.
    \end{aligned}
    \label{eq: j_n_taus}
\end{equation}
Since the holes between any two POPs are uniquely determined by the vertices they connect, we use the vertices $v_iv_j$ as the indices of the angles $\tau_i$. 
Plugging Eq.~\eqref{eq: j_n_taus} into Eq.~\eqref{eq: sum-of-twists} yields Eq.~\eqref{eq: j_n phases}, as we claimed.

Let us illustrate explicitly how the calculation is carried out in the example of $\tau_{v_1v_3}$. The calculation of other twist angles is similar. Consider the seam denoted by the dashed red line in Fig.\ \ref{fig: j_n_pops}. In $\Sigma_{v_1}$ this red seam already goes along the orthogeodesic. In $\Sigma_{v_3}$, we need to homotope this red seam to the orthogeodesic (the horizontal line). In the example drawn with $n=2$ it will then cover an angle $2\pi/3$ on the boundary between $v_1$ and $v_3$. For general $n$, the angle will be $n\pi/(n+1)$. We present another detailed example in App.\ \ref{app: tau-calculation}.

We remark that our calculation relied on a choice of seams, as well as assuming the canonical framing on the manifold $M=S^3$ (such that $Z(M)=D^{-1}$). While this choice is natural, we need to verify this result numerically, as shown later.

\section{Chiral contribution in the lens-space multi-entropy}
\label{sec: LSME}
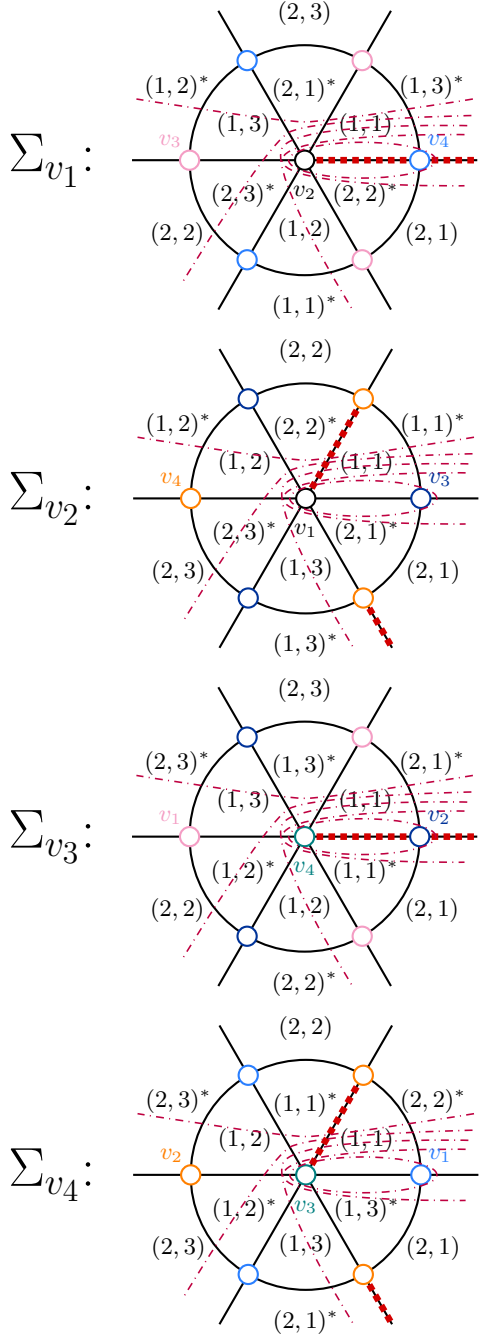
\begin{figure}[h!]
    \centering
    \newcommand{\phithree}[5]{
    \def\rr{1.8}
        \setsepchar{ }
        \readlist\inreg#1
        \readlist\outreg#2
        \readlist\vers#3
        \readlist\cols#4
        \readlist\angs#5
        \begin{tikzpicture}[scale=.85]
        \begin{scope}[every path/.style={thick}]
            \draw (0,0) circle (\rr);
            \foreach \qq in {0,60,...,300}
                \draw (0,0) -- (\qq:2.7);
        \end{scope}

        \begin{scope}[every path/.style={darkred,line width=2.5,dotted}]
            \draw (0,0) -- ({\angs[1]}:\rr);
            \draw (\angs[2]:\rr) -- (\angs[2]:2.7);
        \end{scope}

        \foreach \qq in {0,120,240}{
            \draw[fill=white,draw={\cols[3]},thick] (\qq:\rr) circle (.15);
            \draw[fill=white,draw={\cols[2]},thick] (\qq+60:\rr) circle (.15);
            }
        \draw[fill=white,draw={\cols[1]},thick] (0,0) circle (.15);
        \foreach \nn in {1,2,...,6}{
            \node at (-30+\nn*60:1.1) { $\inreg[\nn]$};
            \node at (-30+\nn*60:2.3) { $\outreg[\nn]$};
        } 
        
        \begin{scope}[every path/.append style={purple,dash dot,semithick}]
        \draw (.9,0) ellipse (1.15 and .28);
        \draw plot [smooth,tension=2] coordinates {(2.5,.40) (-.35,0) (2.5,-.40)};
        \draw plot [smooth,tension=.5] coordinates {(12:2.6) (150:.4) (-72:2.5)};
        \draw plot [smooth,tension=.3] coordinates {(15:2.7) (120:.5) (-135:2.7)};
        \draw plot [smooth,tension=.3] coordinates {(20:2.8) (90:.6) (-200:2.8)};

        \node[{\cols[1]}] at (0,-.5) {\small $v_{\vers[1]}$};
        \node[{\cols[2]}] at (-\rr-.3,.3) {\small $v_{\vers[2]}$};
        \node[{\cols[3]}] at (\rr+.3,.3) {\small $v_{\vers[3]}$};
        \end{scope}
        \end{tikzpicture}
    }
    \begin{tikzpicture}
    \node at (0,0) {\huge $\Sigma_{v_1}$:};
    \node at (3,0) {
    \phithree{{(1,1) (2,1)^* (1,3) (2,3)^* (1,2) (2,2)^*}}{{(1,3)^* (2,3) (1,2)^* (2,2) (1,1)^* (2,1)}}{{2 3 4}}{{black Lavender lightblue}}{{0 0}}};
    \node at (0,-4.5) {\huge $\Sigma_{v_2}$:};
    \node at (3,-4.5) { \phithree{{(1,1) (2,2)^* (1,2) (2,3)^* (1,3) (2,1)^*}}{{(1,1)^* (2,2) (1,2)^* (2,3) (1,3)^* (2,1)}}{{1 4 3}}{{black orange darkblue}}{{60 -60}}};
    \node at (0,-4.5*2) {\huge $\Sigma_{v_3}$:};
    \node at (3,-4.5*2) { \phithree{{(1,1) (1,3)^* (1,3) (1,2)^* (1,2) (1,1)^*}}{{(2,1)^* (2,3) (2,3)^* (2,2) (2,2)^* (2,1)}}{{4 1 2}}{{teal Lavender darkblue}}{{0 0}}};
    \node at (0,-4.5*3) {\huge $\Sigma_{v_4}$:};
    \node at (3,-4.5*3) { \phithree{{(1,1) (1,1)^* (1,2) (1,2)^* (1,3) (1,3)^*}}{{(2,2)^* (2,2) (2,3)^* (2,3) (2,1)^* (2,1)}}{{3 2 1}}{{teal orange lightblue}}{{60 -60}}};
    \end{tikzpicture}
    \caption{The covering surfaces $\Sigma_v$ used to calculate the phase contribution to $\Phi_r$ (here $r=3$). Since the surfaces have more than three punctures in this example, we consider a choice of POP decomposition of $\Sigma_v$ (dashed purple lines). The dotted dark red lines represent a choice of seams for calculating $\tau_i$. Note that the vertices $v_1,v_2$ and $v_3,v_4$ are connected via two punctures: one at 0 and one at $\infty$.}
    \label{fig: phi_r-pops}
\end{figure}
Here we calculate the fractional part of the phase contribution in \eqref{eq: phi_r}. There is an ambiguity in the calculation below, as there is no natural choice of POP decomposition, and there is possibly an additional framing anomaly in $Z(M)$. Therefore, our calculation of $Z(\Sigma)$ will only give $\sum_i \tau_i$ mod $2\pi$. We will then give some physical arguments to support the result \eqref{eq: phi_r}.

We first find the covering maps of $Y_v$. As in the previous case, the same covering map works for all vertices. Considering the permutations in Eq.~\eqref{eq: phi_r-def} and the structure of the vertices in Fig.~\ref{fig: regulating-surface}, each vertex has two holes whose corresponding permutations (which are applied when encircling the hole) are of order 2, and one hole whose corresponding permutation is of order $r$. Therefore, the covering maps should have $r$ preimages for the points $z = 0,1$, each with ramification 2, and 2 preimages of $z = \infty$, with ramification $r$. We can obtain the map by first unfolding the twists at the points $0,1$, then unfolding two $r$-fold twists. We obtain the map
\begin{equation}
    z=f(w)=\frac{(w^r+1)^2}{(w^r+1)^2-(w^r-1)^2}=\frac{(w^r+1)^2}{4w^{r}}.
\end{equation}
We see that the preimages of $0$ are at $w^r=-1$, the preimages of $1$ are at $w^r=1$, and the preimages of $\infty$ are at $0,\infty$. We can therefore draw the covering surfaces as in Fig.\ \ref{fig: phi_r-pops}. 

In contrast to the case of $J_n$, where some holes could be ignored so $\Sigma_{v_i}$ become three-punctured spheres, here $\Sigma_{v_i}$ have more punctures. To calculate the twists, we need to choose some POP decomposition of each of $\Sigma_v$. Unfortunately, there is no ``canonical" choice to be made here. After making such a choice (drawn in Fig.\ \ref{fig: phi_r-pops}), we find that the twists at each of the punctures at $w=0,\infty$ have $\tau_i=\pi(1-\frac{1}{r})$. We have 4 such punctures: two between $v_1,v_2$ and two between $v_3,v_4$. The contributions of all other twist angles are $\pm\pi/2$ and cancel mod $2\pi$. We therefore obtain
\begin{equation}
    \sum_i\tau_i=\frac{4\pi}{r}\mod2\pi,
\end{equation}
giving the second part in the exponent in \eqref{eq: phi_r}.

To obtain the first part of the phase in $\eqref{eq: phi_r}$ (the integer part inside the parenthesis), one has to properly deal with the framing of $M$ and $\Sigma$. We instead argue for the result based on two mild assumptions, and verify our results numerically later. In general, we can write
\begin{equation}
    \Phi_r=e^{-\frac{2\pi ic_-}{24}\qty(\kappa(r)+\frac{2}{r})}\sum_ad_a^2\theta_a^r.
\end{equation}
Where $\kappa(r)$ is some integer-valued function. As we show in App.\ \ref{app: real-phi_r}, hermiticity fixes $\kappa(2)=2$. We now add two additional assumptions: the first assumption will be that we can write
\begin{equation}
    \kappa(r)=ar+b
\end{equation}
for some integer coefficients $a,b$. This reflects the fact that we expect the amount of winding of the framing to be at most linear in $r$. The second assumption is that there is a well-defined $r\to1$ replica limit, in which $\Phi_r\to1$. This fixes $\kappa(1)=1$, and we conclude that $\kappa(r)=r$.

\section{Charged R\'enyi modular commutator}
\label{sec: quantum-hall}
Here we consider wavefunctions with a global $U(1)$ symmetry and prove Eq.~\eqref{eq: s-mu-n-phases}. The main tool for the calculation of $\SS_{\mu,n}$ is a modification of Eq.~\eqref{eq: single-twist-contribution} to incorporate the ground state in the sector with a symmetry defect. We consider a $U(1)$-symmetric chiral CFT defined on a ring, and implement a $U(1)$-twisted boundary condition by requiring
\begin{equation}
    \varphi(2\pi)=e^{iQ\alpha}\varphi(0)e^{-iQ\alpha}
\end{equation}
for an arbitrary field $\varphi$. Here $Q$ is the charge operator, and $\alpha \in [0,2\pi]$ labels the twist. The expectation value \eqref{eq: single-twist-contribution} of the momentum operator then becomes
\begin{equation}
    \mel{0_{\alpha_i}}{e^{i\tau_i P}}{0_{\alpha_i}}=e^{-\frac{i\tau_i c_-}{24}+\frac{i\sigma_{xy}\alpha_i^2\tau_i}{4\pi}}
    \label{eq: twist-with-charge}
\end{equation}
Eq.~\eqref{eq: twist-with-charge} is a consequence of the spectral flow. More physically, let the chiral CFT be the edge of a quantum Hall system. We can create an $\alpha$-twist on the edge by threading magnetic flux inside the bulk, which then pumps momentum (and energy) onto the edge. We explain this result in more detail in App. \ref{app: twisted-sectors}.

The phase contribution to $Z(\Sigma)$ is then given by
\begin{equation}
    Z(\Sigma)\propto\exp(i\sum_i-\frac{\tau_i c_-}{24}+\frac{\sigma_{xy}\alpha_i^2\tau_i}{4\pi}).
    \label{eq: z-sigma-charged}
\end{equation}
An important subtlety is that, for the calculation of $\SS_{\mu,n}$, we need to consider ``imaginary" symmetry defects. This, however, can simply be carried by analytically continuing $\alpha_i$ and setting them to be imaginary.

\begin{figure}[h!]
    \centering
    \newcommand{\smun}[3]{
        \setsepchar{ }
        \readlist\reg#1
        \readlist\vers#2
        \readlist\defs#3
        \begin{tikzpicture}[scale=.75]
        \foreach \nn in {1,2,...,6}{
            \draw[thick] (0,0) -- (\nn*60:2.0);
            \node at (\nn*60-30:1.5) {\large $\reg[\nn]$};
        }
        \begin{scope}[every path/.append style={ForestGreen}]
        \draw[thick,rotate=-60,opacity={\defs[1]}] (0,0) -- (2,0);
        \draw[line width=1.2,dotted,rotate=-60,opacity={\defs[1]}] (0,-0.04) -- (2,-0.04);
        \draw[thick,opacity={\defs[2]}] (0,0) -- (1,0);
        \draw[line width=1.2,dotted,opacity={\defs[2]}] (0,0.04) -- (1,0.04);
        \draw[thick,opacity={\defs[3]}] (1,0) -- (2,0);
        \draw[line width=1.2,dotted,opacity={\defs[3]}] (1,0.04) -- (2,0.04);
        \end{scope}

        \begin{scope}[every node/.append style={font={\large}}]
        \draw[thick,fill=white, draw=red] (0,0) circle (.15) node[red,above=.2] {\small $v_\vers[1]$};
        \draw[thick,fill=white, draw=darkblue] (1.,0) circle (.15) node[darkblue,above=.15]  {\small $v_\vers[2]$};
        \draw[thick, draw=lightblue] (0,0) circle (2.05);
        \node at (150:2.3) {\small \color{lightblue}$v_\vers[3]$};
        \end{scope}
        
        \end{tikzpicture}
    }
    \begin{tikzpicture}
    \node at (0,0) {\huge $\Sigma_{v_1}$:};
    \node at (3,0) {
    \smun{{3 3^* 2 2^* 1 1^*}}{{4 3 2}}{{1 1 0}}
};
    \node at (0,-3.2) {\huge $\Sigma_{v_2}$:};
    \node at (3,-3.2) {
    \smun{{1 2^* 2 3^* 3 1^*}}{{3 4 1}}{{0 0 1}}
};
    \node at (0,-3.2*2) {\huge $\Sigma_{v_3}$:};
    \node at (3,-3.2*2) {
    \smun{{3 3^* 2 2^* 1 1^*}}{{2 1 4}}{{0 0 1}}
};
    \node at (0,-3.2*3) {\huge $\Sigma_{v_4}$:};
    \node at (3,-3.2*3) {
    \smun{{1 2^* 2 3^* 3 1^*}}{{1 2 3}}{{1 1 0}}
};
    \end{tikzpicture}
    \caption{
    The covering surface use to calculate $\mathcal{S}_{\mu,n}$ (here $n=2$). The green lines correspond to the $U(1)$-defect lines. We note that through the puncture connecting $v_1v_4$ we have two defects with opposite orientations, and their contributions cancel. We did not draw punctures that carry no ramification and no defect lines through them.}
    \label{fig: sigma_mun-pops}
\end{figure}
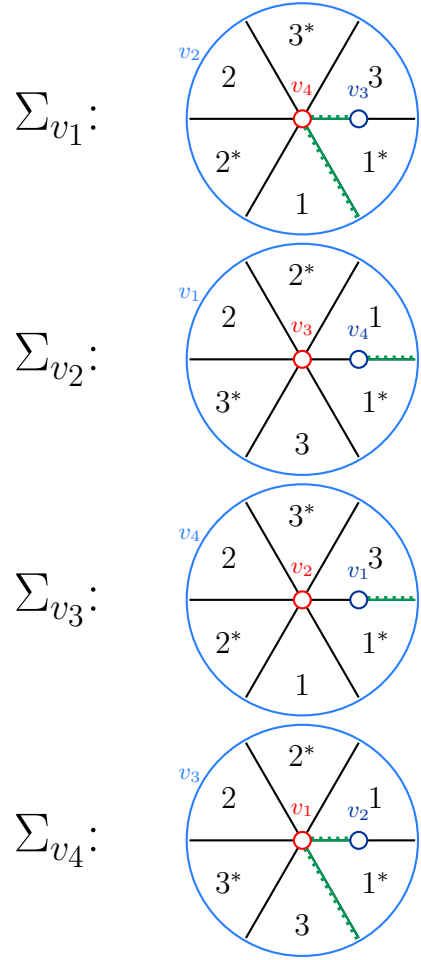

With this result at hand, we can use the method described in the two previous sections to obtain $Z(\Sigma)$. Each vertex has two ramification points of order $n+1$, which we can choose to put at $0,\infty$. The covering map on each vertex is then
\begin{equation}
    f(w)=w^{n+1}.
\end{equation}
In Fig.\ \ref{fig: sigma_mun-pops}, we draw the POPs at each vertex. We note that the surfaces $Y_{v_i}$ include additional $n$ punctures corresponding to the preimages of $1$ under $f$. We ignore them as these are not ramification points, and do not contribute to the twists. Using an imaginary defects, the twist and phase defect contributions are
\begin{equation}
    \begin{aligned}
        \tau_{v_1v_2}&=\tau_{v_3v_4}=-\frac{n\pi}{n+1}, \\
        \tau_{v_1v_4}&=\tau_{v_2v_3}=\frac{n\pi}{n+1}, \\
        \alpha_{v_1v_2}&=\alpha_{v_3v_4}=i\mu, \\
        \alpha_{v_1v_4}&=\alpha_{v_2v_3}=0.
    \end{aligned}
\end{equation}
Using \eqref{eq: z-sigma-charged} we obtain \eqref{eq: s-mu-n-phases}. Notably, we see that the contributions of the twist angles alone cancel. This is expected: when $\mu=0$, the value $\SS_{0,n}$ is the $n$-th moment of the reduced density matrix, and has no phase contribution.

\section{Numerical verification}
\label{sec: numerics}

We have systematically ignored framing anomaly in our analysis. Relatedly, the choice of seam for determining twists also has an intrinsic ambiguities. Therefore, it is necessary to numerically verify our results in microscopic models. Below, we check our results in two examples: the Kitaev honeycomb model, and a $\nu=1/2$ Laughlin wavefunction. The numerical results provide an additional check for the choice of seams, as well as demonstrate the universality of our results. The code required to reproduce the results is available at \cite{GH}.

\subsection{Kitaev Honeycomb Model}

We first verify our results on $J_n$ and $\Phi_r$ using the Kitaev honeycomb model~\cite{kitaev2006anyons}. 
The model consists of spins with $S=1/2$ on the hexagonal lattice. It displays a non-chiral toric code topological phase and a chiral Ising topological phase with $c_-=1/2$. 
Importantly, the model can be mapped exactly to a system of fermions coupled to a static $\mathbb{Z}_2$ gauge field, allowing for efficient calculations in polynomial times~\cite{yao2010entanglement} (see Appendix \ref{app: free-fermions} for details on the numerical procedure).

\begin{figure}
    \centering
    \includegraphics[width=\linewidth]{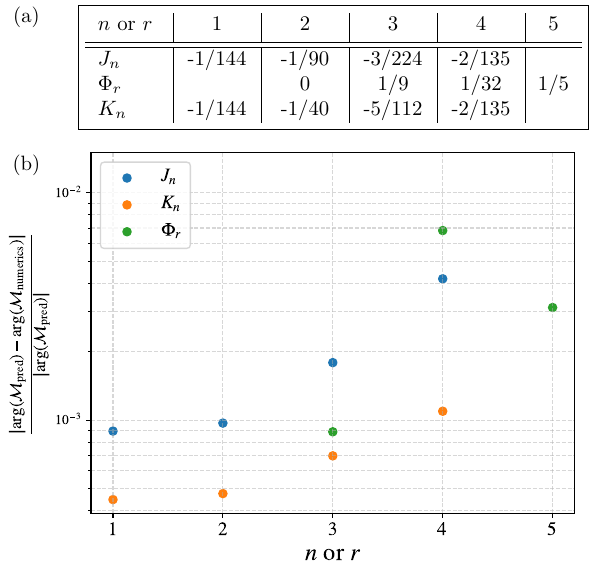}
    \caption{(a) Analytical predictions on the angles of $J_n,\Phi_r$ and $K_n$. These values are calculated from the known topological data of the Ising topological phase: $c_-=1/2$, quantum dimensions $d_0=d_\psi=1$, $d_\sigma=\sqrt{2}$, and spins $\theta_0=1$, $\theta_\psi=-1$, $\theta_\sigma=e^{2\pi i/16}$. (b) The relative error for the three measures $J_n,\Phi_r,K_n$. These are calculated for the Kitaev model on a torus with $n_s\times n_s$ sites, with parameters $n_s=16$, $J_x=J_y=J_z=1$, $K=0.3$. The errors increase with the number of replicas, but are below $0.005$ for all examples considered here. 
    }
    \label{fig: angle-errors}
\end{figure}

To benchmark our analytical method, in App.\ \ref{app: k_n} we define and calculate an additional multi-entanglement measure $K_n$. Similarly to $J_n$, it is defined on $2n+1$ replicas and is sensitive only to the chiral central charge. 
Fig.~\ref{fig: angle-errors}a contains the analytical values of the phases of $J_n,\Phi_r$ and $K_n$ for the Ising phase. These values are calculated based on known anyon data of the Ising phase.

Fig.\ \ref{fig: angle-errors}b shows the normalized errors of the numerical results, where we choose a set of Hamiltonian parameters such that the model is deep inside the Ising phase.
Our analytical predictions (Fig.\ref{fig: angle-errors}~(a)) match very well with the numerical results, with better accuracy for a smaller number of replicas. Qualitatively, this matches the expectation that lower R\'enyi values should suffer smaller finite-size effects \cite{jiang2013accuracy}.

\begin{figure}
    \centering
    \includegraphics[width=1.0\linewidth]{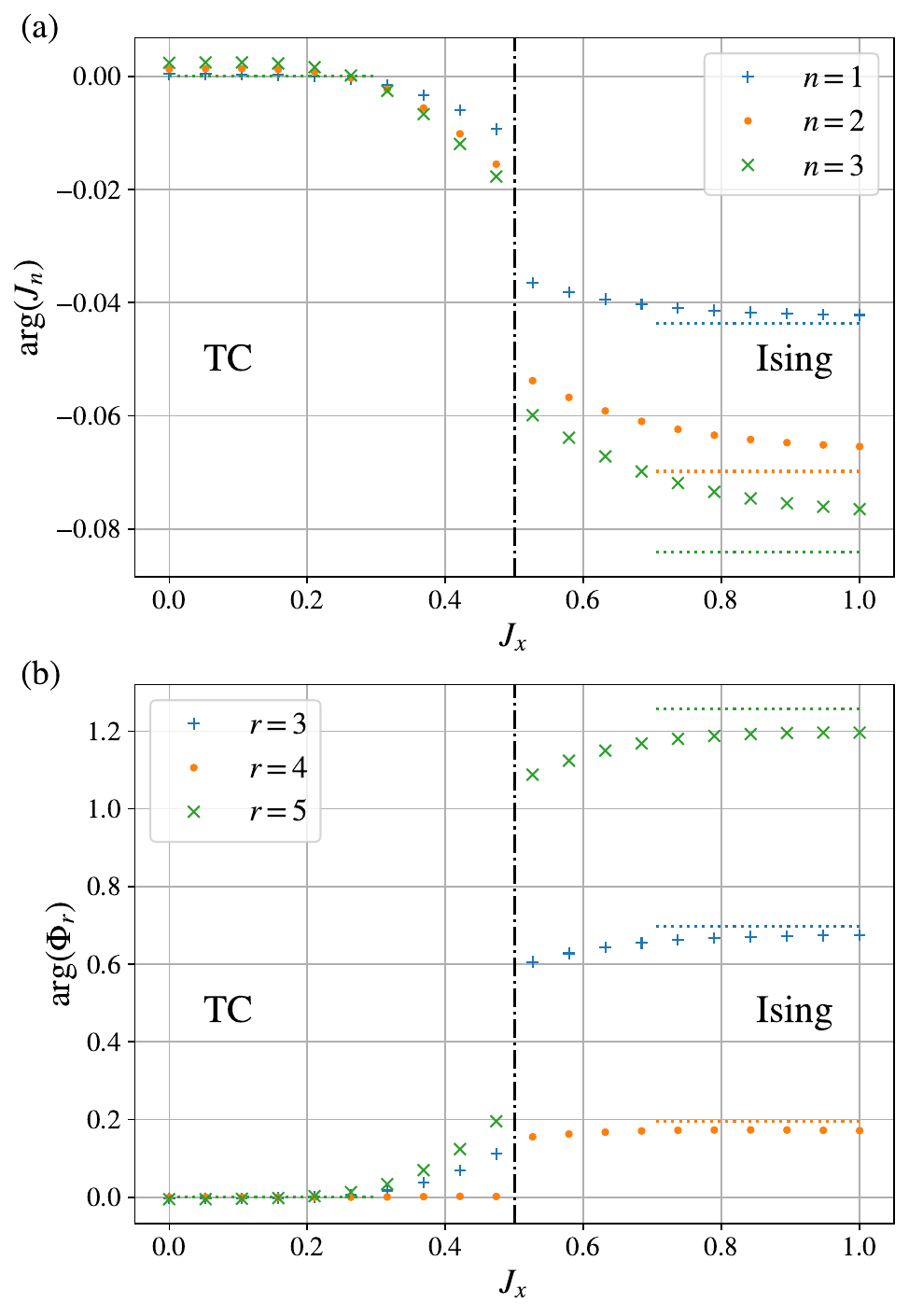}
    \caption{The values of (a) $J_n$ and (b) $\Phi_r$ across the TC-Ising transition tuned by $J_x=J_y$, calculated with $J_z=1,K=0.3,n_s=10$. The transition point is at $J_x=0.5$.
    }
    \label{fig: tc-ising-transition}
\end{figure}

Fig.\ \ref{fig: tc-ising-transition} shows the phases of $J_n,\Phi_r$ across the transition from the toric code (TC) to the Ising phase of the model. 
We find that these phases remain zero inside the toric code phase and jump to the predicted values across the transition, in good agreement with the analytical results. 

\subsection{Chern Insulator}
\begin{figure}
    \centering
    \includegraphics[width=.9\linewidth]{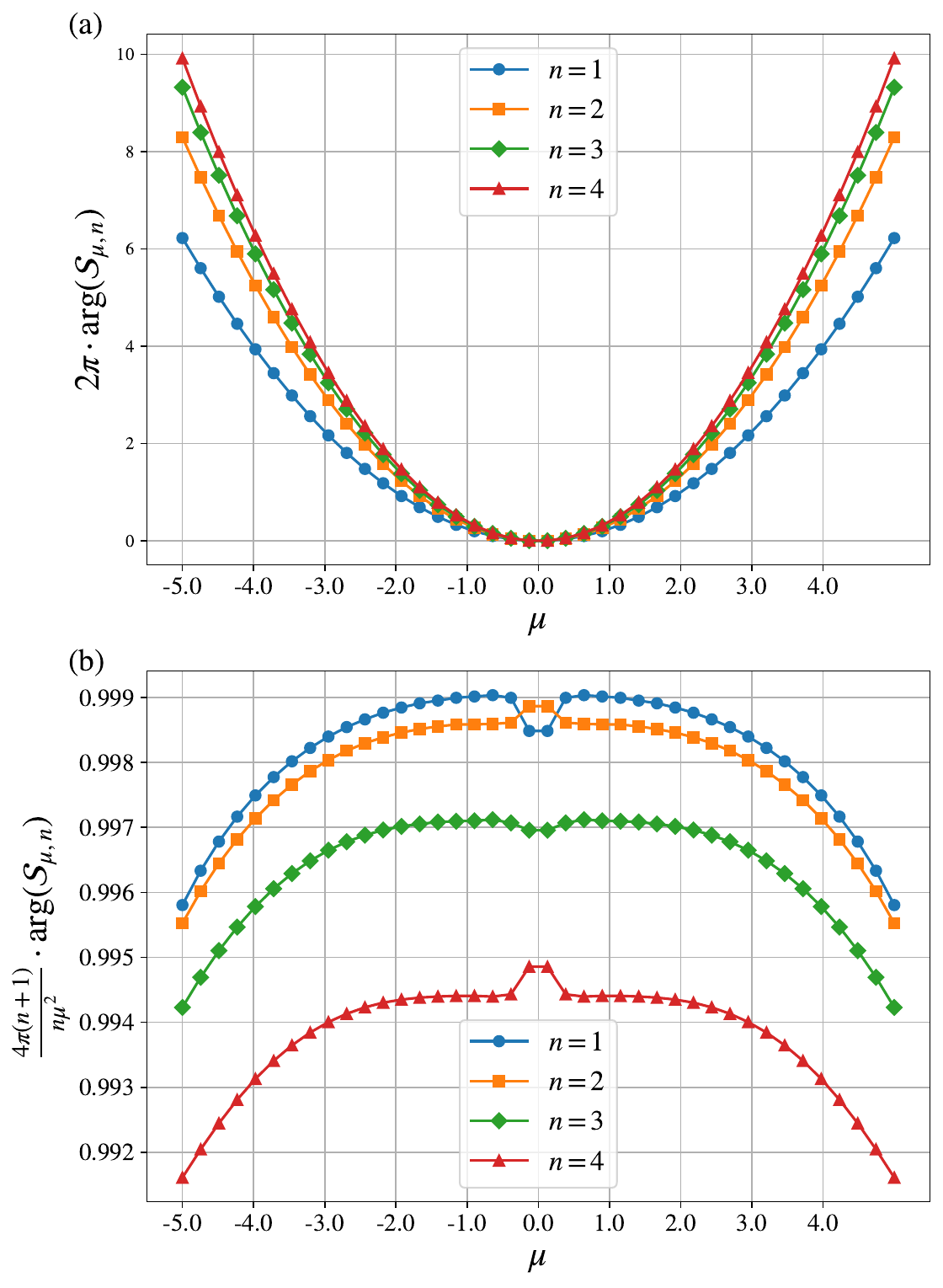}
    \caption{(a) The phase of $\SS_{\mu,n}$ as a function of $\mu$, and different values of $n$. It displays the same functional form predicted in \eqref{eq: s-mu-n-phases}. In (b) we normalized the results by the predicted results, such that the value $1$ indicates perfect matching with our predictions.
    }
    \label{fig:chern insulator}
\end{figure}

We verify our result \eqref{eq: s-mu-n-phases} for the Hall conductance using a $U(1)$ symmetric state, the Chern insulator, which has a unit quantum Hall conductance $\sigma_{xy} = 1/2\pi$.
A standard model is Haldane's model on the honeycomb lattice~\cite{haldane1988model}. 
For convenience of implementation, we choose an equivalent but different microscopic model that is composed of two copies of the free Majorana model used in the Kitaev Honeycomb model (with no coupling to a $\mathbb{Z}_2$ gauge field), with a $U(1)$ symmetry rotating between the two copies.
In Fig.\ \ref{fig:chern insulator}, we present the result for $\arg(\SS_{\mu,n})$ using different values of $\mu$ and $n$, and for $n_s=16$. Our result agrees with the prediction \eqref{eq: s-mu-n-phases}. The deviations are larger for larger $\mu,n$.

\subsection{Bosonic $\nu=1/2$ Laughlin state}
As an additional check, we evaluate our prediction on a trial wavefunction in the phase of the bosonic $\nu=1/2$ Laughlin state~\cite{nielsen2012laughlin}.
This phase has both a unit chiral central charge $c_- = 1$ and a fractional quantum Hall conductance $\sigma_{xy} = 1/4\pi$, and it does not have a free-fermion description.
The same wavefunction was also used to verify the modular commutator \eqref{eq: modular-commutator} by Ref.~\cite{Kim_Shi_Kato_Albert_2022}.

To describe the wavefunction on the sphere, we first consider the planar geometry, where $N$ spins are are placed at points $\qty{z_i}$ and are labeled by $s_i=\pm 1$. 
We then set a spin wavefunction
\begin{equation}
    \ket{\Psi\qty(\qty{z_i})}=\sum_{\qty{s_i}} \delta_\qty{s_i} \prod_{n<m}(z_n-z_m)^{\frac{1}{2}s_ns_m}\ket{\qty{s_i}}
    \label{eq: laughlin-wavefun}
\end{equation}
where $\delta_{\qty{s_i}}$ is 1 if $\sum_is_i=0$ and zero otherwise. 
To define the wavefunction on the sphere, we relate the sphere to the plane via stereographic projection. Namely, we label each point on the sphere by $(\phi_i,\theta_i)$ and assign
\begin{equation}
    z_i=\frac{\sin\theta_i}{1+\cos\theta_i}e^{i\phi_i},
\end{equation}
where $\theta_i\in[0,\pi],\phi_i\in[0,2\pi)$. We used the scheme described in \cite{Kim_Shi_Kato_Albert_2022} to evenly distribute the points on the sphere.

In the absence of a free-fermion description, we evaluate multi-entropy expectation values on $\ket{\Psi}$ using a Monte-Carlo scheme \cite{hastings2010measuring}. The magnitudes of the entanglement measures decrease exponentially with the linear size of the subregion, $\abs{\mathcal{M}}\sim e^{-\alpha \sqrt{N}}$, where $\alpha$ generally increases with the number of replicas. 
As a result, the sampling complexity increases exponentially with $\sqrt{N}$ and the number of replicas. 
Here, we only calculate the most computationally tractable measures $J_1$ and $\SS_{\mu,1}$, as their evaluation requires only three and two replicas. We consider regions of equal area on the sphere, and average the result of each $N$ over random rotations of the sphere, following \cite{kim2022modular}, to reduce finite-size effects.

\begin{figure*}
    \centering
    \includegraphics[width=.95\linewidth]{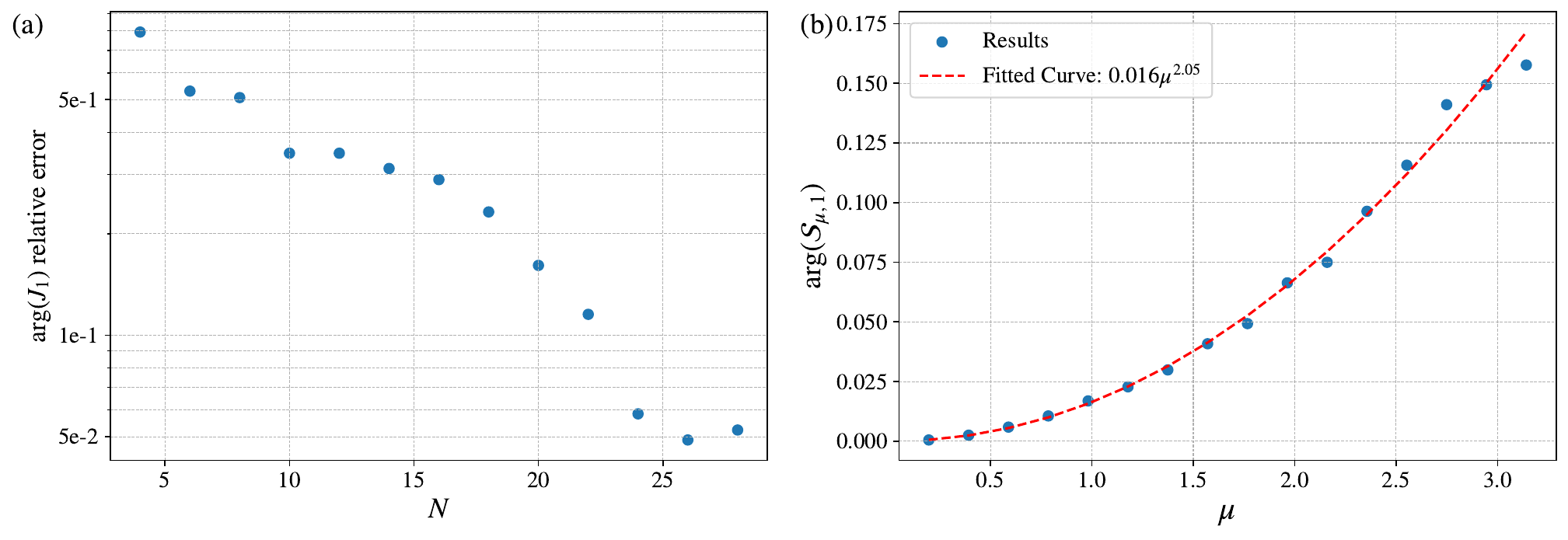}
    \caption{(a) The relative error $\frac{\abs{\arg(J_1)_{\rm numerics}-\arg(J_1)_{\rm analytical}}}{\arg(J_1)_{\rm analytical}}$ calculated for the Laughlin $\nu=1/2$ wavefunction \eqref{eq: laughlin-wavefun}. (b) The argument of $\SS_{\mu,1}$ for the $\nu=1/2$ wavefunction and $N=16$, as a function of $\mu$. Our Monte-Carlo calculation was implemented using NetKet \cite{netket2:2019,netket3:2022}.}
    \label{fig: laughlin_results}
\end{figure*}

In Fig.~\ref{fig: laughlin_results}~(a), we present the normalized error of the numerical result of $\arg(J_1)$ compared to its analytical prediction. 
Curiously, the error decreases faster than the $e^{-b\sqrt{N}}$ scaling we expect based on the analytical argument.  
We further remark that the modular commutator $J$ does exhibit the expected finite-size scaling for the same trial wavefunction~\cite{kim2022modular}, but it can also converge faster in other models, as observed in Ref.~\cite{maity2025identifying}.

Fig.~\ref{fig: laughlin_results}~(b) shows $\SS_{\mu,1}$ as a function of $\mu$ for a fixed system size. Fitting the resulting curve yields $\arg(\SS_{\mu,1})=A\mu^\eta$ with $\eta=2.05$, $A=0.016$. Both numbers are close to the analytical results, $\eta=2$, $A=1/16\pi\approx 0.0199$.

\section{Conclusions}
\label{sec: conclusions}
In this work, we presented a method to calculate multi-entropy measures for chiral topological phases. Using this method, we were able to obtain a R\'enyi version of the modular commutator and the charged modular commutator and calculate the lens-space multi-entropy measure for chiral phases. 

The main omission in our calculation is the framing anomaly. A cleaner derivation of our results should include a treatment of the framing of $\Sigma$ and $M$. Specifically, we need a better understanding of how the framing of $M$ obtained from the procedure presented in Sec.~\ref{sec: bulk-edge} compares with other framing choices~\cite{ atiyah1990framings,freed1991computer, kirby1999canonical}. While we used the FN coordinates to describe $\Sigma$ and obtain its phase contribution, it might be possible to achieve this through other means~\cite{dhoker_geometry_1988}. We note that the framing ambiguity should not be seen as an ambiguity of the final result, but rather as a shortcoming of our scheme of calculation. Indeed, the free-fermion calculation in \cite{gass2025r} does not suffer from this ambiguity.

While we expect our results to hold for ``generic" wavefunctions, they can suffer from spurious contributions, similar to the entanglement entropy~\cite{Zou_Haah_2016, williamson2016spurious} and the modular commutator~\cite{Gass_Levin_2024}. Indeed, as we show in App.\ \ref{app: spurious}, the example constructed in Ref.~\cite{Gass_Levin_2024} can spoof our measures as well. In a similar vein, while we considered the entanglement measures in the limit $L/\xi\gg 1$, we note that it is more accurate to use $\xi$ as the ``replica correlation length"~\cite{Zou_Haah_2016} rather than the correlation length (in practice, we expect the two to be the same in most cases). However, It is interesting to note that while the example in Ref.~\cite{Gass_Levin_2024} gives spurious contributions to $J_n$, the $n$ dependence of the resulting phase is different from the one predicted in Eq.~\eqref{eq: j_n phases}. It might thus be possible to obtain a more robust measure of $c_-$ if we could incorporate information from many values of $n$, and ``project out" spurious contributions. We note that recent work \cite{Sopenko:2024index,Sopenko:2025reflection} presented an index which extracts that chiral central charge from bulk entanglement and does not suffer from spurious contributions, based on the spin of a twist defect of the CFT \cite{PhysRevB.94.085116}.

It is interesting to compare our discussion and Ref.~\cite{li_strict_2024}, where the authors show that, for a finite-dimensional Hilbert space, the existence of a nonzero $c_-$ implies that the wavefunction cannot be ``strict area law" entangled. That is, there must be an $o(1)$ correction to the entanglement area law $S=\alpha L-\gamma$. 
The ``extended Hilbert space" picture considered in this work violates the assumption of a finite-dimensional local Hilbert space but still has exponentially decaying corrections to the area-law, as can be seen in \eqref{eq: separated-pf}. It is thus interesting to ask whether a nonzero $c_-$, or more generally, ungappable edges \cite{levin_protected_2013}, are incompatible with a strict area-law even for infinite-dimensional Hilbert spaces.

The two main quantities of interest here, $c_-$ and $\sigma_{xy}$, both measure the anomalies of the edge theory. The chiral central charge indicates that the edge theory cannot be regularized on a 1d lattice \cite{alvarez1984gravitational, di_francesco_conformal_1997, hellerman2021quantum}, while the Hall conductance signifies that it cannot be regularized on a 1d lattice while retaining the $U(1)$ symmetry. It would be interesting to extend our methods to probe additional instances of edge anomalies. Importantly, a universal entanglement quantity could be obtained if we replace the action $e^{\mu Q}$ in \eqref{eq: s-mu-n-def} with the action of a $\mathbb{Z}_2$ local symmetry (For $\mathbb{Z}_2$ the universality of the result can be argued along the lines of App.\ \ref{app: topological-invariance}). Could this method probe, for example, the gravitational anomaly in $\mathbb{Z}_2\times\mathbb{Z}_2$ symmetric fermionic systems \cite{ryu2012interacting}? We can also consider entanglement measures that probe finer structures of symmetry-enriched topological phases, such as the ``higher Hall conductance", which were obtained from partial rotations on a ground-state wavefunction in \cite{kobayashi2024higher}.

We can further consider the generalization of the entanglement measures defined in this work to broader classes of systems, including gapless ones. For example, we found  that multi-entropy measures analogous to the LeSME can also be defined for 1+1D CFTs, where they measure the partition functions on tori of modulus $il+p/q$ where $p,q\in \mathbb{Z}$ depend on the chosen entanglement measure, and $l$ can be tuned arbitrarily. This result is described in App.~\ref{app: 1+1d lesme}. 

We also comment on the numerical evaluation of the measures we propose here. In Sec. \eqref{sec: numerics} we presented a Monte-Carlo evaluation of $J_1,S_{\mu,1}$ for a trial wavefunction. Since the magnitude of the expectation value decreases exponentially with $L$ and the number of replicas \footnote{A detailed analysis of the scale of $\Phi_r$ for string-net models was carried in \cite{sheffer2025extracting}. the number of samples needed to reach a given accuracy should scale exponentially as well. A similar scaling of hardness with $L$ and the number of replicas should be expected for tensor-network methods, as these parameters control the number of legs across a cut. Note, however, that the measures themselves converge exponentially fast to their thermodynamic-limit values as $L$ increases,  the required $L$ grows logarithmically with the inverse of the required precision $\epsilon$. We therefore expect that the measures with lowest replica number, $J_1$ and $S_{\mu,1}$, could be effectively employed in numerics.}

Finally, it is intriguing to use the entanglement measures presented here in experimental settings. As an example, $J_1$ can be used to verify that certain wavefunctions prepared on quantum devices \cite{evered2025probing} indeed carry a nontrivial chiral central charge. Such measurements can be carried out using random measurement protocols \cite{huang2020predicting} or by measuring the expectation values of the operators directly using the Hadamard test. Note that, while in principle the Hadamard test measures the expectation value of a unitary operator, it can be modified to measure the expectation value of a product of a unitary and a Hermitian operator, as in \eqref{eq: s-mu-n-def}. These approaches carry the same scaling of hardness as the classical Monte-Carlo schemes, making experimental realization demanding, but possibly not prohibitive, for moderate system sizes.


\section*{Acknowledgements}
We thank Yaar Vituri, Tom Manovitz, Ryan Thorngren, Ryohei Kobayashi, Michael Levin, and Julian Gass for useful conversations. RF thanks Daniel Jafferis and Xi Yin for discussing the replica trick and cutting map method in the course of previous projects. YS thanks Noa Zilberman and Alek Bedroya for their hospitality during part of this work. This work was supported by grants from the DFG (CRC/Transregio 183, EI 519/71, projects C02 and C03), and the ISF Quantum Science and Technology program (Grant no. 2478/24). This research was supported in part by grant NSF PHY-2309135 to the Kavli Institute for Theoretical Physics (KITP). YS is supported by the Chaim Mida Prize and the Adams Fellowships Program of the Israel Academy of Sciences and Humanities.
RF is supported by the Gordon and Betty Moore Foundation (Grant GBMF8688).
SR is supported by the National Science Foundation under Award No. DMR-2409412. 

\textbf{Data availability:} the code used to generate all results of the numerical section is available at \cite{GH}.

\textbf{Note added:} Shortly after the initial publication of this work, a related work \cite{gass2025r} presented a formula for the R\'enyi modular commutator \eqref{eq: j_n phases} using different methods, via exact free-fermion calculations. Our results are in agreement with theirs.

\appendix

\section{Topological invariance of multi-entropy measures}
\label{app: topological-invariance}

\tocignoredsubsection{Invariance of $\mathcal{M}$}

Here we follow Ref.~\cite{sheffer2025extracting} to argue that the phase of $\mathcal{M}$, as defined in \eqref{eq: multi-ent}, is topologically invariant for ``generic" perturbations. The argument is similar to the original argument on the invariance of the topological entanglement entropy by \cite{Kitaev_Preskill_2006}.

Throughout this section, we will find it useful to separate between the abstract permutations, and the operators acting within a specific region. We will therefore denote by $\pi_{i,J}$ the permutation $\pi_i$ acting in region $J$. We want to argue for the topological invariance of expectation values of the form
\begin{equation}
    \mathcal{M}=\mel{\psi^{\otimes R}}{\pi_{1,A}\pi_{2,B}\pi_{3,C}}{\psi^{\otimes R}}\equiv \pizza{.6}{\pi_1}{\pi_2}{\pi_3}.
\end{equation}
We use the diagram as a shorthand for the expectation value. We will assume that $\pi_1=\pi_2\pi_3$, and that there exist permutations $\sigma,\tau$ satisfying 
\begin{equation}
\begin{aligned}
    \sigma^2&=\tau^2=1, \\
    \sigma \pi_2\sigma &= \pi_3 ,\\
    \tau \pi_2 \tau&=\pi_2^{-1},\\
    \tau \pi_3 \tau&=\pi_3^{-1}.\\
\end{aligned}
\label{eq: perms-for-topo-invariance}
\end{equation}
That is, $\sigma$ switches between $\pi_2$ and $\pi_3$, and $\tau$ inverts them. Such permutations $\sigma,\tau$ exist for all $\Phi_r,J_n$.

Under this assumption, we notice that the expectation values of permutation operators acting on only two regions are real-valued, that is
\begin{equation}
    \pizza{.7}{ }{\pi_2}{\pi_3}=
    \pizza{.7}{ }{\tau\pi_2\tau}{\tau\pi_3\tau}=
    \pizza{.7}{ }{\pi_2^{-1}}{\pi^{-1}_3}=
    \overline{\pizza{.7}{ }{\pi_2}{\pi_3}}
    \label{eq: pizza-real}
\end{equation}
where, in the first equality, we used the fact that $\tau_{ABC\Lambda}\ket{\psi^{\otimes R}}=\ket{\psi^{\otimes R}}$ where $\tau_{ABC\Lambda}$ is defined as $\tau$ acting on all regions. The same can be shown similarly on expectation values involving $\pi_1,\pi_2$ or $\pi_1,\pi_3$.

We now make the physical assumption that modifications of the region boundaries far from a region $I$ do not depend on the permutation of region $I$, namely
\begin{equation}
\frac{
\vcenter{\hbox{\begin{tikzpicture}[every node/.style={font={\scriptsize}}]
    \def\r{.6};
    \draw (0,0) ++ (52:\r) arc (52:70-322:\r);
    \draw (80:\r) ++ (-23:.3) arc (-23:180+6:.3);
    \draw (0,0) -- (90:\r+.28);
    \draw (0,0) -- (210:\r);
    \draw (0,0) -- (330:\r);
    \node at (-90:\r*.6) {$\pi_1$};
    \node at (+30:\r*.6) {$\pi_2$};
    \node at (180-30:\r*.6) {$\pi_3$};
\end{tikzpicture}}}}{
\vcenter{\hbox{\begin{tikzpicture}[every node/.style={font={\scriptsize}}]
    \def\r{.6};
    \draw (0,0) ++ (52:\r) arc (52:70-322:\r);
    \draw (80:\r) ++ (-23:.3) arc (-23:180+6:.3);
    \draw (0,0) -- (90:\r+.28);
    \draw (0,0) -- (210:\r);
    \draw (0,0) -- (330:\r);
    \node at (+30:\r*.6) {$\pi_2$};
    \node at (180-30:\r*.6) {$\pi_3$};
\end{tikzpicture}}}}
\approx\frac{\pizza{.6}{\pi_1}{\pi_2}{\pi_3}}{\pizza{.6}{ }{\pi_2}{\pi_3}}
\label{eq: pizza-invariance}
\end{equation}
where corrections are expected to be exponentially small in $L$. We note that this assumption is not true for systems with spurious contributions to the entanglement entropy \cite{williamson2016spurious,Zou_Haah_2016}, but is expected to be true for ``generic" systems \cite{Kitaev_Preskill_2006, Levin_Wen_2006}. Since the arguments of the denominators are quantized to be $0$ or $\pi$, it follows that the phase of the numerators must be universally quantized as well and cannot change under deformations of the external boundary.

To argue that the inside corner can be moved as well, we notice that
\begin{equation}
    \pizza{.6}{\pi_1}{\pi_2}{\pi_3}=
\begin{tikzpicture}[every node/.style={font={\scriptsize}},baseline=-2]
    \def\r{.6};
    \draw (0,0) circle (\r);
    \draw (0,0) -- (90:\r);
    \draw (0,0) -- (210:\r);
    \draw (0,0) -- (330:\r);
    \node at (120:\r*1.3) {$\pi_1^{-1}$};
    \node at (+30:\r*.6) {$\pi_3^{-1}$};
    \node at (180-30:\r*.6) {$\pi_2^{-1}$};
\end{tikzpicture}
\label{eq: pizza-on-lambda}
\end{equation}
The right hand side is obtained from the left hand side by applying $\pi_1^{-1}$ in all regions on the ket of the expectation value. Note that in the right hand side of \eqref{eq: pizza-on-lambda}, $\pi_1^{-1}$ is applied in $\Lambda$. The inner corner is now external, and the argument from above can be carried similarly.

\tocignoredsubsection{Invariance of the charged modular commutator}
Here we show that the phase of the charged modular commutator \eqref{eq: s-mu-n-phases} is invariant under deformations of the region boundaries. The analogous relation to Eq.\ \eqref{eq: pizza-invariance} holds when some permutation operators are replaced by $e^{\mu Q}$. It remains to show that, for any two regions, the action of the operators acting on only two regions gives a real result. For the regions $BC$, we have
\begin{equation}
    \pizza{.6}{}{\pi}{e^{\mu Q}}=
    \pizza{.6}{}{\pi^{-1}}{e^{\mu Q}}=
    \overline{\pizza{.6}{}{\pi}{e^{\mu Q}}}
    \label{eq: charged-perm-real}
\end{equation}
where $\pi$ is the cyclic permutation on $n+1$ replicas, and we used the fact that $\pi$ can be inverted by conjugating it with a permutation $\tau$ that stabilizes the first replica. We also used the fact that $e^{\mu Q}$ is Hermitian. Notice that, as a result, if we replace $e^{\mu Q}$ with a unitary partial symmetry (for example, with $e^{i\mu Q}$ or with a $\mathbb{Z}_n$ symmetry), the argument that the phase is universal does not hold.

The argument for the other pairs of regions follows similarly, for example, for $A,B$, we have
\begin{equation}
    \pizza{.6}{e^{\mu Q}\pi}{\pi}{}=
    \begin{tikzpicture}[every node/.style={font={\scriptsize}},baseline=-2]
        \def\r{.6};
        \draw (0,0) circle (\r);
        \draw (0,0) -- (90:\r);
        \draw (0,0) -- (210:\r);
        \draw (0,0) -- (330:\r);
        \node at (120:\r*1.3) {$\pi^{-1}$};
        \node at (-90:\r*.6) {$e^{\mu Q}$};
        \node at (180-30:\r*.6) {$\pi^{-1}$};
    \end{tikzpicture},
\end{equation}
which is shown to be real using \eqref{eq: charged-perm-real}.

\vspace{5mm}

\vspace{5mm}

\section{Proof that $\Phi_2$ is real}%
\label{app: real-phi_r}

We now prove that Eq.\ \eqref{eq: phi_r} is consistent with the requirement that $\Phi_2$ is real. That is, defining the sum
\begin{equation}
    s_2\equiv e^{-\frac{2\pi i c_-}{8}}\sum_ad_a^2\theta_a^2
\end{equation}
we need to prove that $s_2\in\mathbb{R}$. To do so, we use the result that for every manifold there is a unique canonical framing \cite{atiyah1990framings}. The partition function of the lens-space $L(r,1)$, in the canonical framing, was shown in \cite{freed1991computer} to be
\begin{equation}
    Z(L(r,1))=\frac{1}{D^2}e^{-\frac{2\pi i c_-}{8}}\sum_ad_a^2\theta_a^r.
\end{equation}
Since the canonical framing is unique, the partition function of a manifold in the canonical framing should satisfy
\begin{equation}
    Z(M)=\overline{Z(\bar{M})},
\end{equation}
where $\bar{M}$ is the mirror image of $M$. On the other hand, we know that $L(2,1)=\mathbb{RP}^3$ and is identical to its mirror image. We therefore conclude that $s_2=\overline{s_2}$ as required.

A second proof, relying solely on properties of the modular matrices $S,T$ can be given as follows \footnote{We thank Michael Levin for pointing this out to us.}: Using the identity \cite{kitaev2006anyons}
\begin{equation}
    S^\dagger TSTST=e^{\frac{2\pi i c_-}{8}}
\end{equation}
rearranging and squaring both sides gives
\begin{equation}
    S^\dagger T^2S=e^{\frac{2\pi i c_-}{4}}(T^\dagger S^\da T^\da)^2,
\end{equation}
taking the $(1,1)$ matrix elements and using the fact that $T_{1i}=\delta_{i1},S_{i,1}=d_i/D$ gives $s_2=\overline{s_2}$.

\section{The topology of $M$ and $\Sigma$}
\label{app: m-manifold-topology}

As shown in the main text, the topological multi-entropy measure is mapped to a partition function on a three-dimensional spacetime region $M\backslash W$, where $M$ is the space obtained by gluing the regions $A,B,C$ according to the replica permutations and$W$ is an excised handlebody surrounding the permutation defects. The gapless degrees of freedom reside on the two-dimensional surface $\Sigma = \partial(M\backslash W) = \partial W$.
In this appendix, we explain how the permutations $\pi_{A,B,C}$ determine the topology of $M$ and $\Sigma$. 

We first analyze $M$ using gluing operations within the standard TQFT picture.
Following Ref.~\cite{sheffer2025extracting}, we represent the density matrix $\rho_{ABC}$ as a three-ball whose upper and
lower hemispheres correspond, respectively, to the ket and bra of the state: 
\begin{equation}
  \rho_{ABC,i}=D\cdot\vcenter{\hbox{
  \begin{tikzpicture}[every path/.append style={thick}, every node/.style={font={\tiny}}]
    \coordinate (a) at (0,0);
    \coordinate (b) at (1.4,.3);
    \coordinate (c) at (.8,-.2);
    \coordinate (d) at (.7,1.0);
    \coordinate (e) at (.7,-1.0);
    \draw[thin,->] (0,.5) to[out=100, in=120] (.6,.5);
    \draw[thin,->] (.1,-.8) to[out=-80, in=-120] (.6,-.6);
    \filldraw[fill=white,fill opacity=.65] (a) -- (d) -- (b) -- (e) -- cycle;
    \draw (a) -- (c) -- (b);
    \draw[dashed] (a) -- (b);
    \draw (d) -- (c) -- (e);
    \node at (.5,.2) {$A_i$};
    \node at (.5,-.4) {$A_i^*$};
    \node at (1.05,.35) {$B_i$};
    \node at (1.1,-.4) {$B_i^*$};
    \node at (.0,.4) {$C_i$};
    \node at (.0,-.7) {$C_i^*$};
  \end{tikzpicture}
  }},
\end{equation}
where the factor $D=1/Z(S^3)$ ensures that $\Tr\rho_{ABC,i}=1$. To obtain $M$ for a R\'enyi-$R$ quantity, we take $R$ copies of such balls, one for each replica, and glue their faces according to the specified permutations. For example, the face $A$ of the $i$-th ball is glued to the face $A^*$ of the $\pi_A(i)$-th ball. The edges and vertices from different replicas are identified accordingly.

By construction, $M$ is a simplicial complex: a space obtained by gluing tetrahedra on their faces. 
Its vertices and edges come from the vertices (triple junctions) $v_i$ and the boundaries of the regions $A,B,C,\Lambda$, respectively.
Specifically, given two regions $I$ and $J$, there are $2R$ boundaries (each replica contains two: bra and ket) before gluing.
After gluing, each bra edge is identified with some ket edge, and any two ket edges related by the action $\pi_{IJ} = \pi_I \pi_{J}^{-1}$ are also identified.
Therefore, the boundary between two regions $IJ$ contributes $|\pi_{IJ}|$ edges in $M$, where $|\pi_{IJ}|$ is the number of cycles.
Similarly, The vertex between three regions $IJK$ can give rise to multiple vertices in $M$, the number of which is the number of orbits of the group $\langle \pi_{IJ}, \pi_{JK} \rangle$, denoted by $o_{IJK}$.

However, $M$ is generally not a manifold. The obstruction is that
the neighborhood of a vertex may not look locally like $\mathbb{R}^3$. Specifically, consider a single vertex $v$ in $M$ and its enclosing surface $X_v$. 
If $M$ is a manifold, then $X_v$ must be topologically $S^2$ so that the neighborhood of $v$, or equivalently the interior of $X$, is simply a three-ball.  
Otherwise, if $X_v$ has genus $g\ge1$, the interior is topologically the cone $CX_v$, which is not a manifold.

Suppose $M$ is a manifold, and thus $X_v$ is a sphere given any vertex $v$. 
The corresponding regularization surface $\Sigma_v \subset Y$, obtained by removing small disks from $X_v$ around the defect lines emanating from $v$, is a punctured sphere. As a result, $\Sigma$ is degenerate, i.e., any nontrivial genus arises only from
the long, thin handles connecting different $\Sigma_v$'s,
and there are no additional ``small'' handles localized near the
vertices. In this case, the chiral CFT partition function
\(Z(\Sigma)\) depends on the CFT only through its central charge
\cite{lunin_correlation_2001}, a fact which we use in the main text.

We now determine whether $M$ is a topological manifold, or equivalently, whether $X_v=S^2$ holds for any $v$, by calculating the Euler characteristic $\chi_{X_v}$ of $X_v$. The Euler characteristic can,
in turn, be deduced from the permutations, as we now demonstrate. 

\begin{figure}
    \centering
    \begin{tikzpicture}[scale=1.5]
    \begin{scope}[every path/.style={thick}]
        \draw (-1,0,0) -- (1,0,0);
        \draw (0,-1,0) -- (0,1,0);
        \draw (0,0,-1) -- (0,0,1);
    \end{scope}
    \filldraw[gray,opacity=.1] (-1,-1,0) -- (-1,1,0) -- (1,1,0) -- (1,-1,0);
    \filldraw[gray,opacity=.1] (-1,0,-1) -- (-1,0,1) -- (1,0,1) -- (1,0,-1);
    \filldraw[gray,opacity=.1] (0,-1,-1) -- (0,-1,1) -- (0,1,1) -- (0,1,-1);
    \begin{scope}[scale=.8, every path/.style={thick,dashed,violet,fill={violet},fill opacity=.07}]
        \filldraw (-1,0,0) -- (0,1,0) -- (0,0,1) -- cycle;
        \filldraw (-1,0,0) -- (0,1,0) -- (0,0,-1) -- cycle;
        \filldraw (-1,0,0) -- (0,-1,0) -- (0,0,1) -- cycle;
        \filldraw (-1,0,0) -- (0,-1,0) -- (0,0,-1) -- cycle;
        \filldraw (+1,0,0) -- (0,-1,0) -- (0,0,1) -- cycle;
        \filldraw (+1,0,0) -- (0,-1,0) -- (0,0,-1) -- cycle;
        \filldraw (+1,0,0) -- (0,1,0) -- (0,0,1) -- cycle;
        \filldraw (+1,0,0) -- (0,1,0) -- (0,0,-1) -- cycle;
    \end{scope}
    \node[left,violet] at (0,1,0) {\large $X_v$};
    \node[right] at (0,-.15,0) {$v_1$};
    \begin{scope}[scale=1.1,every node/.style={font={\small}}]
        \node at (-1,-1,-1) {$(1,1)$};
        \node at (+1,-1,-1) {$(1,2)^*$};
        \node at (-1,+1,-1) {$(2,1)^*$};
        \node at (-1,-1,+1) {$(2,2)^*$};
        \node at (+1,+1,+1) {$(1,1)^*$};
        \node at (+1,+1,-1) {$(2,2)$};
        \node at (+1,-1,+1) {$(2,1)$};
        \node at (-1,+1,+1) {$(1,2)$};
    \end{scope}
    \end{tikzpicture}
    \caption{The neighborhood of the vertex $v$ in $M$, illustrated for the vertex $v_1$ and the permutation operators of $\Phi_2$. The gray surfaces and dark lines correspond to the faces and edges of $\rho_{ABC,i}$. The surface $X_v$ surrounding the vertex is drawn here in purple. It has 8 faces, 12 edges, and 6 vertices, giving $\chi_{X_v}=2$. It is thus topologically $S^2$, and its interior is $B^3$. 
    }
    \label{fig: x_v-surface}
\end{figure}
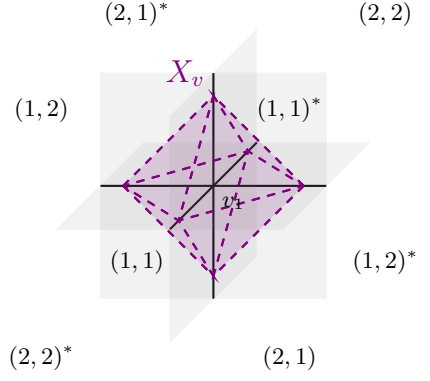

Without loss of generality, we focus on the vertex $v_1$ between the regions $A,B,C$.
We assume that it gives rise to a single vertex $v$ in the simplicial complex $M$ to simplify our discussion (that is, $o_{ABC}=1$). This is true when the group $\ev{\pi_{AB},\pi_{CD}}$ has a single orbit. 
The number of faces, edges, and vertices of $X_v$ is related to the number of balls, faces, and edges in each $\rho_{ABC}$ that are attached to the vertex $v$, respectively (see Fig. \ref{fig: x_v-surface} for an example of the surface $X_v$). 
The number of such balls is $2R$ (each replica gives two balls, one from the bra and one from the ket). 
The number of such faces is $3R$ (each replica contributes 6 faces, each bra face is identified with a ket face).
Finally, each boundary between two regions $IJ$ contributes $\abs{\pi_{IJ}}$ edges as explained before. Therefore, $X_{v} = S^2$ requires
\begin{equation}
    \chi_{X_{v}}=(2R)-(3R)+(\abs{\pi_{AB}}+\abs{\pi_{BC}}+\abs{\pi_{AC}})=2. 
    \label{eq: chi-x_v}
\end{equation}
A similar calculation holds for other vertices. We conclude that $M$ is a manifold if and only if, for any three regions $I,J,K$, we have
\begin{equation}
    \abs{\pi_{IJ}} + \abs{\pi_{JK}} + \abs{\pi_{IK}} = 2+R.
\end{equation}
Since the RHS obtains its maximal value (as $\chi_X\ge2$), this is true only if
\begin{equation}
    \sum_{I,J\in \qty{A,B,C,\Lambda};\;I\neq J}\abs{\pi_{IJ}}=8+4R.
    \label{eq: permutation-sum}
\end{equation}
One can check that this is true for all multi-entropy measures defined in this paper \footnote{It does not hold, for example, when $\pi_A$ is a cyclic permutation and $\pi_B,\pi_C$ are trivial. This is because, in that case, the vertex $v_2$ has $R$ copies in $M$, violating the assumption used in deriving this equation. In that case, $M$ is still a manifold despite the violation of \eqref{eq: permutation-sum}.}. The calculation above can similarly be carried out using the Riemann-Hurwitz formula, as was used, for example, in \cite{lunin_correlation_2001, liu_multipartite_2024}. We find that this derivation presents a clearer connection between the topology of $M$ and $\Sigma$.

To determine the genus $g$ of $\Sigma$, we consider the enclosed handlebody $W$.
The deformation retraction of $W$ is a 1D complex whose vertices and edges are exactly those of $M$. We thus have 
\begin{equation}
    \chi_{W}=-\sum_{I,J\in\qty{A,B,C,\Lambda};I\neq J}\abs{\pi_{IJ}}+\sum_{v_i} o_{IJK}.
\end{equation}
On the other hand, gluing two copies of $W$ along the boundary $\Sigma$, we obtain a closed three-manifold $H$. Since any closed odd-dimensional manifold has a vanishing Euler characteristic, we have
\begin{equation}
    2\chi_{W}-\chi_{\Sigma}=\chi_{H}=0.
\end{equation}
This allows us to conclude that
\begin{equation}
\begin{aligned}
    g-1 = & -\chi_\Sigma/2 = \\
    &\sum_{I,J\in\qty{A,B,C,\Lambda};I\neq J} \abs{\pi_{IJ}} - \sum_{v_i} o_{IJK}.
\end{aligned}
\end{equation}
This result is used for Eq.\ \eqref{eq: sigma-genus} in the main text.

\section{Dependence of $Z(\Sigma)$ on the regularization}
\label{app: sigma-complex-structure}
As shown in Sec.~\ref{sec: bulk-edge}, the cutting map is implemented via a Euclidean path integral that imposes boundary conditions on the tubular neighborhoods surrounding the region boundaries. There is no canonical choice for these conditions, which is an inherent ambiguity in the regularization scheme. However, the value of the edge partition function $Z(\Sigma)$ is sensitive to the details of this choice. In this appendix, we discuss this dependence and justify that the universal information about the chirality extracted from the measure is independent of these choices. Specifically, we show that changes in the regularization scheme give only local, real-valued contributions that do not affect the phase $\arg(Z(\Sigma))$. 

For concreteness and simplicity, we consider the $U(1)_k$ Chern-Simons theory as an example. The bulk action reads
\begin{equation}
    S=\frac{k}{4\pi}\int d^3x AdA\,,
\end{equation}
where $A$ is the dynamical gauge field. To define the cutting map, we assign a boundary condition for the gauge field $A$ at the neighborhood of region boundaries, e.g., along the blue arc drawn in Eq.~\eqref{eq: i_xi-map}. It follows from the method in Sec.~\ref{sec: bulk-edge} that the edge contribution $Z(\Sigma)$ is given by the Chern-Simons path integral inside a solid handlebody with the boundary $\Sigma$. The boundary condition on $\Sigma$ is not arbitrary; it is inherited from the one we use in the definition of the cutting map, a property we implicitly use below.

Specifically, the regularization scheme first imposes a metric on $\Sigma$ to specify its geometry, such as the linear size $L$ of the region boundary and the cutoff scale $\epsilon$. This choice is not unique. Given such a metric, there is a unique complex structure locally satisfying $ds^2=e^\phi dwd\bar{w}$, where $w$ is a complex coordinate, locally defined as $w=x+iy$ for some real coordinates $x,y$ ~\footnote{Our convention is to use $w$ for the complex coordinates in $\Sigma$, and $z$ for the complex coordinates in $Y$.}. With this coordinate defined, we define the boundary condition $A_{\bar{w}}=A_x+iA_y=0$. Since, this condition is invariant under conformal transformations $w\mapsto f(w)$, the resulting partition function depends only on the given complex structure, not the full metric data. In this way, we regard the choice of boundary condition and the choice of complex structure as physically equivalent, using the terms interchangeably below.

We consider the bipartite entanglement to gain some intuition about the above definition. In this case, $\Sigma$ is topologically a torus, and we use $\sigma \in [0,L]$ to parameterize its longitudinal direction and $\tau \in [0,n\epsilon]$ for the orthogonal direction. To obtain the result in Sec.~\ref{sec: bulk-edge}, we make the most natural choice: a uniform boundary condition $A_\sigma + i A_\tau = 0$, or equivalently, a flat metric $ds^2 = dw d\bar{w}$ and uniform complex structure $w=\sigma+i\tau$. 
A more general choice is of the form $A_\sigma+iv(\sigma,\tau)A_\tau=0$ for any complex function $v\in \mathbb{C}$, which amounts to local deformations of the complex structure. The multipartition case involves higher-genus surfaces, for which it is harder to explicitly define the complex structure and write down a global description for it. Instead, we turn to a local perspective and describe the effect of infinitesimal changes of the structure.

Our goal is to determine how local variations in the boundary condition, as a consequence, the complex structure, affect the final result. Below, we show that under an arbitrary local infinitesimal deformation of the complex structure, the change of the edge partition function $Z(\Sigma)$ is a real number. Consequently, the phase $\arg(Z(\Sigma))$ is intact, confirming that it is a universal quantity robust against the microscopic details of regularization.

Let us consider an infinitesimal deformation of the complex structure on the surface $\Sigma$, given by
\begin{equation}
    dw'=dw+\mu(w,\bar{w})d\bar{w},
\end{equation}
for some complex function $\mu$. This deformation is equivalent to replacing the boundary condition $A_{\bar{w}}=0$ with $A_{\bar{w}^{\prime}}=0$. As explained in \cite{friedan1987analytic}, the change in the complex structure alters the partition function as 
\begin{equation}
    \delta_\mu\log Z(\Sigma)=\frac{1}{2\pi}\int\ev{T(w)}_\Sigma\mu(w,\bar{w})dwd\bar{w}\,,
    \label{eq: delta-mu-z}
\end{equation}
where $T$ is the chiral stress tensor (for simplicity, we consider only the chiral sector).
This confirms our earlier assertion: changes in the regularization contribute only locally to the edge partition function.

To evaluate Eq.~\eqref{eq: delta-mu-z}, we begin by analyzing the simpler case of the embedded wavefunction $\langle i_\epsilon \psi | i_\epsilon \psi \rangle$. The corresponding edge contribution $Z(Y)$ is on the regularization surface $Y$, e.g., the genus-3 surface in Fig.~\ref{fig: regulating-surface}. 
Deforming the boundary condition changes $Z(Y)$ in the same way as in Eq.~\eqref{eq: delta-mu-z} with the right-hand side involving $\ev{T(z)}_Y$ and $\mu(z,\bar{z})$ on $Y$. Using the POP decomposition of $Y$, we can reduce the integral over $Y$ to local computations on the vertex surfaces $Y_v$.
In the limit $\epsilon/L \rightarrow 0$, operator insertions at the punctures of $Y_v$ are exponentially suppressed in its conformal weight. Consequently, the vacuum sector dominates, and the variation decomposes as:
\begin{equation}
    \delta_\mu\log Z(Y)\approx\frac{1}{2\pi}\sum_v\int\ev{T(z)}_{Y_v} \mu(z,\bar{z})dzd\bar{z}.
    \label{eq: delta ZY}
\end{equation}
where $\ev{}_{Y_v}$ is the expectation value on $Y_v$ with boundary conditions in the vacuum sector.
Effectively, we evaluate $T(z)$ on a sphere obtained by closing the punctures of $Y_v$. We parametrize each $Y_v$ as the complex plane $\mathbb{C}$ with punctures located at $0,1,\infty$.

We must verify that $\delta_\mu\log Z(Y)$ yields a real-valued result to be consistent the reality of the norm $\langle i_\epsilon \psi | i_\epsilon \psi \rangle$. The physical constraint ensuring this is time-reversal symmetry. Since $Y$ represents the overlap between the bra and the ket, the surface $Y_v$ must be symmetric under $z\to\bar z$.
Accordingly, the stress tensor and any physical deformation of the complex structure and must also respect this symmetry, satisfying $\langle T(z) \rangle = \overline{\langle T(\bar{z}) \rangle}$, $\mu(z, \bar{z})=\overline{\mu(\bar{z}, z)}$, where $T(\bar{z})$ means the chiral stress tensor at the time-reversal conjugate position of $z$. These conditions ensure that the integral is real, consistent with the requirement.

The analysis of Eq.~\eqref{eq: delta-mu-z} proceeds analogously. 
We use the POP decomposition of $\Sigma$ to localize the computation of the integral onto vertex surfaces $\Sigma_v$, each of which is a covering over the corresponding base surface $Y_v$.
In the limit $\epsilon/L \to 0$, we assume vacuum dominance and evaluate $T(w)$ on the capped vertex surface $\Sigma_v$. 
We can reduce the computation from the cover to the base by using the holomorphic covering map $f:\Sigma_v\to Y_v$, $f(w) = z$ in Sec.~\ref{sec: fn_coordinates}. According to the conformal transformation law, we have
\begin{equation}
    \ev{T(w)}_{\Sigma_v}=(\partial_wz)^2\ev{T(z)}_{Y_v}+\frac{c}{12}\qty{z,w},
\end{equation}
where $c$ is the chiral central charge and
\begin{equation}
    \qty{z,w}=\frac{\partial_wf\partial_w^3f-\frac{3}{2}(\partial_w^2f)^2}{(\partial_wf)^2}
\end{equation}
is the Schwarzian derivative~\cite{di_francesco_conformal_1997}. The change of the partition function $Z(\Sigma)$ localized to each capped vertex surface $\Sigma_v$ becomes 
\begin{equation}
\begin{aligned}
    &\frac{1}{2\pi}\int_{\Sigma_v}\ev{T(w)}_{\Sigma_v}\mu(w)dwd\bar{w}=\\
    &\frac{1}{2\pi}\int_{Y_v}\sum_{w\in f^{-1}(z)}\qty(\ev{T(z)}_{Y_v}-\frac{c}{12}(\partial_wz)^{-2}\qty{z,w})\mu(z)dzd\bar z,
    \label{eq: Y_v-mu-deformation}
\end{aligned}
\end{equation}
where we used the fact that $\mu$ transforms as
\begin{equation}
    \mu(z)=\frac{\partial_wz}{\overline{\partial_wz}}\mu(w).
\end{equation}
The first term in Eq.~\eqref{eq: Y_v-mu-deformation} is the same as what we get in the analysis of $Z(Y)$, and thus is real.
In the rest of this section, we show that the second term is also real and complete our argument.

To establish the reality of the Schwarzian contribution, we impose an additional condition on the permutation operators, which we call local time-reversal symmetry.
Let $\pi_{IJ}=\pi_I \pi_{J}^{-1}$ denote the relative permutation.
Given a vertex $v$ and its three surrounding regions $I,J,K$, we demand that there exists a $v$-dependent permutation $\tau_v$ satisfying $\tau_v^2=1$ and
\begin{align}
    \tau_v \pi_{IJ}\tau_v = \pi_{IJ}^{-1}\,,\quad  \tau_v\pi_{JK}\tau_v = \pi_{JK}^{-1}.
    \label{eq: local-t-sym}
\end{align}
This condition follows from the assumptions in \eqref{eq: perms-for-topo-invariance}. It ensures that $\Sigma_v$ is time-reversal symmetric in the following sense: generally, the cover map corresponding to the opposite permutations in the ramification points ($\pi_{IJ}^{-1}$) is just $z\mapsto\overline{f(\bar z)}$. As a result, if $w_i$ are the $R$ preimages of $z$ under $f$, then \eqref{eq: local-t-sym} ensures that 
\begin{equation}
    f^{-1}(\bar w_i)=\overline{f^{-1}(w_{\tau(i)})},
\end{equation}
or, in terms of the derivatives
\begin{equation}
    \partial_{\bar w_i}^n\bar z=\overline{\partial_{w_{\tau_v(i)}}^nz}.
\end{equation}
Essentially, this condition means that while $\Sigma$ is not time reversal symmetric, each of $\Sigma_v$ is (with a different time reversal operation corresponding to $\tau_v$). As a result, the imaginary part of the second term in \eqref{eq: Y_v-mu-deformation} cancels between $w_i$ and $\bar w_{\tau_v(i)}$, ensuring that $\delta_\mu \log Z(\Sigma)$ is real up to corrections that are exponentially small in $L$. 

For concreteness, consider the example of $J_1$ and the vertex $v_1$ as visualized in Fig.\ \ref{fig: covering-map}. The permutation $\tau_v$ exchanges $1\leftrightarrow3$, and corresponds to a $\pi$ rotation of $\Sigma$ in the drawing. This symmetry ensures that the contribution from $\mu$ will cancel between a point in region $1$ and the corresponding point in $3^*$.

\section{Example calculation of the angle $\tau$}
\label{app: tau-calculation}
\begin{figure}
    \centering
    \begin{tikzpicture}
        \node at (-4.7,1) {(a)};
        \node at (-4.5,0) {\Large $\Sigma_{v_1}$:};
        \node at (-2.2,0) {
        \begin{tikzpicture}[scale=1]
            \draw[semithick, lightblue] (0,0) ellipse (1.5 cm and 1cm);
            \draw[thick] (-1.4,0) -- (1.4,0);
            \foreach \fl in {-1,1}
                \foreach \fh in {-1,1}
                    \draw[thick] plot [smooth,tension=1.2] coordinates {(-.8*\fl,.8*\fh) (-.5*\fl,.4*\fh) (-.4*\fl,0)};
                
            \draw[line width=2.5,dotted,darkblue] plot [smooth,tension=1.2] coordinates {(.8,.8) (.5,.4) (.4,0)};
            \draw[line width=2.5,dotted,Lavender] plot [smooth,tension=1.2] coordinates {(-.8,-.8) (-.5,-.4) (-.4,0)};
            \draw[fill=white, draw=lightblue,semithick] (-.4,0) circle (.1) node[lightblue,above left] {$v_3$};
            \draw[fill=white, draw=lightblue,semithick] (.4,0) circle (.1) node[lightblue,above right] {$v_4$};
            \node[lightblue] at (-1.3,.8) {$v_2$};
            \begin{scope}[every node/.style={font={\normalsize}}]
                \node at (0,-.5) {$3$};
                \node at (0,.5) {$1^*$};
                \node at (1.,-.3) {$3^*$};
                \node at (+1.,.3) {$2$};
                \node at (-1.,-.3) {$2^*$};
                \node at (-1.,.3) {$1$};
            \end{scope}
        \end{tikzpicture}
        };
        \node at (-4.5,-2.7) {\Large $\Sigma_{v_2}$:};
        \node at (-2.2,-2.7) {
        \begin{tikzpicture}[scale=1]
            \draw[semithick, lightblue] (0,0) ellipse (1.5 cm and 1cm);
            \draw[thick] (-1.4,0) -- (1.4,0);
            \foreach \fl in {-1,1}
                \foreach \fh in {-1,1}
                    \draw[thick] plot [smooth,tension=1.2] coordinates {(-.8*\fl,.8*\fh) (-.5*\fl,.4*\fh) (-.4*\fl,0)};
                
            \draw[line width=2.5,dotted,Lavender] plot [smooth,tension=1.2] coordinates {(.8,.8) (.5,.4) (.4,0)};
            \draw[line width=2.5,dotted,darkblue] plot [smooth,tension=1.2] coordinates {(-.8,-.8) (-.5,-.4) (-.4,0)};
            \draw[fill=white, draw=lightblue,semithick] (-.4,0) circle (.1) node[lightblue,above left] {$v_4$};
            \draw[fill=white, draw=lightblue,semithick] (.4,0) circle (.1) node[lightblue,above right] {$v_3$};
            \node[lightblue] at (-1.3,.8) {$v_1$};
            \begin{scope}[every node/.style={font={\normalsize}}]
                \node at (0,-.5) {$2$};
                \node at (0,.5) {$2^*$};
                \node at (1.,-.3) {$3^*$};
                \node at (+1.,.3) {$3$};
                \node at (-1.,-.3) {$1^*$};
                \node at (-1.,.3) {$1$};
            \end{scope}
        \end{tikzpicture}
        };
        \node at (2.2,0) {
        \begin{tikzpicture}[scale=1]
            \draw[semithick, lightblue] (0,0) ellipse (1.5 cm and 1cm);
            \draw[thick] (-1.4,0) -- (1.4,0);
            \foreach \fl in {-1,1}
                \foreach \fh in {-1,1}
                    \draw[thick] plot [smooth,tension=1.2] coordinates {(-.8*\fl,.8*\fh) (-.5*\fl,.4*\fh) (-.4*\fl,0)};
                
            \draw[line width=2.5,dotted,darkblue] (.4,0) -- (1.5,0);
            \draw[line width=2.5,dotted,darkblue] plot [smooth,tension=1.2] coordinates {(1.5,0) (1.3,.45) (.8,.8)} node [right=.3,darkblue] {$\tau_1$};
            \begin{scope}[scale=-1]
            \draw[line width=2.5,dotted,Lavender] (.4,0) -- (1.5,0);
            \draw[line width=2.5,dotted,Lavender] plot [smooth,tension=1.2] coordinates {(1.5,0) (1.3,.45) (.8,.8)};
            \end{scope}
            \draw[fill=white, draw=lightblue,semithick] (-.4,0) circle (.1) node[lightblue,above left] {$v_3$};
            \draw[fill=white, draw=lightblue,semithick] (.4,0) circle (.1) node[lightblue,above right] {$v_4$};
            \node[lightblue] at (-1.3,.8) {$v_2$};
            \begin{scope}[every node/.style={font={\normalsize}}]
                \node at (0,-.5) {$3$};
                \node at (0,.5) {$1^*$};
                \node at (1.,-.3) {$3^*$};
                \node at (+1.,.3) {$2$};
                \node at (-1.,-.3) {$2^*$};
                \node at (-1.,.3) {$1$};
            \end{scope}
        \end{tikzpicture}
        };
        \node at (-4.5,-2.7) {\Large $\Sigma_{v_2}$:};
        \node at (2.2,-2.7) {
        \begin{tikzpicture}[scale=1.1]
            \draw[semithick, lightblue] (0,0) ellipse (1.5 cm and 1cm);
            \draw[thick] (-1.4,0) -- (1.4,0);
            \foreach \fl in {-1,1}
                \foreach \fh in {-1,1}
                    \draw[thick] plot [smooth,tension=1.2] coordinates {(-.8*\fl,.8*\fh) (-.5*\fl,.4*\fh) (-.4*\fl,0)};
                
            \draw[line width=2.5,dotted,Lavender] (.4,0) -- (1.5,0);
            \draw[line width=2.5,dotted,Lavender] plot [smooth,tension=1.2] coordinates {(1.5,0) (1.3,.45) (.8,.8)} node [right=.3,darkblue] {$\tau_2$};
            \begin{scope}[scale=-1]
            \draw[line width=2.5,dotted,darkblue] (.4,0) -- (1.5,0);
            \draw[line width=2.5,dotted,darkblue] plot [smooth,tension=1.2] coordinates {(1.5,0) (1.3,.45) (.8,.8)};
            \end{scope}
            \draw[fill=white, draw=lightblue,semithick] (-.4,0) circle (.1) node[lightblue,above left] {$v_4$};
            \draw[fill=white, draw=lightblue,semithick] (.4,0) circle (.1) node[lightblue,above right] {$v_3$};
            \node[lightblue] at (-1.3,.8) {$v_1$};
            \begin{scope}[every node/.style={font={\normalsize}}]
                \node at (0,-.5) {$2$};
                \node at (0,.5) {$2^*$};
                \node at (1.,-.3) {$3^*$};
                \node at (+1.,.3) {$3$};
                \node at (-1.,-.3) {$1^*$};
                \node at (-1.,.3) {$1$};
            \end{scope}
        \end{tikzpicture}
        };
        \node at (0,-1.5) {\Huge $\Rightarrow$};
        \node at (0,-1) {deform seams};
    \end{tikzpicture}
    \includegraphics{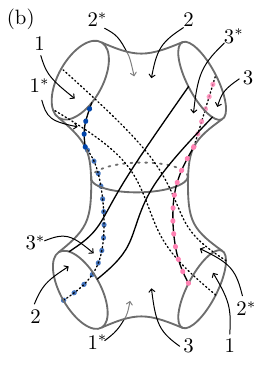}
    \caption{(a) The POPs $\Sigma_{v_{1,2}}$ used to calculate $J_1$. The seams (blue and pink dashed lines) are deformed such that they coincide with the orthogeodesics (the horizontal line) inside each POP. We draw only the seams connecting $v_1$ and $v_2$. (b) Visualization of the gluing of the two POPs with a twist angle. 
    }
    \label{fig: tau-calculation-example}
\end{figure}
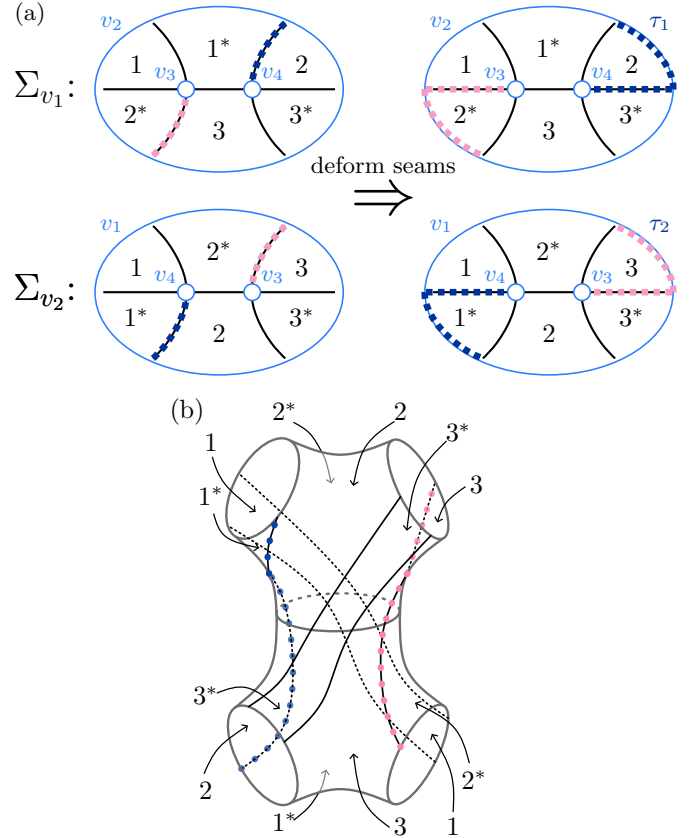

All calculations of the twist angles $\tau$ between two POPs in this paper follow a similar pattern. Here we work out one example in details: the angle between $v_1,v_2$ for the measure $J_1$. In Fig.\ \ref{fig: tau-calculation-example}a we draw the surfaces $\Sigma_{v_1},\Sigma_{v_2}$. The punctures without ramification (drawn in Fig.\ \ref{fig: covering-map}) are ignored, as we expect that they do not contribute any twists. Fig.\ \ref{fig: tau-calculation-example}b visualizes the gluing of the two POPs. 
To calculate the angle, we consider the seams connecting the two POPS, chosen according to the prescription in Sec.\ \ref{sec: fn_coordinates}. We first deform the seams so that, inside the POPs, they coincide with the orthogeodesics of the POP (the RHS of Fig.\ \ref{fig: tau-calculation-example}a). By symmetry, these orthogeodesics lie on the real (horizontal) line. The angle then obtains a contribution from the two POPs, and we write
\begin{equation}
    \tau_{12}=\tau_1+\tau_2
\end{equation}
where $\tau_1$ is the angle traversed by the seam in the drawing in $\Sigma_{v_2}$. Since each region covers an equal angle at infinity, we have
\begin{equation}
    \tau_1=\tau_2=-\frac{2\pi}{6}
\end{equation}
where the sign was chosen following the convention that the sign is positive if the deformed seam turns right. This gives $\tau_{12}=-2\pi/3$, matching the value for general $n$ in \eqref{eq: j_n_taus}.

\section{Definition and calculation of the measures $K_n$}
\label{app: k_n}
\begin{figure}
    \centering
    \newcommand{\kn}[3]{
        \setsepchar{ }
        \readlist\reg#1
        \readlist\vers#2
        \readlist\cols#3
        \begin{tikzpicture}[scale=.8]
        \begin{scope}[every path/.style={thick}]
            \draw (-4,0) -- (4,0);
            \foreach \xx in {-2.5,-1.5,...,2.5}
                \draw (\xx,-1.5) -- (\xx,1.5);
        \end{scope}
        \foreach \xx in {-2.5,-.5,1.5}
            \draw[fill=white,draw={\cols[1]},thick] (\xx,0) circle (.15);
        \foreach \xx in {-1.5,.5,2.5}
            \draw[fill=white,draw={\cols[2]},thick] (\xx,0) circle (.15); 
        \node at (-2.8,.3) {\color{\cols[1]} $v_{\vers[1]}$};
        \node at (2.8,.3) {\color{\cols[2]} $v_{\vers[2]}$};
        \foreach \ss in {1,-1}{
        \begin{scope}[xscale=\ss,every path/.style={thick,dash dot,purple}]
            \draw (-2,0) ellipse (.76 and .25);
            \draw (-1.5,0) ellipse (1.4 and .35);
        \end{scope}
        }
        \begin{scope}[every node/.style={font={\large}}]
            \node at (3,1) {$\reg[1]$};
            \node at (2,1) {$\reg[2]$};
            \node at (1,1) {$\reg[3]$};
            \node at (0,1) {$\reg[4]$};
            \node at (-1,1) {$\reg[5]$};
            \node at (-2,1) {$\reg[6]$};
            \node at (-3,1) {$\reg[7]$};
            \node at (-3,-1) {$\reg[8]$};
            \node at (-2,-1) {$\reg[9]$};
            \node at (-1,-1) {$\reg[10]$};
            \node at (0,-1) {$\reg[11]$};
            \node at (1,-1) {$\reg[12]$};
            \node at (2,-1) {$\reg[13]$};
            \node at (3,-1) {$\reg[14]$};
        \end{scope}
        \end{tikzpicture}
    }
    \begin{tikzpicture}
    \node at (0,0) {\huge $\Sigma_{v_1}$:};
    \node at (4,0) {
    \kn{{1 1^* 2 3^* 4 5^* 6 7^* 7  6^* 5 4^* 3 2^* 1}}{{4 3 2}}{{lightblue ForestGreen}}
    };
    
    \node at (0,-3) {\huge $\Sigma_{v_2}$:};
    \node at (4,-3) {
    \kn{{1 2^* 3 4^* 5 6^* 7 7^* 6 5^* 4 3^* 2 1^*}}{{3 4 1}}{{orange magenta}}
    };
    
    \node at (0,-3*2) {\huge $\Sigma_{v_3}$:};
    \node at (4,-3*2) {
    \kn{{1 2^* 2 4^* 4 6^* 6 7^* 7 5^* 5 3^* 3 1^*}}{{2 1 4}}{{orange ForestGreen}}
    };
    
    \node at (0,-3*3) {\huge $\Sigma_{v_4}$:};
    \node at (4,-3*3) {
    \kn{{1 1^* 3 3^* 5 5^* 7 7^* 6 6^* 4 4^* 2 2^*}}{{1 2 3}}{{lightblue magenta}}
    };
    \end{tikzpicture}
    \caption{The vertex surfaces $\Sigma_{v_i}$ for calculating $K_n$, drawn for $n=3$. The dashed purple lines are a POP decomposition, chosen to preserve the symmetries $x\to-x,y\to-y$ in each POP.}
    \label{fig: k_n-pops}
\end{figure}
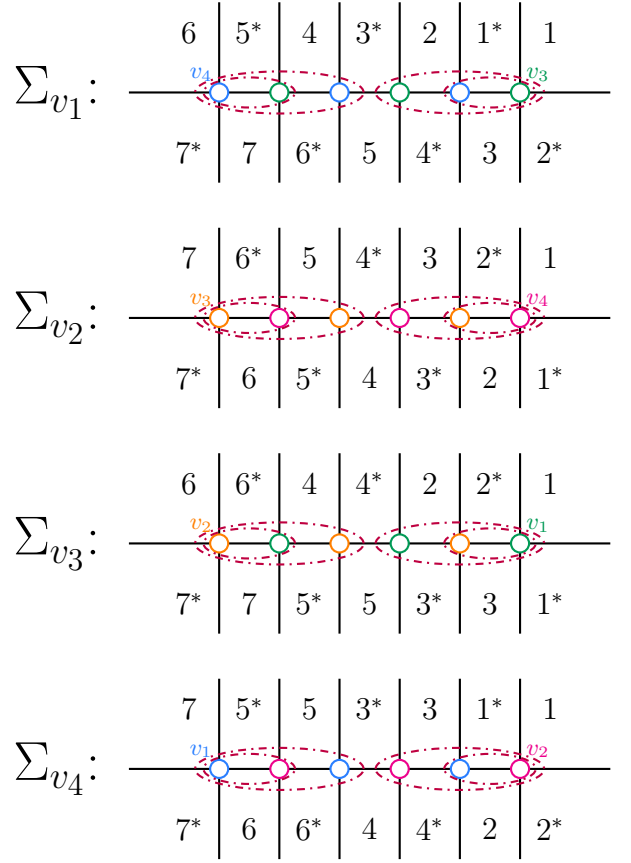

Here we consider an additional set of entanglement measures, defined using the permutations on $R=2n+1$ replicas given by
\begin{equation}
    K_n:\begin{cases}
        \pi_A&=(2,4,\cdots,2n,2n+1,2n-1,2n-3,\cdots1)\\
        \pi_B&=(1,2)(3,4)\cdots (2n-1,2n) \\
        \pi_C&=(2,3)(4,5)\cdots (2n,2n+1) \\
    \end{cases}.
\end{equation}
We see that $\pi_A$ is a cycle of length $2n+1$, while $\pi_B,\pi_C$ are comprised of $n$ transpositions each. We can therefore write the covering map $f$ as
\begin{equation}
    f'(w)=C(w-\omega_1)\cdots(w-\omega_n)(w-\lambda_1)\cdots(w-\lambda_n),
\end{equation}
where $f(\omega_i)=f(0)=0$, $f(\lambda_i)=f(1)=1$ and we require $f'(0),f'(1)\neq 0$. These requirements can be solved numerically for a given $n$. The pattern that emerges is that the points $\lambda_i,\omega_i$ are on the real line, with 
\begin{equation}
    0<\lambda_1<\omega_1<\lambda_2<\omega_2< \cdots <1.
\end{equation}
Similar to the calculation of the measures in the main text, we draw the vertex surfaces $\Sigma_{v_i}$ and choose a POP decomposition. We can choose the POP decomposition such that it preserves the reflection symmetries of $\Sigma_{v_i}$ along the real and imaginary lines. The surfaces are drawn in Fig.\ \ref{fig: k_n-pops}. We find
\begin{equation}
\begin{aligned}
    \tau_{v_1v_2}&=\tau_{v_3v_4} = \frac{2\pi n}{2n+1}, \\
    \tau_\alpha &= -\frac{\pi}{2},
\end{aligned}
\end{equation}
where $\tau_\alpha$ are the angles of the punctures with ramification 2. There are $4n$ such punctures, so we find
\begin{equation}
    K_n \propto\exp(\frac{2 \pi ic_-}{24} \frac{n(2n-1)}{2n+1}).
    \label{eq: k_n_phase}
\end{equation}

\section{Ground-state momentum in a $U(1)$-twisted sector}
\label{app: twisted-sectors}
Here we obtain \eqref{eq: twist-with-charge} based on an explicit calculation for the chiral boson theory. For simplicity, we assume that the bulk theory is given as a $1/m$ Laughlin state. The action on the edge is given by
\begin{equation}
    S=\frac{m}{4\pi}\int dtd\sigma [\partial_t\phi \partial_\sigma\phi-(\partial_\sigma\phi)^2]
\end{equation}
where $\phi$ is a bosonic field. We consider the theory on a circle of circumference $2\pi$. The boundary condition on the $\mu$-twisted sector is given by
\begin{align}
    \phi(0)&=\phi(2\pi)+\frac{\mu}{m}+\frac{2\pi k}{m}; & k&\in \Z.
\end{align}
The allowed integer values of $k$ allow for the state to be in any of the sectors corresponding to quasiparticle excitations on the boundary. The shift in the vacuum energy and momentum is then given by the shift of the energy and momentum of the ``zero modes" in the twisted sectors, which are given by
\begin{equation}
    \phi(\sigma)=\frac{\mu+2\pi k}{m}\frac{\sigma}{2\pi}.
\end{equation}
The energy and momentum shifts (compared to the untwisted states) are then
\begin{equation}
    \delta E=\delta P =\frac{m}{2}\qty(\frac{\mu+2\pi k}{2\pi m})^2=\frac{\sigma_{xy}}{4\pi}(\mu+2\pi k)^2.
    \label{eq: em-mu-shifts}
\end{equation}
Let us now consider the contribution to the partition function from the multiple zero modes. We focus on imaginary defect, and replace $\mu\to i\mu$ below. We obtain 
\begin{equation}
    Z_{\rm zm}=\sum_k e^{\frac{1}{8\pi^2 m}(-\beta +i\tau)(i\mu+2\pi k)^2}\approx e^{\frac{\sigma_{xy}\mu^2}{4\pi}(\beta-i\tau)}
\end{equation}
where the approximation holds for $\mu\tau\ll\beta$, otherwise we obtain additional contributions from $k\neq0$. For our application, this means $\mu\ll L/\xi$. 

The contribution of the chiral central charge in \eqref{eq: twist-with-charge} is more subtle and is obtained from the Casimir energy of the ground state \cite{di_francesco_conformal_1997}. Note that for multiple bosonic modes, we will obtain \eqref{eq: em-mu-shifts} as a sum of all modes, so the relation \eqref{eq: twist-with-charge} holds beyond the simple model we considered here (in the simplest generalization, for a multicomponent Luttinger liquid with general $K$ and $V$ matrices).

\section{Free-fermion numerical calculations}
\begin{figure}[h]
    \centering
    \includegraphics{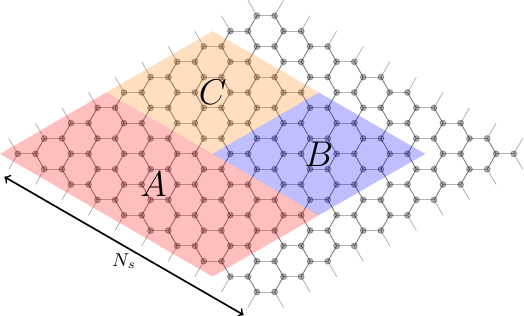}
    \caption{The geomtry used for the free-fermion numerical calculations. The sizes of the regions $B,C$ is set as $\frac{\sqrt{3}}{4}N_s$}.
    \label{fig: kitaev-geometry}
\end{figure}

Here we describe the free-fermion calculations on the Kitaev model \cite{kitaev2006anyons}. We begin by reviewing the model, then describe our numerical procedure. The calculations are carried out on $N_s\times N_s$ sites with periodic boundary conditions and regions as defined in Fig. \ref{fig: kitaev-geometry}.

\label{app: free-fermions}
\tocignoredsubsection{Review of the Kitaev model}
The model is defined on the hexagonal lattice, with a spin-1/2 degree of freedom at each vertex. The Hamiltonian is given by
\begin{equation}
    H=-J_x\sum_{x-\rm link}\sigma_i^x\sigma_j^x -J_y\sum_{y-\rm link}\sigma_i^y\sigma_j^y-J_z\sum_{z-\rm link}\sigma_i^z\sigma_j^z
\end{equation}
where the $x,y,z$ links correspond to the three different directions of edges on the Honeycomb lattice. The model can be solved exactly by mapping it to free Majorana fermions. One defines
\begin{equation}
    \sigma_i^\alpha = ic_ib_i^\alpha. 
\end{equation}
The mapping preserves the commutation relation of the operators, but one has to impose an additional gauge constraint on the Hilbert space, namely
\begin{equation}
    G_i=b_i^xb_i^yb_i^zc_i=1.
\end{equation}
Using this transformation, the Hamiltonian becomes
\begin{equation}
\begin{aligned}
    H&=\frac{i}{4}\sum_{ij}\hat{A}_{ij} c_i c_j\\
    \hat{A}_{ij}&=2J_{\alpha_{ij}}\hat{u}_{ij}; & \hat{u}_{ij}&=b_i^{\alpha_{ij}}b_j^{\alpha_{ij}}
\end{aligned}
\end{equation}
where the sum is carried over all adjacent pairs $ij$, and we have $\hat{u}_{ij}=-\hat{u}_{ji}$. The operators $\hat{u}_{ij}$ (the gauge field) commute with the Hamiltonian, so the ground state is found by fixing a certain configuration of the $u_{ij}$, then imposing the gauge constraint by summing over all configurations related by the action of $G_i$. We choose $u_{ij}=1$ when $i$ belongs to the even sublattice and $j$ belongs to the odd sublattice. This fixes the Hamiltonian for the $c$ fermions as
\begin{equation}
\begin{aligned}
    H&=\frac{i}{4}\sum_{ij}A_{ij} c_i c_j,\\
    A_{ij}&=\pm2J_{\alpha_{ij}}u_{ij} ,
    \label{eq: h-majorana}
\end{aligned}
\end{equation}
where the $\pm$ sign depends on whether $i$ is in the odd or even sublattice. The ground state of the model is given by summing over all equivalent gauge configurations for the ground state of \eqref{eq: h-majorana}, that is,
\begin{equation}
    \ket{\psi}=\frac{1}{N}\qty(\prod_i(1+G_i))\ket{\psi_0}
    \label{eq: psi-gauge}
\end{equation}

To obtain the Ising phase of the model, we need to break time reversal symmetry by adding a ``magnetic field" term. This is most simply achieved by adding next-nearest-neighbor interactions in the fermionic Hamiltonian. That is, we add
\begin{equation}
    H_{\rm mag}=-\frac{i}{2}\sum_{ij\textrm{ nnn}}\pm K c_i c_j
\end{equation}
where the gauge has been fixed as described above, and the sign convention is as depicted in Eq.\ (48) in \cite{kitaev2006anyons}.

\tocignoredsubsection{Implementation of the multi-entropy measures}
We want to measure operators of the form \eqref{eq: multi-ent} on the ground state of the Kitaev model. This can be achieved by representing the wavefunction as a sum over gauge configurations using \eqref{eq: psi-gauge} and by implementing the permutation operators as fermionic operators. To do so, we first consider the operator that exchanges two spins. It is given by
\begin{equation}
    P_{12}=\frac{1}{2}(1+{\bm \sigma}_1\cdot{\bm \sigma}_2).
\end{equation}
In terms of the fermion operators, we have
\begin{equation}
    P_{12}=\exp(\frac{\pi}{4}c_1c_2+\frac{\pi}{4}\sum_\alpha b_1^\alpha b_2^\alpha).
    \label{eq: perm-fermions}
\end{equation}
Since any permutation operator can be written as a product of exchanges of two spins, we can use \eqref{eq: perm-fermions} to write the operator $\pi_A\pi_B\pi_C$ as a product of exponents of fermionic bilinears. We can further commute the projection operators through the permutations to obtain
\begin{equation}
    \mathcal{M}=\frac{\mel{\psi_0^{\otimes R}}{\qty(\prod_{i,r}\frac{1+G_{i,r}}{2})\pi_A\pi_B\pi_C}{\psi_0^{\otimes R}}}{\qty(\mel{\psi_0}{\qty(\prod_{i,r}\frac{1+G_{i,r}}{2})}{\psi_0})^R}
    \label{eq: m-fermions1}
\end{equation}
where $R$ is the number of replicas, and the gauge transformations are labeled by the site and replica indices $i,r$. Importantly, gauge configurations in which $G_i$ acts only on some of the sites inside a region, but not on others, do not contribute. We can then define for a region
\begin{equation}
    G_I=\prod_{i\in I} G_i,
\end{equation}
such that \eqref{eq: m-fermions1} becomes, after canceling the denominator,
\begin{equation}
    \mathcal{M}=\mel{\psi_0^{\otimes R}}{\qty(\prod_{I,r}(1+G_{I,r}))\pi_A\pi_B\pi_C}{\psi_0^{\otimes R}}.
    \label{eq: m-fermions2}
\end{equation}
The evaluation of $\mathcal{M}$ becomes then a sum of gaussian terms, such that the number of summands depends on $R$, but not on the system size. This allows us to carry out the calculation in time polynomial in the system size. The calculation of \eqref{eq: m-fermions2} is still exponential in $R$, as we need to consider $2^{4R}$ summands. We can reduce this number by applying the following considerations:
\vspace{5mm}

\begin{enumerate}
    \item Since $G_AG_BG_CG_\Lambda\ket{\psi}=\ket\psi$, we can ignore summands where $G_{\Lambda,r}$ are applied, since any such configuration is equivalent to one where $G_{\Lambda,r}$ are not applied. This immediately reduces the number of summands to evaluate to $2^{3R}$.
    \item A less trivial property is that the summand vanishes under the following condition: denoting $\pi_{IJ}=\pi_I\pi_J^{-1}$ for the ``boundary permutation" between the regions $I,J$, and considering the orbits $o_{\pi_{IJ}}(r)$ to be the orbit of replica $r$ under $\pi_{IJ}$. If, for any orbits, we have
    \begin{equation}
        \sum_{r'\in o_{\pi_{IJ}}(r)} \delta_{G_{I,r'}\textrm{ is applied}}+ \delta_{G_{J,r'}\textrm{ is applied}}\neq0\mod 2,
        \label{eq: vanishing summands}
    \end{equation}
    then the summand vanishes. We found that employing this condition significantly reduces the evaluation time of \eqref{eq: m-fermions2}. We prove this property at the end of this section.
\end{enumerate}
\tocignoredsubsection{Evaluation of fermionic gaussian operators}
To evaluate the summands in \eqref{eq: m-fermions2}, we use two properties of fermionic states. The first is that the product of a fermionic gaussian operator satisfies
\begin{equation}
    e^{\frac{1}{4}c^TAc}e^{\frac{1}{4}c^TBc}=e^{\frac{1}{4}c^TMc}\,, \quad e^Ae^B=e^M\,.
    \label{eq:expmat}
\end{equation}
To show this, we use the fact that, for any anti-symmetric matrices $A$, $B$, 
\begin{equation}
\left[\frac{1}{4}c^T A c,\frac{1}{4}c^T B c\right] = \frac{1}{4}c^T [A,B]c.
\label{eq:comm}
\end{equation}
Applying the Baker-Campbell-Hausdorff formula to the left hand side of \eqref{eq:expmat} and using Eq.\ \eqref{eq:comm} for the resulting repeated commutators, we obtain Eq.~\eqref{eq:expmat}. 
This allows us to write each summand of \eqref{eq: m-fermions2} in the form $\ev{e^{\frac{1}{4}c^TMc}}$ for some matrix $M$. The second property is that if $\psi_0$ is the ground state of $H=\frac{1}{4}\sum_{ij}H_{ij}c_ic_j$, we have (see derivation below)
\begin{equation}
    \mel{\psi}{e^{\frac{1}{4}c^TMc}}{\psi}=\qty[\det(U^\dagger e^MU)_{11}]^{1/2}
    \label{eq: eval-m}
\end{equation}
where $U$ diagonalizes $H$, and the subscript $(*)_{11}$ indicates that we only take the block of the matrix that corresponds to the negative eigenvalues of $H$. 

We note that Eq. (\ref{eq: eval-m}) contains an ambiguity, since the sign on the RHS is undetermined. In some cases, the sign ambiguity can be fixed using the methods of Ref. \cite{han2024pfaffian}. However, we found that the operators we evaluate here violate the assumptions of \cite{han2024pfaffian}. Namely, the authors assumed that all the gaussian fermionic operators have non-vanishing trace, an assumption that is violated by the gauge operators $G_i$. Instead, here we fix the sign by carrying out every calculation in the limit $J_x=J_y=1,J_z\ll1$, where the system is in the toric code phase and the answer is known, and adiabatically moving to the point of interest. Practically, we found that, since the phase factors we are interested in are small, the default numerical choice of sign was always correct.

\bigskip
\noindent{\emph{Proof of \eqref{eq: eval-m}}} We begin with the following result on Majorana operators 
\begin{equation}
    \Tr(e^{\frac{1}{4}c^TMc})=\det(1+e^M)^{1/2}.
\end{equation}
The value $C=\ev{e^{\frac{1}{4}c^TMc}}$ can then be evaluated as
\begin{equation}
\begin{aligned}
    C&=\lim_{\beta\to\infty}\frac{\Tr(e^{\frac{1}{4}c^TMc}e^{\frac{\beta}{4}c^THc})}{\Tr(e^{\frac{1}{4}c^TMc})} \\
    &=\frac{\det(1+e^Me^{-\beta H})^{1/2}}{\det(1+e^{-\beta H})^{1/2}}
\end{aligned}
\end{equation}
To evaluate the numerator we diagonalize $H$ as 
\begin{equation}
   \begin{aligned}
       H&=UDU^\dagger, \\
       D&={\rm diag}(-E_1,\ldots, -E_N,E_1,\ldots,E_N)
   \end{aligned}
\end{equation}
with $E_i>0$. In the limit $\beta\to\infty$ we have
\begin{equation*}
\begin{aligned}
    \det(1+e^Me^{-\beta H})&=\det(1+(U^\dagger e^MU)e^{-\beta D})\\
    &\approx\begin{pmatrix}
        \mathbb{1}+(U^\dagger e^MU)_{11}e^{-\beta D_{11}}& 0\\
        (U^\dagger e^MU)_{21}e^{-\beta D_{11}} & \mathbb{1} 
    \end{pmatrix} \\
\end{aligned}
\end{equation*}
where the index $(*)_{11}$ denotes the block of the first $N$ eigenvalues of $H$ (the negative ones). We then find
\begin{equation}
    \det(1+e^Me^{-\beta H})=\det(U^\dagger e^M U)\det(e^{-\beta D_{11}}).
\end{equation}
The second determinant cancels with the denominator, and we obtain \eqref{eq: eval-m}.

\bigskip

\tocignoredsubsection{Vanishing of summands satisfying \eqref{eq: vanishing summands}}
We finish by proving that summands in \eqref{eq: m-fermions2} satisfying \eqref{eq: vanishing summands}. We simplify the problem by assuming that $\pi_J=\rm id$, this is because we can multiply the wavefunction by $\pi_J^{-1}$ acting on all regions. We can further assume that no gauge transformations are applied on region $J$, using the fact that $G_AG_BG_CG_\Lambda\ket{\psi_0}=\ket{\psi_0}$. Finally, we can consider the case in which the permutation $\pi_I$ is a cyclic permutation on $R$ replicas, since the condition \eqref{eq: vanishing summands} can be checked for each cycle of $\pi_I$ separately.

To obtain the condition \eqref{eq: vanishing summands}, we consider two gauge degrees of freedom that are cut between the regions $I,J$
\begin{equation}
\begin{aligned}
    u_{IJ,1}&=ib_{I,1}b_{J,1}\\
    u_{IJ,2}&=ib_{I,2}b_{J,2}
\end{aligned}
\end{equation}
We assume that $\ket{\psi_0}$ satisfies $u_{IJ,\alpha}=+1$ (we will consider $\ket{\psi_0}$ to be a state only on these degrees of freedom; this is sufficient to give the required condition). Following \cite{yao2010entanglement}, we consider the ``hybridized" degrees of freedom $w_I,w_J$, given by
\begin{equation}
    w_{I}=ib_{I,1}b_{I,2}
\end{equation}
and similar for $w_J$. In terms of the eigenstates of $w_I,w_J$, we have
\begin{equation}
    \ket{\psi_0}=\frac{1}{\sqrt{2}}\qty(\ket{+1,+1}+\ket{-1,-1}).
\end{equation}
In addition, the gauge transformation $G_I$ acts as
\begin{equation}
    G_I\ket{w_I,w_J}=(-1)^{w_I}\ket{w_I,w_J}.
\end{equation}
We then have
\begin{equation}
\begin{aligned}
    &\mel{\psi_0^{\otimes R}}{\qty(\prod_{r=1}^RD_{I,r}^{\alpha_r})\pi_I}{\psi_0^{\otimes R}} \\
    =&\frac{1}{2^R}\qty(\bra{1,1}\ket{1,1}+(-1)^{\sum_r\alpha_r}\bra{-1,-1}\ket{-1,-1}) \\
    =&\frac{1}{2^{R-1}}\delta_{\sum_r\alpha_r}
\end{aligned}
\end{equation}
where the indices $\alpha_r$ indicate whether $G_{I,r}$ is applied, and the delta requires that $\sum_r\alpha_r=0\mod 2$. This completes the argument.

\bigskip 

\section{Spurious contributions to $J_n$}
\label{app: spurious}
Here we use the toy model in Ref.~\cite{Gass_Levin_2024} to assess the spurious contribution to $J_n$. There, the authors constructed a model by decorating a two-dimensional model with a one-dimensional modified cluster state on the edge of the region $ABC$. They showed that the model can be constructed such that its reduced density matrices are given by
\begin{equation}
    \begin{aligned}
        \rho_{AB}&=\frac{1}{2^{N_{AB}}}(\mathbb{1}+\alpha P_{AB}) \\
        \rho_{AB}&=\frac{1}{2^{N_{AC}}}(\mathbb{1}+\alpha P_{BC}) \\
        \rho_{ABC}&=\frac{1}{2^{N_{ABC}}}(\mathbb{1}+\alpha P_{AB}+\beta P_{BC}+i\gamma P_{AB}P_{AC})
    \end{aligned}
\end{equation}
where $P_{AB},P_{AC}$ are Pauli operators supported on $AB,AC$ and satisfying $\qty{P_{AB},P_{AC}}=0$ and $\alpha,\beta,\gamma$ are constants which, in their construction, are given by $\alpha=\beta=\gamma=\frac{1}{\sqrt{3}}$.

Defining
\begin{equation}
\begin{aligned}
    \lambda_n(\alpha)&=\frac{1}{2}((1+\alpha)^n+(1-\alpha)^n)\\
    \kappa_n(\alpha)&=\frac{1}{2}((1+\alpha)^n-(1-\alpha)^n), 
\end{aligned}
\end{equation}
we obtain
\begin{equation}
    \rho_{AB}^n=\frac{1}{2^{nN_{AB}}}(\lambda_n(\alpha)\mathbb{1}+\kappa_n(\alpha)P_{AB}),
\end{equation}
and similarly for $P_{AC}$. In particular, we see that the construction of \cite{Gass_Levin_2024} gives a nontrivial contribution to the phase of $J_n$:
\begin{equation}
\begin{aligned}
    \tan & (\arg(J_n)) = \\
    &\frac{\gamma\kappa_n(\alpha)\kappa_n(\beta)}{\lambda_n(\alpha)\lambda_n(\beta)+\alpha\lambda_n(\beta)\kappa_n(\alpha)+\beta\lambda_n(\alpha)\kappa_n(\beta)}
\end{aligned}
\end{equation}
Interestingly, the spurious contribution has a different functional form than that in \eqref{eq: j_n phases}.

\section{LeSME in 1+1D}
\label{app: 1+1d lesme}
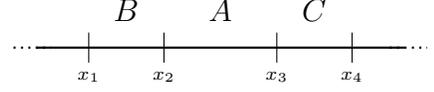
\begin{figure}
    \centering
    \begin{tikzpicture}
        \draw[thick](-.7,0) -- ++ (4.9,0);
        \draw (0,-.2) node[below,font={\scriptsize}] {$x_1$} -- ++(0,.4);
        \draw (1,-.2) node[below,font={\scriptsize}] {$x_2$} -- ++(0,.4);
        \draw (2.5,-.2) node[below,font={\scriptsize}] {$x_3$} -- ++(0,.4);
        \draw (3.5,-.2) node[below,font={\scriptsize}] {$x_4$} -- ++(0,.4);
        \draw[thick,dotted] (-1.,0) -- (-.5,0);
        \draw[thick,dotted] (4,0) -- ++ (.5,0);
        \node[font={\large}] at (.5,.5) {$B$};
        \node[font={\large}] at (1.75,.5) {$A$};
        \node[font={\large}] at (3.,.5) {$C$};
    \end{tikzpicture}
    \caption{The geometry used for the calculation of the LeSME $\Phi_{r,l}$ in 1+1D CFT}
    \label{fig: 1d-geom}
\end{figure}
Here we consider the analog of the LeSME \eqref{eq: phi_r-def} for the ground state of a conformal field theory in 1+1D. We consider the geometry outlined in Fig.\ \ref{fig: 1d-geom} on the infinite line, we consider a slight generalization of the permutations defined in \eqref{eq: phi_r-def}, where $\pi_{A,B,C}$ are given by
\begin{equation}
    \Phi_{r,l}:\;\begin{cases}
    \pi_A(1,t) = (1,t-1); & \pi_A(2,t) = (2,t+1), \\
    \pi_B(s,t) = (s+1,t), \\ 
    \pi_C(1,t) = (2,t-l); & \pi_C(2,t) = (1,t+l)    
    \end{cases},
    \label{eq: phi_r,l-def}
\end{equation}

The general scheme for calculating such quantities was presented in \cite{calabrese2004entanglement}: we consider the vacuum expectation value of the theory defined on $R=2r$ replicas of the twist operators that exchange the replicas, located at the points $x_i$. For any $l=0,\ldots,r-1$ we have
\begin{equation}
    \mathcal{O}_{r,l}(x_1,x_2,x_3,x_4)=\ev{\s_1(x_1) \s_2(x_2)\s_3(x_3)\s_4(x_4)}
    \label{eq: cft-twist-def}
\end{equation}
where the permutation operators on the twists are defined as
\begin{equation}
    \begin{aligned}
        \s_1&=\per{\b}, \\
        \s_2&= \per{\a}\per{\b}^{-1} ,\\
        \s_3&=\per{\a}^l\per{\b}\per{\a}^{-1}=\per{\a}^{l+1}\per{\b} ,\\
        \s_4&=\per{\b}\per{\a}^{-l}.
    \end{aligned}
    \label{eq: sigma-defs}
\end{equation}
We note for later that each of $\sigma_i$ is composed of $r$ disjoint cycles of length 2. The general scheme for calculating such correlators was presented in \cite{lunin_correlation_2001}. We should first find the Riemann surface $\Sigma$ that is obtained as a branched cover of the sphere with ramification points at $x_i$ corresponding to the permutation operators \eqref{eq: sigma-defs}. A point $p$ of the sphere other than $x_i$ has $2r$ pre-images $p_j$ on $\Sigma$. The cover should be defined such that a curve that circles $x_i$ on the sphere, starting and ending in $p$ has a pre-image in $\Sigma$ that starts at $p_j$ and ends in $p_{\sigma_i(j)}$. The genus of $\Sigma$ can be obtained using the Riemann–Hurwitz formula as \cite{lunin_correlation_2001,liu_multipartite_2024}
\begin{equation}
    g=\sum_{p\in \Sigma}(e_p-1)-R+1
\end{equation}
where $R$ is the covering number ($2r$ in our case), and $e_p$ is the ramification index at $p$. That is, the number such that the mapping from $\Sigma$ looks like $(z-p)^{e_p}$ locally around $p$. In our case, since all of $\s_i$ are comprised of $r$ disjoint 2-cycles, we have $4r$ points with $e_p=2$. This gives $g=1$, so $\Sigma$ is a torus, making the calculation tractable for general values of $x_i$. 


For simplicity, we use conformal invariance to write
\begin{align}
    \label{eq: Oeta}
        \mathcal{O}_{r,l}(\eta)&=\ev{\s_1(0)\s_2(1)\s_3(\eta)\s_4(\infty)}, \\
    \label{eq: eta-def}
        \eta &=\frac{(x_3-x_1)(x_4-x_2)}{(x_4-x_3)(x_2-x_1)}.
\end{align}
The general value is obtained by conformal transformations, and $1<\eta<\infty$. Locally around the preimages of the points $0,1,\eta$, the map $f$ from $\Sigma$ to the sphere looks like $(z-p)^2$. This means that $f$ satisfies a differential equation of the form
\begin{equation}
    f'^2(z)=Cf(z)\qty(f(z)-1)\qty(f(z)-\eta)
\end{equation}
where $C$ is some proportionality constant that we can set to $1$ by rescaling $z$. This differential equation has a unique solution (up to rescaling of $z$) given by 
\begin{equation}
\begin{aligned}
    f(z)&=\wp(z|g_2,g_3)+\frac{l+1}{3}, \\
    g_2 &= \frac{4}{3}(\eta^2-\eta+1), \\
    g_3 &= \frac{4}{27}(2\eta^3-3\eta^2-3\eta+2),
\end{aligned}
\end{equation}
where $\wp$ is the Weierstrass elliptic function specified by the invariants $g_2,g_3$, which can be written in terms of $\omega_1,\omega_2$. The map $f$ is doubly-periodic on a lattice with lattice vectors $\w_1,\w_2$, and therefore maps the torus $\mathbb{C}/(\Z\omega_1+\Z\omega_2)$ to the sphere. The lattice can be chosen such that $\w_1$ is real and $\w_2$ is purely imaginary. The modulus $\tau_0$ satisfies \cite{donaldson_riemann_2011}
\begin{equation}
    J(\tau_0)=\frac{(\eta^2-\eta+1)^3}{\eta^2(\eta-1)^2},
\end{equation}
where $J$ is the Klein absolute invariant, as defined in \cite{weisstein2002klein}. In particular, $\tau_0$ can be chosen to be pure imaginary for $\eta>1$. We mention the following asymptotic behavior of the above result. We have
\begin{equation}
    -i\tau_0=\begin{cases}
        \frac{1}{2\pi}\log(256\eta^2) & \eta\to\infty,\\
        \frac{2\pi}{\log(\frac{256}{(\eta-1)^2})} & \eta\to 1.
    \end{cases}
\end{equation}
Note that the surface $\Sigma$ is \textit{not} the torus given by $\mathbb{C}/(\Z\w_1+\Z\w_2)$, since that is only a double cover of the sphere under the map $f$. Instead, we need to have 
\begin{equation}
\begin{aligned}
    \Sigma&=\mathbb{C}/\Lambda, \\
    \Lambda&=\Z(a\w_1+b\w_2)+\Z(c\w_1+d\w_2),
\end{aligned}
\label{eq: lattice}
\end{equation}
where $ad-bc=r$. The choice of $a,b,c,d$ should be such that the twist operators behave as in \eqref{eq: sigma-defs}. The expectation value \eqref{eq: Oeta} will be proportional to the CFT partition function on the torus with modulus $\tau=(a\tau_0+b)/(c\tau_0+d)$. To determine $\tau$, we need to consider the preimages in $\Sigma$ of paths surrounding the twist defects $x_i$. 

\tocignoredsubsection{Determining the modulus}
\renewcommand{\e}{\epsilon}
As the next step, we need to obtain the lattice $\Lambda$ in \eqref{eq: lattice}. To do so, we must consider how paths on $\CC$ that surround the defects lift under $f^{-1}$ to paths on $\CC/\Lambda$. For small $\epsilon'>0$, we have
\begin{align}
    f^{-1}(-\epsilon')&\approx\w_2/2\pm i\epsilon+n\w_1+k\w_2, & n,k&\in\Z,
    \label{eq: preimages}
\end{align}
where $\epsilon\propto\sqrt{\epsilon'}$. The permutations $\pi_{(i)}$ act on the points in \eqref{eq: preimages} as follows: We consider a path $\Gamma$ that starts in one point $p$, goes through the branch cuts $A,B$ or $C$, and returns to $p$, then set $\pi_{(i)}$ to be the end point of $f^{-1}(\Gamma)$. We must eventually identify each replica $(s,t)$ with a set of points in \eqref{eq: preimages}, such that the actions of the replica permutation match the action via the lifting of paths.

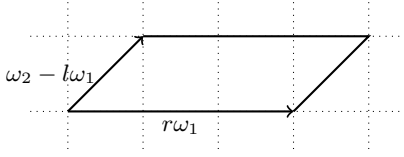
\begin{figure}
    \centering
    \begin{tikzpicture}
        \draw[thick,->] (0,0) -- node[left] {$\w_2-l\w_1$} (1,1);
        \draw[thick,->] (0,0) -- node[below] {$r\w_1$} (3,0);
        \draw[thick] (1,1) -- (4,1) -- (3,0);
        \foreach \y in {0,1}
            \draw[dotted] (-.5,\y) -- (4.5,\y);
        \foreach \x in {0,...,4}
            \draw[dotted] (\x,-.5) --++ (0,2);
        
    \end{tikzpicture}
    \caption{The cover of the Riemann sphere obtained by gluing $A,B,C$ along the permutation defects.}
    \label{fig: sheared-torus}
\end{figure}

The action of the permutation operators can be checked to be (see Fig. \ref{fig: lifting})
\begin{equation}
\begin{aligned}
    \pi_B(\w_2/2\pm i\e)&=(\w_2/2\mp i\e), \\
    \pi_A(\w_2/2\pm i\e)&=(\w_2/2\pm i\e\mp\w_1), \\
    \pi_C(\w_2/2\pm i\e)&=(\w_2/2\mp i\e\mp\w_2). \\
    \label{eq: path-perm-action}
\end{aligned}
\end{equation}
We see that to satisfy the algebra of $\pi_i$, we must identify
\begin{equation}
\begin{aligned}
    z+r\w_1&\sim z, \\
    z+ \w_2&\sim z+l\w_1.
\end{aligned}
\end{equation}
This means that the lattice is given by (see Fig.\ \ref{fig: sheared-torus})
\begin{equation}
    \Lambda=\mathbb{Z}r\w_1+\mathbb{Z}(\w_2-l\w_1).
\end{equation}
This determines the modulus of $\Sigma$ as
\begin{equation}
    \tau=\frac{\tau_0}{r}-\frac{l}{r}.
\end{equation}
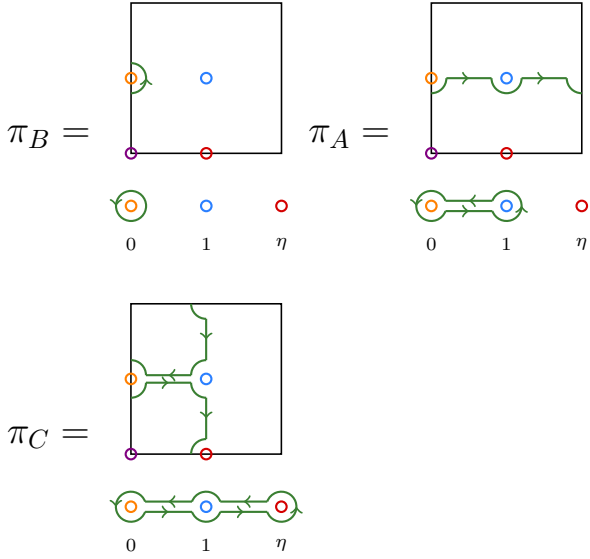
\begin{figure}
    \centering
    \begin{tikzpicture}
        \def\shf{-1.7};
        \newcommand{\sqr}{
            \draw[semithick] (-1,-1) -- (-1,1) -- (1,1) -- (1,-1) -- cycle;
            \draw[thick,violet] (-1,-1) circle (.07);
            \draw[thick,orange] (-1,0) circle (.07);
            \draw[thick,lightblue] (0,0) circle (.07);
            \draw[thick,darkred] (0,-1) circle (.07);

            \draw[thick,orange] (-1,\shf) circle (.07) node[black,below=.25] {\scriptsize $0$};
            \draw[thick,lightblue] (-0,\shf) circle (.07) node[black,below=.25] {\scriptsize $1$};
            \draw[thick,darkred] (1,\shf) circle (.07) node[black,below=.25] {\scriptsize $\eta$};
            }
        \node at (0,0) {
        \Large $\pi_B=\vcenter{\hbox{
        \begin{tikzpicture}
        \sqr
        \draw[OliveGreen,thick,midarr] (-1,-.2) arc (-90:90:.2);
        \draw[midarr,OliveGreen,thick] (-1,\shf) circle (.2);
        \end{tikzpicture}}}$
        };
        \node at (4,0) {
        \Large $\pi_A=\vcenter{\hbox{
        \begin{tikzpicture}
            \sqr 
            \begin{scope}[every path/.append style={thick,OliveGreen}]
            \draw (-1,-.20) arc (-90:0:.2);
            \draw (-.20,-.00) arc (180:360:.2);
            \draw (1-.2,-.00) arc (180:270:.2);
            \draw[midarr] (-.8,0) -- (-.2,0);
            \draw[midarr] (.2,0) -- (.8,0);
            \draw[midarr] (-1,\shf)++(20:.2) arc (20:340:.2);
            \draw[midarr] (0,\shf)++(180+20:.2) arc (180+20:360+180-20:.2);
            \draw[midarr] (-1,\shf)++(-20:.2) -- ($ (0,\shf)+(200:.2) $);
            \draw[midarr] (0,\shf)++(160:.2) -- ($ (-1,\shf)+(20:.2) $);
            \end{scope}
        \end{tikzpicture}}}$
        };
        
        \node at (0,-4) {
        \Large $\pi_C=\vcenter{\hbox{
        \begin{tikzpicture}
            \sqr 
            \begin{scope}[every path/.append style={thick,OliveGreen}]
            \draw (-1,-.25) arc (-90:0:.2) ;
            \draw (-.20,-.05) arc (180:270:.2) ;
            \draw (0,-1+.2) arc (90:180:.2);
            \draw (-.2,1) arc(180:270:.2) ; 
            \draw (0,.25) arc(90:180:.2); 
            \draw (-1+.2,.05) arc(0:90:.2);
            \draw[midarr] (-.8,-.05) --++ (.6,0);
            \draw[midarr] (-.2,.05) --++ (-.6,0);
            \draw[midarr] (0,-.25) -- (0,-.8);
            \draw[midarr] (0,.8) -- (0,.25);

            \draw[midarr] (-1,\shf)++(20:.2) arc (20:340:.2);
            \draw[midarr] (1,\shf)++(180+20:.2) arc (180+20:360+180-20:.2);
            \draw (0,\shf) ++ (20:.2) arc(20:160:.2);
            \draw (0,\shf) ++ (-20:.2) arc(-20:-160:.2);
            \draw[midarr] (-1,\shf)++(-20:.2) -- ($ (0,\shf)+(200:.2) $);
            \draw[midarr] (0,\shf)++(160:.2) -- ($ (-1,\shf)+(20:.2) $);
            \draw[midarr] (0,\shf)++(-20:.2) -- ($ (1,\shf)+(200:.2) $);
            \draw[midarr] (1,\shf)++(160:.2) -- ($ (0,\shf)+(20:.2) $);
            \end{scope}
        \end{tikzpicture}}}$
        };
    \end{tikzpicture}
    \caption{Lifting of paths surrounding the defects $\sigma_i$ under the mapping $f^{-1}$. Each point on the Riemann sphere has $2r$ pre-images on the torus $\CC/\Lambda$. Paths that surround $\sigma_i$ permute the $2r$ copies, via the actions \eqref{eq: path-perm-action}, obtained by the lifting of paths.}
    \label{fig: lifting}
\end{figure}

\tocignoredsubsection{Final result}
Following \cite{lunin_correlation_2001} (see also \cite{harper_multi-entropy_2024}), the final result is obtained by combining the partition function on the torus with modulus $\tau$ and the terms arising from the Weyl anomaly. Since the torus with modulus $\tau_0$ is covered $r$ times, we get
\begin{equation}
    \mathcal{O}_{r,l}(\eta)=\qty[\frac{2^8\eta}{\eta-1}]^{\frac{rc}{12}}Z\qty(\frac{\tau_0}{r}-\frac{l}{r}),
\end{equation}
where $Z(\tau)$ is the partition function of the CFT on the torus with modulus $\tau$, and $c$ is the central charge of the chiral sector of the theory (we have implicitly assumed that $c_-=0$). Notice that the $l$ dependence enters only through the contribution to $Z$. This is because it does not modify the conformal weights of the defects nor the OPE coefficients. By conformal transformation, we obtain the general result
\begin{equation}
\begin{aligned}
\mathcal{O}_{r,l}(x_1,x_2,x_3,x_4)=2^{-\frac{4}{3}c}(x_{12}x_{13}x_{14}x_{23}x_{24}x_{34})^{-\frac{rc}{12}}\\
\times Z\qty(\frac{\tau_0}{r}-\frac{l}{r}).
\end{aligned}
\label{eq: final-cft-o}
\end{equation}
where $x_{ij}=x_j-x_i$ and $\tau_0$ are obtained from the cross-ratio \eqref{eq: eta-def}. Note that \eqref{eq: final-cft-o} depends on a choice of normalization of the operators $\sigma_i$. When comparing to numerical calculations, the result will be corrected by a constant (real and positive) multiplicative factor, which is nonuniversal. The functional dependence on $x_i$ will, however, remain the same.

\bibliography{bibliography}

\end{document}

%% file: M_with_g3_boundary.pdf_tex
\begingroup%
  \makeatletter%
  \providecommand\color[2][]{%
    \errmessage{(Inkscape) Color is used for the text in Inkscape, but the package 'color.sty' is not loaded}%
    \renewcommand\color[2][]{}%
  }%
  \providecommand\transparent[1]{%
    \errmessage{(Inkscape) Transparency is used (non-zero) for the text in Inkscape, but the package 'transparent.sty' is not loaded}%
    \renewcommand\transparent[1]{}%
  }%
  \providecommand\rotatebox[2]{#2}%
  \newcommand*\fsize{\dimexpr\f@size pt\relax}%
  \newcommand*\lineheight[1]{\fontsize{\fsize}{#1\fsize}\selectfont}%
  \ifx\svgwidth\undefined%
    \setlength{\unitlength}{96.39660572bp}%
    \ifx\svgscale\undefined%
      \relax%
    \else%
      \setlength{\unitlength}{\unitlength * \real{\svgscale}}%
    \fi%
  \else%
    \setlength{\unitlength}{\svgwidth}%
  \fi%
  \global\let\svgwidth\undefined%
  \global\let\svgscale\undefined%
  \makeatother%
  \begin{picture}(1,0.71102518)%
    \lineheight{1}%
    \setlength\tabcolsep{0pt}%
    \put(0,0){\includegraphics[width=\unitlength,page=1]{M_with_g3_boundary.pdf}}%
    \put(0.42358346,0.63788482){\color[rgb]{0,0,0}\transparent{0.04313732}\makebox(0,0)[lt]{\lineheight{1.25}\smash{\begin{tabular}[t]{l}$M$\end{tabular}}}}%
  \end{picture}%
\endgroup%

%% file: g_3_boundary_with_charge.pdf_tex
\begingroup%
  \makeatletter%
  \providecommand\color[2][]{%
    \errmessage{(Inkscape) Color is used for the text in Inkscape, but the package 'color.sty' is not loaded}%
    \renewcommand\color[2][]{}%
  }%
  \providecommand\transparent[1]{%
    \errmessage{(Inkscape) Transparency is used (non-zero) for the text in Inkscape, but the package 'transparent.sty' is not loaded}%
    \renewcommand\transparent[1]{}%
  }%
  \providecommand\rotatebox[2]{#2}%
  \newcommand*\fsize{\dimexpr\f@size pt\relax}%
  \newcommand*\lineheight[1]{\fontsize{\fsize}{#1\fsize}\selectfont}%
  \ifx\svgwidth\undefined%
    \setlength{\unitlength}{80.06528839bp}%
    \ifx\svgscale\undefined%
      \relax%
    \else%
      \setlength{\unitlength}{\unitlength * \real{\svgscale}}%
    \fi%
  \else%
    \setlength{\unitlength}{\svgwidth}%
  \fi%
  \global\let\svgwidth\undefined%
  \global\let\svgscale\undefined%
  \makeatother%
  \begin{picture}(1,0.25119109)%
    \lineheight{1}%
    \setlength\tabcolsep{0pt}%
    \put(0,0){\includegraphics[width=\unitlength,page=1]{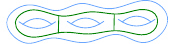}}%
    \put(0.00000026,0.10096182){\color[rgb]{0,0.50196078,0}\makebox(0,0)[lt]{\lineheight{1.25}\smash{\begin{tabular}[t]{l}\scriptsize$a_1$\end{tabular}}}}%
    \put(0.35972922,0.09188534){\color[rgb]{0,0.50196078,0}\makebox(0,0)[lt]{\lineheight{1.25}\smash{\begin{tabular}[t]{l}\scriptsize$a_2$\end{tabular}}}}%
    \put(0.45349711,0.19496055){\color[rgb]{0,0.50196078,0}\makebox(0,0)[lt]{\lineheight{1.25}\smash{\begin{tabular}[t]{l}\scriptsize$a_3$\end{tabular}}}}%
    \put(0.45487271,0.0259964){\color[rgb]{0,0.50196078,0}\makebox(0,0)[lt]{\lineheight{1.25}\smash{\begin{tabular}[t]{l}\scriptsize$a_4$\end{tabular}}}}%
    \put(0.80013899,0.07001436){\color[rgb]{0,0.50196078,0}\makebox(0,0)[lt]{\lineheight{1.25}\smash{\begin{tabular}[t]{l}\scriptsize$a_6$\end{tabular}}}}%
    \put(0.65017154,0.09617152){\color[rgb]{0,0.50196078,0}\makebox(0,0)[lt]{\lineheight{1.25}\smash{\begin{tabular}[t]{l}\scriptsize$a_5$\end{tabular}}}}%
  \end{picture}%
\endgroup%

%% file: M_with_charges.pdf_tex
\begingroup%
  \makeatletter%
  \providecommand\color[2][]{%
    \errmessage{(Inkscape) Color is used for the text in Inkscape, but the package 'color.sty' is not loaded}%
    \renewcommand\color[2][]{}%
  }%
  \providecommand\transparent[1]{%
    \errmessage{(Inkscape) Transparency is used (non-zero) for the text in Inkscape, but the package 'transparent.sty' is not loaded}%
    \renewcommand\transparent[1]{}%
  }%
  \providecommand\rotatebox[2]{#2}%
  \newcommand*\fsize{\dimexpr\f@size pt\relax}%
  \newcommand*\lineheight[1]{\fontsize{\fsize}{#1\fsize}\selectfont}%
  \ifx\svgwidth\undefined%
    \setlength{\unitlength}{85.45650993bp}%
    \ifx\svgscale\undefined%
      \relax%
    \else%
      \setlength{\unitlength}{\unitlength * \real{\svgscale}}%
    \fi%
  \else%
    \setlength{\unitlength}{\svgwidth}%
  \fi%
  \global\let\svgwidth\undefined%
  \global\let\svgscale\undefined%
  \makeatother%
  \begin{picture}(1,0.59696834)%
    \lineheight{1}%
    \setlength\tabcolsep{0pt}%
    \put(0,0){\includegraphics[width=\unitlength,page=1]{M_with_charges.pdf}}%
    \put(0.3966363,0.51446456){\color[rgb]{0,0,0}\transparent{0.04313732}\makebox(0,0)[lt]{\lineheight{1.25}\smash{\begin{tabular}[t]{l}$M$\end{tabular}}}}%
    \put(0,0){\includegraphics[width=\unitlength,page=2]{M_with_charges.pdf}}%
    \put(0.04154482,0.23749042){\color[rgb]{0,0.50196078,0}\makebox(0,0)[lt]{\lineheight{1.25}\smash{\begin{tabular}[t]{l}\scriptsize$a_1$\end{tabular}}}}%
    \put(0.3557064,0.22571897){\color[rgb]{0,0.50196078,0}\makebox(0,0)[lt]{\lineheight{1.25}\smash{\begin{tabular}[t]{l}\scriptsize$a_2$\end{tabular}}}}%
    \put(0.44355873,0.32229145){\color[rgb]{0,0.50196078,0}\makebox(0,0)[lt]{\lineheight{1.25}\smash{\begin{tabular}[t]{l}\scriptsize$a_3$\end{tabular}}}}%
    \put(0.44484754,0.16398679){\color[rgb]{0,0.50196078,0}\makebox(0,0)[lt]{\lineheight{1.25}\smash{\begin{tabular}[t]{l}\scriptsize$a_4$\end{tabular}}}}%
    \put(0.7683319,0.20522778){\color[rgb]{0,0.50196078,0}\makebox(0,0)[lt]{\lineheight{1.25}\smash{\begin{tabular}[t]{l}\scriptsize$a_6$\end{tabular}}}}%
    \put(0.61802276,0.24607264){\color[rgb]{0,0.50196078,0}\makebox(0,0)[lt]{\lineheight{1.25}\smash{\begin{tabular}[t]{l}\scriptsize$a_5$\end{tabular}}}}%
  \end{picture}%
\endgroup%

%% file: pops_w_seams_1.pdf_tex
\begingroup%
  \makeatletter%
  \providecommand\color[2][]{%
    \errmessage{(Inkscape) Color is used for the text in Inkscape, but the package 'color.sty' is not loaded}%
    \renewcommand\color[2][]{}%
  }%
  \providecommand\transparent[1]{%
    \errmessage{(Inkscape) Transparency is used (non-zero) for the text in Inkscape, but the package 'transparent.sty' is not loaded}%
    \renewcommand\transparent[1]{}%
  }%
  \providecommand\rotatebox[2]{#2}%
  \newcommand*\fsize{\dimexpr\f@size pt\relax}%
  \newcommand*\lineheight[1]{\fontsize{\fsize}{#1\fsize}\selectfont}%
  \ifx\svgwidth\undefined%
    \setlength{\unitlength}{69.15418309bp}%
    \ifx\svgscale\undefined%
      \relax%
    \else%
      \setlength{\unitlength}{\unitlength * \real{\svgscale}}%
    \fi%
  \else%
    \setlength{\unitlength}{\svgwidth}%
  \fi%
  \global\let\svgwidth\undefined%
  \global\let\svgscale\undefined%
  \makeatother%
  \begin{picture}(1,1.42566522)%
    \lineheight{1}%
    \setlength\tabcolsep{0pt}%
    \put(0,0){\includegraphics[width=\unitlength,page=1]{pops_w_seams_1.pdf}}%
    \put(0.39911819,0.6149684){\color[rgb]{0.16470588,0.49803922,1}\transparent{0.94509798}\makebox(0,0)[lt]{\lineheight{1.25}\smash{\begin{tabular}[t]{l}$\tau_i$\end{tabular}}}}%
    \put(-0.00600155,1.29472218){\color[rgb]{0,0,0}\makebox(0,0)[lt]{\lineheight{1.25}\smash{\begin{tabular}[t]{l}$l_1$\end{tabular}}}}%
    \put(0.78781363,1.25872984){\color[rgb]{0,0,0}\makebox(0,0)[lt]{\lineheight{1.25}\smash{\begin{tabular}[t]{l}$l_2$\end{tabular}}}}%
    \put(-0.00600155,0.12993972){\color[rgb]{0,0,0}\makebox(0,0)[lt]{\lineheight{1.25}\smash{\begin{tabular}[t]{l}$l_3$\end{tabular}}}}%
    \put(0.81176004,0.09394771){\color[rgb]{0,0,0}\makebox(0,0)[lt]{\lineheight{1.25}\smash{\begin{tabular}[t]{l}$l_4$\end{tabular}}}}%
    \put(0.10163648,0.69107635){\color[rgb]{0,0,0}\makebox(0,0)[lt]{\lineheight{1.25}\smash{\begin{tabular}[t]{l}$l_5$\end{tabular}}}}%
  \end{picture}%
\endgroup%

%% file: pops_w_seams_2.pdf_tex
\begingroup%
  \makeatletter%
  \providecommand\color[2][]{%
    \errmessage{(Inkscape) Color is used for the text in Inkscape, but the package 'color.sty' is not loaded}%
    \renewcommand\color[2][]{}%
  }%
  \providecommand\transparent[1]{%
    \errmessage{(Inkscape) Transparency is used (non-zero) for the text in Inkscape, but the package 'transparent.sty' is not loaded}%
    \renewcommand\transparent[1]{}%
  }%
  \providecommand\rotatebox[2]{#2}%
  \newcommand*\fsize{\dimexpr\f@size pt\relax}%
  \newcommand*\lineheight[1]{\fontsize{\fsize}{#1\fsize}\selectfont}%
  \ifx\svgwidth\undefined%
    \setlength{\unitlength}{59.06841488bp}%
    \ifx\svgscale\undefined%
      \relax%
    \else%
      \setlength{\unitlength}{\unitlength * \real{\svgscale}}%
    \fi%
  \else%
    \setlength{\unitlength}{\svgwidth}%
  \fi%
  \global\let\svgwidth\undefined%
  \global\let\svgscale\undefined%
  \makeatother%
  \begin{picture}(1,1.6703573)%
    \lineheight{1}%
    \setlength\tabcolsep{0pt}%
    \put(0,0){\includegraphics[width=\unitlength,page=1]{pops_w_seams_2.pdf}}%
  \end{picture}%
\endgroup%